\documentclass[aps, prl, twocolumn, superscriptaddress, amsmath, showpacs, tightenlines, footinbib, longbibliography]{revtex4-1}
\usepackage[colorlinks, breaklinks, citecolor=blue, linkcolor=blue,  urlcolor={blue}]{hyperref}
\usepackage{breakurl}
\usepackage{amsfonts}
\usepackage{amsmath}
\usepackage{amssymb}
\usepackage{mathrsfs}
\usepackage{epsfig}
\usepackage{graphicx}
\usepackage{bm}

\usepackage{colortbl}
\definecolor{mygray}{gray}{.9}
\definecolor{darkblue}{rgb}{1,1,.70}
\definecolor{lightblue}{rgb}{1,1,.90}

\begin{document}

\title{Nonreciprocal Photon Blockade}
\author{Ran Huang}
\affiliation{Key Laboratory of Low-Dimensional Quantum Structures and Quantum Control of Ministry of Education, Department of Physics and Synergetic Innovation Center for Quantum  Effects and Applications, Hunan Normal University, Changsha 410081, China}
\author{Adam Miranowicz}
\affiliation{Theoretical Quantum Physics Laboratory, RIKEN Cluster for Pioneering Research, Wako-shi, Saitama 351-0198, Japan}
\affiliation{Faculty of Physics, Adam Mickiewicz University, 61-614 Pozna\' n, Poland}
\author{Jie-Qiao Liao}
\affiliation{Key Laboratory of Low-Dimensional Quantum Structures and Quantum Control of Ministry of Education, Department of Physics and Synergetic Innovation Center for Quantum  Effects and Applications, Hunan Normal University, Changsha 410081, China}
\author{Franco Nori}
\affiliation{Theoretical Quantum Physics Laboratory, RIKEN Cluster for Pioneering Research, Wako-shi, Saitama 351-0198, Japan}
\affiliation{Physics Department, The University of Michigan, Ann Arbor, Michigan 48109-1040, USA}
\author{Hui Jing}
\email{jinghui73@foxmail.com}
\affiliation{Key Laboratory of Low-Dimensional Quantum Structures and Quantum Control of Ministry of Education, Department of Physics and Synergetic Innovation Center for Quantum  Effects and Applications, Hunan Normal University, Changsha 410081, China}
\date{\today}

\begin{abstract}
We propose how to create and manipulate one-way nonclassical light
via photon blockade in rotating nonlinear devices. We refer to
this effect as nonreciprocal photon blockade (PB). Specifically, we
show that in a spinning Kerr resonator, PB happens
when the resonator is driven in one direction but not the other.
This occurs because of the {Fizeau drag,} leading to a full
split of the resonance frequencies of the countercirculating
modes. Different types of purely quantum correlations, such as
single- and two-photon blockades, can emerge in different
directions in a well-controlled manner, and the transition from PB to
photon-induced tunneling is revealed as well. Our work opens up a new
route to achieve quantum nonreciprocal devices, which are crucial
elements in chiral quantum technologies or topological photonics.
\end{abstract}


\maketitle

Nonreciprocal devices, allowing the flow of light from one side
but blocking it from the other, are indispensable in a wide range
of practical applications, such as invisible sensing or cloaking,
and noise-free information processing~\cite{sounas2017non}. To
avoid the difficulties of conventional magnet-based devices (e.g.,
bulky and quite lossy at optical frequencies), nonreciprocal
optical devices have been demonstrated in recent experiments based
on nonlinear optics~\cite{Fan2012an,cao2017experimental},
optomechanics~\cite{manipatruni2009optical,shen2016experimental,bernier2017nonreciprocal},
atomic gases~\cite{wang2013optical,ramezani2018nonreciprocal}, and
non-Hermitian
optics~\cite{bender2013observation,peng2014parity,chang2014parity}.
Similar advances have also been achieved in making acoustic and
electronic one-way
devices~\cite{liang2010acoustic,popa2014non,kim2017dynamically,fleury2014sound,
Shabir2017mechanical,torrent2018nonreciprocal}. However, previous
studies have mainly focused on the \emph{classical} regimes, i.e.,
one-way control of transmission rates instead of quantum noises.
Nonreciprocal quantum devices have been explored very recently,
including one-way quantum
amplifiers~\cite{metelmann2015nonreciprocal,Lecocq2017nonreciprocal,
Kamal2017minimal,Peterson2017demonstration,malz2018quantum,shen2018reconfigurable,gu2017microwave}
and routers of thermal noises~\cite{Shabir2018manipulating}. Such
devices can find applications for quantum control of light in
chiral and topological quantum
technologies~\cite{bliokh2015quantum,bliokh2015transverse,lodahl2017chiral}.

Here we propose how to induce and control nonreciprocal
\emph{quantum} effects with rotating nonlinear devices.
Specifically, we show that photon blockade (PB), which is a purely
quantum effect, can emerge nonreciprocally in a spinning Kerr
resonator. We note that single-photon blockade (1PB), i.e.,
blockade of the subsequent photons by absorbing the first
one~\cite{tian1992quantum, leonski1994possibility,
adam96,imamoglu1997strongly}, has been demonstrated experimentally
in diverse systems from cavity or circuit
QED~\cite{Birnbaum2005,Faraon2008, reinhard2012strongly,
muller2015coherent,Snijders18,Lang2011,Hoffman2011,Vaneph18} to
cavity-free devices~\cite{peyronel2012quantum}. In view of its
important role in achieving single-photon devices, optomechanical
PB ~\cite{rabl2011photon,
nunnenkamp2011single,liao2013photon,liao2013correlated} have also
been explored, offering a way to test, e.g., the quantumness of
massive
objects~\cite{Liu10,Didier11,adam16,XinWang2016,MeiWang18}. In a
very recent experiment~\cite{hamsen2017two}, two-photon blockade
(2PB)~\cite{shamailov2010multi,adam2013two,adam14,carmichael2015breakdown,Zhu2017,Leonski1996,adam96,adam01,Leonski01}
has also been observed, opening a route for creating two-photon
devices. Thus, nonreciprocal PB devices, as studied here, together
with other nonreciprocal quantum
devices~\cite{metelmann2015nonreciprocal,Lecocq2017nonreciprocal,Kamal2017minimal,Peterson2017demonstration,malz2018quantum,shen2018reconfigurable,Shabir2018manipulating},
are expected to play a key role in quantum
engineering~\cite{harris1998photon, chang2007single,
kubanek2008two},
metrology~\cite{fattal2004entanglement,buluta2009quantum,georgescu2014quantum},
and quantum information
processing~\cite{bennett2000quantum,buluta2011natural} at the
single- or few-photon levels.

In a very recent experiment~\cite{maayani2018flying}, an optical
diode with $99.6\%$ isolation has been demonstrated by using a
spinning resonator. Inspired by this
experiment~\cite{maayani2018flying}, here we study nonreciprocal
PB in a spinning Kerr resonator. We find that {light with
\emph{sub-} or \emph{super-Poissonian} photon-number statistics}
can emerge when driving the resonator from its left or right side.
Also, by varying the parameters of the system, different quantum
correlations (i.e., 1PB or 2PB) can be achieved for the clockwise
(CW) or counterclockwise (CCW) modes, for a resonator spinning
along the CCW direction. {We note that the main idea of
nonreciprocal PB is analogous to the classical nonreciprocity
induced by the Doppler effect, which has been studied extensively
in various areas of physics (see, e.g.,
Refs.~\cite{DemokritovBook,wang2013optical,ramezani2018nonreciprocal}).
Here we focus on \emph{quantum} nonreciprocity induced by the
Fizeau light-dragging effect.} This opens up the prospect of
engineering nonreciprocal PB devices for applications in, e.g.,
unidirectional quantum sensing and quantum optical
communications~\cite{lodahl2017chiral}.

\begin{figure}[tb]
\centering
\includegraphics[width=0.48 \textwidth]{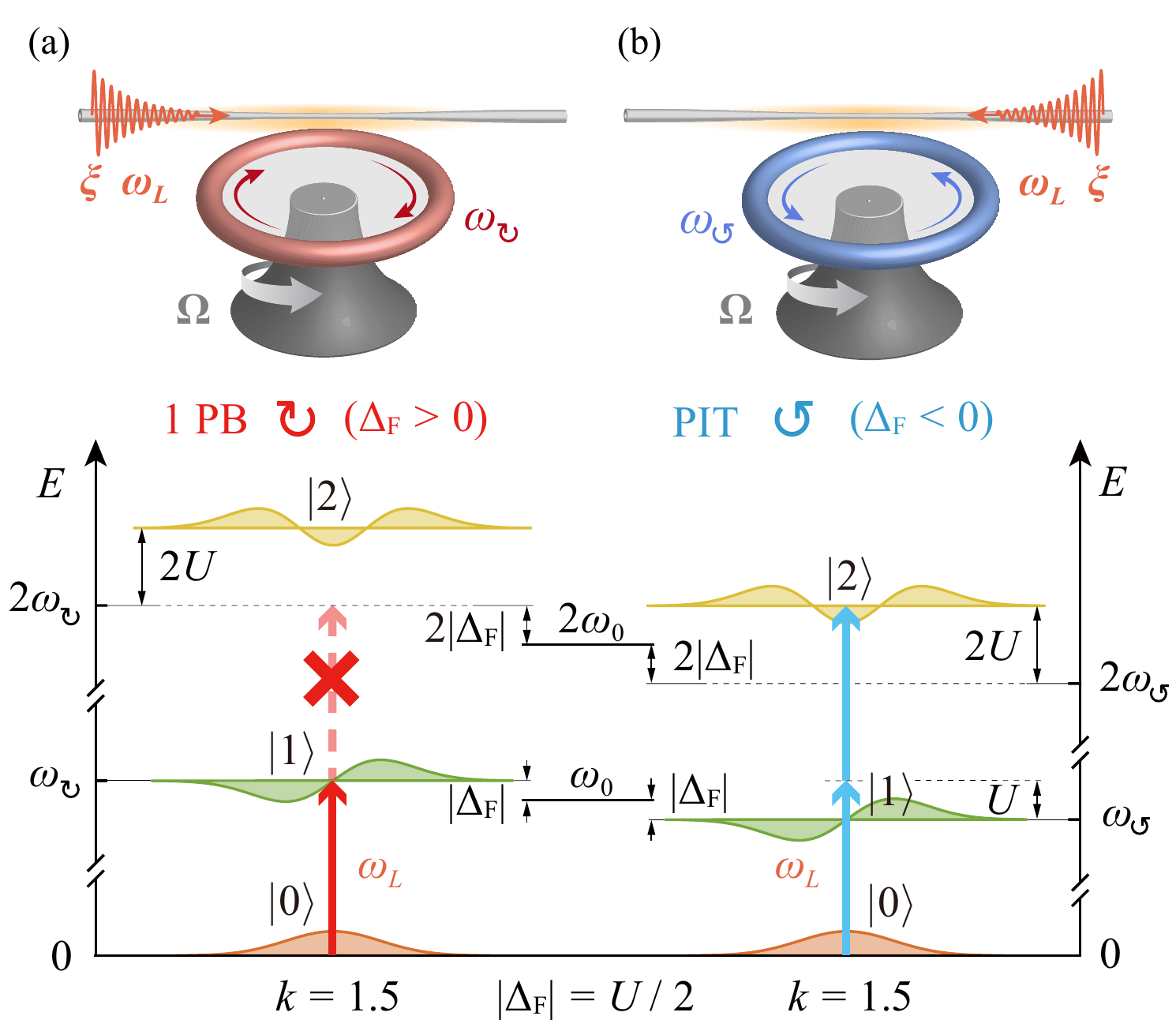}
\caption{Nonreciprocal 1PB in a spinning Kerr resonator. 1PB
arises due to the anharmonic spacing of the energy levels
$|n\rangle$. Here we take $n= 0,1,2$, and $\hbar=1$, for
simplicity. By fixing the CCW rotation of the resonator (the
angular speed $\Omega$ fulfills the condition
$\Delta_\mathrm{F}=\pm U/2$), under the same driving power
$P_\mathrm{in}=2~\mathrm{fW}$ and the same detuning
$\Delta_L=-U/2$, i.e., $k=1-\Delta_L/U=1.5$, (a) 1PB emerges by
driving the device from its left side ($\Delta_{F}>0$), while (b)
PIT caused by two-photon resonance occurs by driving from the
right side ($\Delta_{F}<0$). This PIT exhibits $g^{(\mu)}(0)>1$
($\mu=2,3,4$)~\cite{SM}.} \label{Fig1}
\end{figure}

\emph{Model.}---We consider a spinning optical Kerr resonator as
shown in Fig.~\ref{Fig1}. As a generic PB
model~\cite{leonski1994possibility,imamoglu1997strongly,adam2013two},
Kerr interactions can also be experimentally achieved in
cavity-atom systems~\cite{Birnbaum2005,schmidt1996giant}, or
magnon devices~\cite{wang2018bistability}, and theoretically in
optomechanical systems~\cite{rabl2011photon,nunnenkamp2011single}.
For a resonator spinning at an angular velocity $\Omega$, the
light circulating in the resonator experiences a {Fizeau shift},
i.e., $\omega_0\to\omega_0+\Delta_{F}$,
with~\cite{malykin2000sagnac}
\begin{equation}\label{Sagnac}
\Delta_{F}=\pm\frac{nr\Omega\omega_0}{c}\left(1-\frac{1}{n^2}-\frac{\lambda}{n}\frac{dn}{d\lambda}\right),
\end{equation}
where $\omega_0$ is the resonance frequency of a nonspinning
resonator, $n$ is the refractive index, $r$ is the resonator
radius, and $c$ ($\lambda$) is the speed (wavelength) of light in
vacuum. Usually, the dispersion term $dn/d\lambda$, characterizing
the relativistic origin of the Sagnac effect, is relatively small
(up to $\sim 1\%$)~\cite{malykin2000sagnac,maayani2018flying}. We
fix the CCW rotation of the resonator; hence $\Delta_{F}>0$
($\Delta_{F}<0$) corresponds to the situation of driving the
resonator from its left (right) side, i.e., the CW and CCW mode
frequencies are
$\omega_{\circlearrowright,\circlearrowleft}\equiv\omega_0\pm|\Delta_{F}|$,
respectively.

In a frame rotating at driving frequency $\omega_L$, the effective
Hamiltonian of the system can be written at the simplest level
as~\cite{SM}
\begin{equation}\label{Hsystem}
\hat{H}=\hbar(\Delta_k+\Delta_{F})\hat{a}^{\dag}\hat{a}+\hbar
U\hat{a}^{\dag}\hat{a}(\hat{a}^{\dag}\hat{a}-k)+\hbar\xi(\hat{a}^{\dag}+\hat{a}),
\end{equation}
where $\Delta_k=\Delta_L+U(k-1)$, $\Delta_L=\omega_0-\omega_L$,
the tuning parameter $k$ is simply $k=1-\Delta_L/U$ for
$\Delta_k=0$, $\hat{a}$ ($\hat{a}^\dag$) is the annihilation
(creation) operator of the cavity field, and $\xi=\sqrt{\gamma
P_\text{in}/(\hbar\omega_L)}$, with the cavity loss rate $\gamma$
and the driving power $P_\text{in}$. The Kerr parameter
is~\cite{marin2017microresonator}
$U={\hbar\omega_0^2cn_2}/(n_0^2V_{\text{eff}})$, where $n_0$
($n_2$) is the linear (nonlinear) refraction index, and
$V_{\text{eff}}$ is the effective mode volume. The Kerr coupling
is also attainable by using other kinds of
devices~\cite{rabl2011photon,nunnenkamp2011single,Birnbaum2005,schmidt1996giant,wang2018bistability}.
Note that the term $\Delta_{F}$ makes Eq.~(\ref{Hsystem})
fundamentally different from that used for studying conventional
PB~\cite{adam2013two}.

The energy eigenstates of this system are the Fock states
$|n\rangle$ ($n= 0,1,2,...$) with eigenenergies
\begin{equation}\label{energy}
E_{n}=n\hbar\Delta_L+(n^2-n)\hbar U\pm n\hbar|\Delta_{F}|,
\end{equation}
where $n$ is the cavity photon number. The second term, with $U$,
leads to an anharmonic energy-level structure. The last term, with
$\pm|\Delta_{F}|$, describing upper or lower shifts of energy
levels with an amount being proportional to $\Omega$, is the
origin of nonreciprocal implementations of PB. When
$|\Delta_{F}|=U/2$ and the probe with frequency
$\omega_0+|\Delta_{F}|$ ($k=1.5$) comes from the left side, the
light is resonantly coupled to the transition
$|0\rangle\to|1\rangle$. As shown in Fig.~\ref{Fig1}(a), the
transition $|1\rangle\to|2\rangle$ is detuned by $2\hbar U$ and,
thus, suppressed for $U>\gamma$; i.e., once a photon is coupled
into the resonator, it suppresses the probability of the second
photon with the same frequency going into the resonator. In
contrast, by driving from the right side, there is a two-photon
resonance with the transition $|0\rangle\to|2\rangle$; hence the
absorption of the first photon favors also that of the second or
subsequent photons, i.e., resulting in photon-induced tunneling
(PIT), as defined below and shown in Fig.~\ref{Fig1}(b). This is a
clear signature of nonreciprocal 1PB; i.e., {\emph{sub-Poissonian}
light} emerges by driving the system from one side, while
{\emph{super-Poissonian} light} emerges by driving from the other
side.

\emph{Analytical results.}---To confirm this intuitive picture, we
study the $\mu$th-order ($\mu=2,3$) correlation function with
zero-time delay, i.e., $g^{(\mu)}(0) \equiv{\langle
\hat{a}^{\dag\mu}\hat{a}^\mu\rangle}/{\langle\hat{n}\rangle}^\mu$,
with $\hat{n}=\hat{a}^{\dag}\hat{a}$. {The condition
$g^{(2)}(0)>1$ [$g^{(2)}(0)<1$] characterizes
PIT~\cite{Faraon2008,Majumdar2012} (1PB) via super-Poissonian
(sub-Poissonian) photon-number statistics or photon bunching
(antibunching)~\cite{scully1997quantum,adam10}.} The latter terms
can also refer to different (i.e., two-time) optical correlation
effects~\cite{Zou90,adam10}, which are, however, not studied here.
{We stress that, although PIT has a classical-like property of
super-Poissonian photon-number
statistics~\cite{Majumdar2012,Xu2013photon,majumdar2012probing},
it is a purely quantum effect~\cite{Faraon2008}.} The analysis of
higher-order correlation functions $g^{(\mu)}(0)>1$ with $\mu>2$
can reveal the relation of a particular PIT and
multi-PB~\cite{SM}. Thus, more refined criteria for PIT are
sometimes applied~\cite{Xu2013photon,Rundquist14,MeiWang18}, and
we refer here to PIT if the conditions $g^{(\mu)}(0)>1$ for
$\mu=2,3,4$ are satisfied~\cite{SM}. {We also note that partially
coherent mixtures of the vacuum, and single- and multiphoton
states, as generated here, can be described by $\mu$th-order
super-Poissonian correlations, i.e., $g^{(\mu)}(0)>1$, for
specific values of $\mu$~\cite{VogelBook}. Particularly,
$g^{(3)}(0)<1$ [$g^{(3)}(0)>1$] is a signature of third-order
sub-Poissonian (super-Poissonian) statistics, which is also
interpreted as three-photon antibunching (bunching) in recent
experiments on multi-PB~\cite{hamsen2017two} and
PIT~\cite{Rundquist14}. Thus, $g^{(3)}(0)$, which is usually
measured with extended Hanbury Brown and Twiss interferometers,
provides a more refined test and classification of the
nonclassical character of light, including 2PB (as studied below)
or unconventional PB~\cite{Radulaski2017}.}

According to the quantum-trajectory
method~\cite{plenio1998quantum}, the optical decay can be included
in the effective Hamiltonian
$\hat{H}_{\text{s}}=\hat{H}-(i\hbar\gamma/2)\hat{a}^{\dag}\hat{a}$,
where $\gamma=\omega_0/Q$ is the cavity dissipation rate and $Q$
is the quality factor. In the weak-driving regime
($\xi\ll\gamma$), by truncating the Hilbert space to $n=2$, the
state of this system is written as
$|\varphi(t)\rangle=\sum_{n=0}^2C_n(t)|n\rangle$, with probability
amplitudes $C_n$. Then we have the following equations of motion
\begin{align}
\dot{C}_{0}(t) & =-i\nu_{0}C_{0}(t)-i\xi C_{1}(t),\nonumber \\
\dot{C}_{1}(t) & =-i(\nu_{1}-i\frac{\gamma}{2})C_{1}(t)-i\xi C_{0}(t)-i\xi\sqrt{2}C_{2}(t),\label{MEq}\\
\dot{C}_{2}(t) &
=-i(\nu_{2}-i\gamma)C_{2}(t)-i\xi\sqrt{2}C_{1}(t),\nonumber
\end{align}
with $\hbar\nu_{n}=E_{n}$, $C_0(0)=1$, $C_1(0)=C_2(0)=0$. Solving
these equations (and dropping higher-order terms) leads to the
steady-state solutions
\begin{equation}
C_{1}(\infty)
=\frac{-\xi}{(\nu_{1}-\nu_{0}-i\frac{\gamma}{2})},~~C_{2}(\infty)
=\frac{-\sqrt{2}\xi C_1(\infty)}{(\nu_{2}-\nu_{0}-i\gamma)}.
\label{solution}
\end{equation}

Denoting the probability of finding $m$ photons in the resonator
by $P(m)=|C_m|^2$, we have
\begin{equation}\label{g2ana}
g^{(2)}(0) =\frac{2P_2}{(P_1+2P_2)^2} \simeq
\frac{(\Delta_L+\Delta_F)^2+\gamma^2/4}{(\Delta_L+\Delta_{F}+U)^2+\gamma^2/4}.
\end{equation}
1PB and PIT correspond to the minimum and the maximum of
$g^{(2)}(0)$, respectively, i.e., when $U>\gamma$,
$g^{(2)}_\mathrm{min}(0)=1/[4(U/\gamma)^2+1]<1$ for
$\Delta_L=-\Delta_F$, and
$g^{(2)}_\mathrm{max}(0)=4(U/\gamma)^2+1>1$ for
$\Delta_L=-\Delta_F-U$.

\begin{figure}[tb]
\centering
\includegraphics[width=0.48 \textwidth]{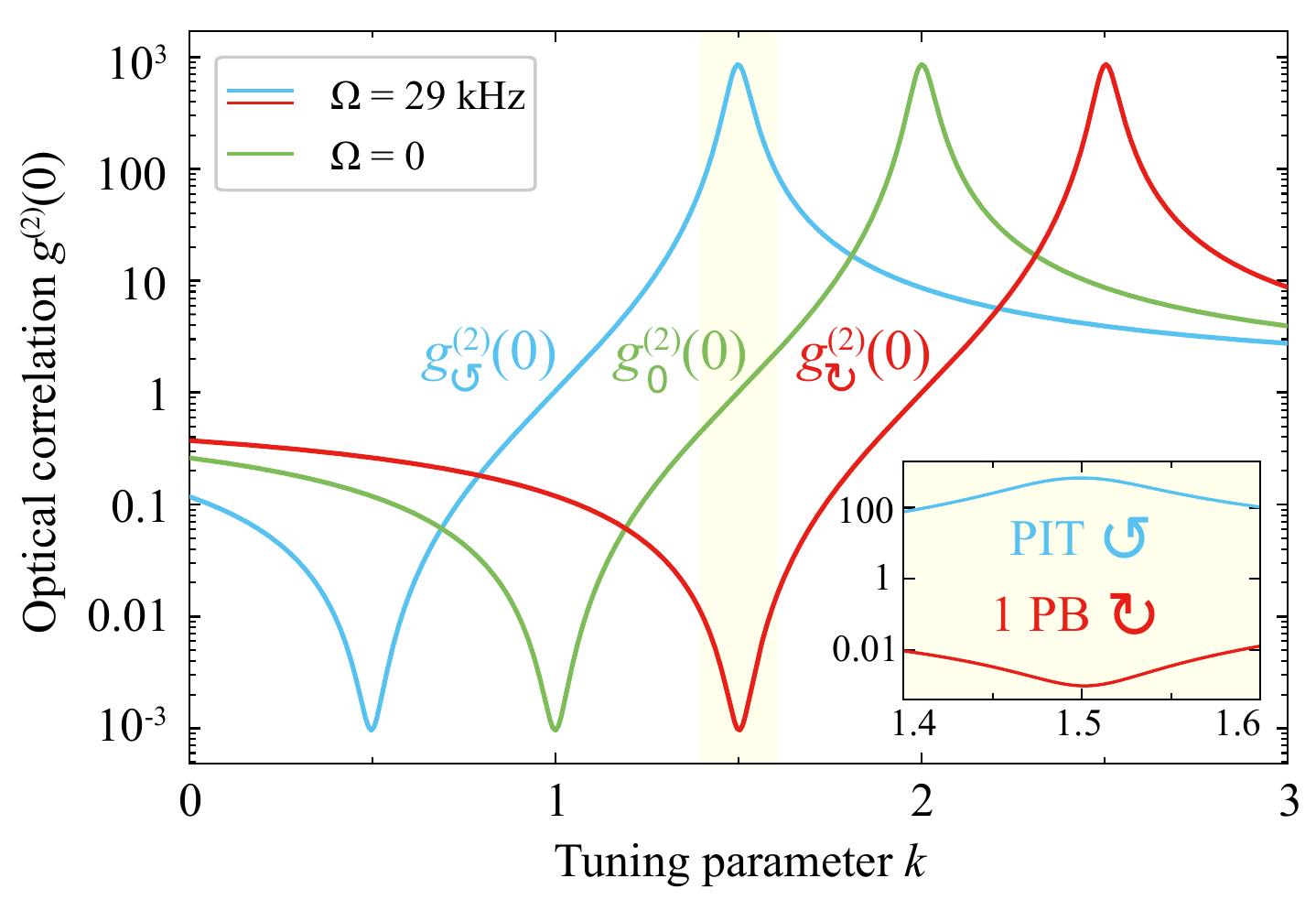}
\caption{The second-order correlation function $g^{(2)}(0)$ versus
the tuning parameter $k$ for different input directions. At
$k=1.5$, 1PB (red curve) or PIT (blue curve) occurs by driving the
device from the left or right side, with the same strength. Here
$P_\mathrm{in}=2~\mathrm{fW}$, $\Omega=29~\mathrm{kHz}$ for the
spinning resonator, and $g_0^{(2)}(0)$ corresponds to a
nonspinning resonator (green). {Note that $\Omega$ is related to
$\Delta_\mathrm{F}$ by Eq.~(\ref{Sagnac}).} For the other
parameter values, see the main text. {On the scale of this figure,
there are no differences between our numerical and (approximate)
analytical results~\cite{SM}.}} \label{Fig2}
\end{figure}

\emph{Numerical results.}---In order to confirm our analytical
results, now we numerically study the full quantum dynamics of the
system. We introduce the density operator $\hat{\rho}(t)$ and then
solve the master
equation~\cite{johansson2012qutip,johansson2013qutip2}:
\begin{equation}\label{QME}
\dot{\hat{\rho}}=\frac{i}{\hbar}[\hat{\rho},\hat{H}]+\frac{\gamma}{2}(2\hat{a}\hat{\rho}
\hat{a}^{\dag}-\hat{a}^{\dag}\hat{a}\hat{\rho}-\hat{\rho}
\hat{a}^{\dag}\hat{a}).
\end{equation}
The photon-number probability $P(n)=\langle n
|\hat{\rho}_\text{ss}|n\rangle$ can be obtained for the
steady-state solutions $\hat{\rho}_\text{ss}$ of the master
equation. The experimentally accessible parameters are chosen
as~\cite{vahala2003optical,spillane2005ultrahigh,pavlov2017pulse,huet2016millisecond,zielinska2017self}:
$V_{\text{eff}}=150\,\mu\mathrm{m}^3$, $Q=5\times10^9$,
$n_2=3\times10^{-14}\,\mathrm{m}^2/\mathrm{W}$, $n_0=1.4$,
$P_{\text{in}}=2\,\mathrm{fW}$, $r=30\,\mu\mathrm{m}$, and
$\lambda=1550\,\mathrm{nm}$. $V_{\text{eff}}$ is typically
$10^2$--$10^4\,\mu\mathrm{m}^3$~\cite{vahala2003optical,spillane2005ultrahigh},
$Q$ is typically
$10^9$--$10^{12}$~\cite{pavlov2017pulse,huet2016millisecond}, and
$g^{(2)}(0)$ as low as $\sim0.13$ was achieved
experimentally~\cite{Birnbaum2005}. {Moreover, in Fig.~\ref{Fig2},
we set $\Omega=29~\mathrm{kHz}$; a similar property of quantum
nonreciprocity is also confirmed for $\Omega=6.6~\mathrm{kHz}$
(see the Supplemental Material~\cite{SM}). These values of
$\Omega$ are experimentally feasible~\cite{maayani2018flying}.
Very recently, spinning objects have reached much higher
velocities, reaching the $\mathrm{GHz}$
regime~\cite{Reimann2018,Ahn2018}; such systems could also be
applied to study the nonreciprocal PB via Kerr-like optomechanical
interactions~\cite{Chang2010,Fonseca2016}.} We note that the Kerr
coefficient can be $n_2\sim10^{-14}\,\mathrm{m}^2/\mathrm{W}$ for
materials with potassium titanyl
phosphate~\cite{zielinska2017self}, and $n_2$ can be further
enhanced with various
techniques~\cite{higo2018large,lu2013quantum,bartkowiak2014quantum,xinyou2015squeezed,zhang2017quantum,rossi2018normal},
e.g., feedback control~\cite{zhang2017quantum,rossi2018normal} or
quadrature
squeezing~\cite{bartkowiak2014quantum,xinyou2015squeezed}.

An excellent agreement between our analytical results and the
exact numerical results is seen in Fig.~\ref{Fig2}. Here we use
$g^{(2)}_0(0)$, $g^{(2)}_\circlearrowright(0)$, and
$g^{(2)}_\circlearrowleft(0)$ to denote the cases with
$\Delta_F=0$, $\Delta_F>0$, and $\Delta_F<0$, respectively. For a
nonspinning resonator, regardless of the driving direction,
$g^{(2)}_0(0)$ always has a dip at $k=1$ (i.e., $\Delta_L=0$) or a
peak at $k=2$ (i.e., $\Delta_L=-U$), corresponding to 1PB or PIT,
respectively. In contrast, for a spinning device, by driving from
the left (right) side, we have $\Delta_{F}>0$ ($\Delta_{F}<0$)
and, thus, a redshift (blueshift) for $g^{(2)}(0)$, leads to 1PB
(PIT) at $k=1.5$, i.e., $g^{(2)}_\circlearrowright(0)\sim0.001$,
$g^{(2)}_\circlearrowleft(0)\sim673$. This quantum nonreciprocity,
with up to $6$ orders of magnitude difference of $g^{(2)}(0)$ for
opposite directions, is fundamentally different from the classical
transmission-rate nonreciprocity.

\begin{figure}[tb]
\centering
\includegraphics[width=0.48 \textwidth]{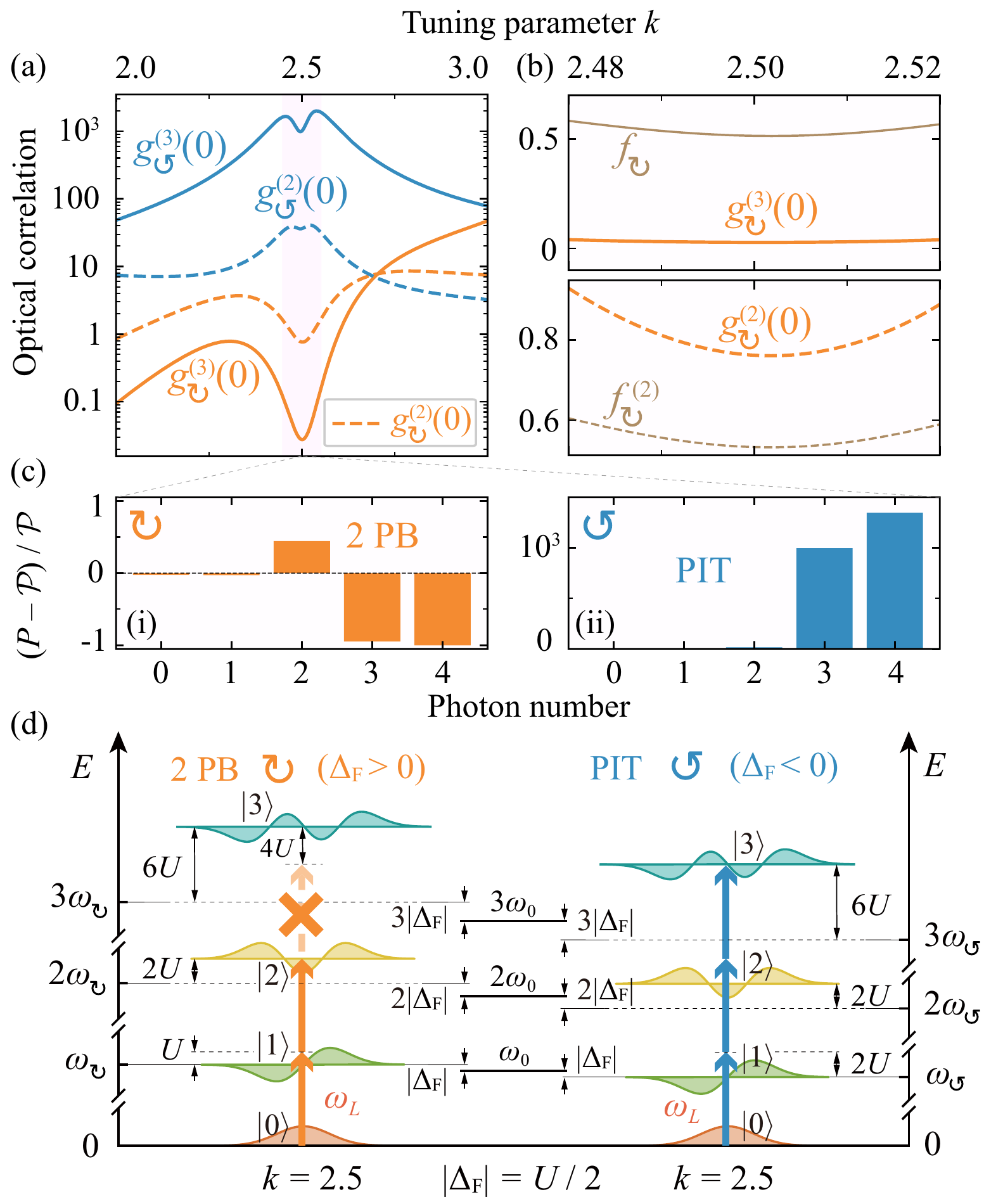}
\caption{(a) The correlation functions $g^{(3)}(0)$ (solid curves)
and $g^{(2)}(0)$ (dashed curves) versus the tuning parameter $k$
for different driving directions. Note that at $k=2.5$, 2PB can
emerge by driving the system from the left side (orange), while
PIT occurs by driving from the right side (blue). In (b), 2PB is
confirmed by the criteria given in Eq.~(\ref{2PBC}) for the CW
mode. (c) This nonreciprocal 2PB can also be recognized from the
deviations of the photon distribution to the standard Poisson
distribution with the same mean photon number. (d) The
energy-level diagram shows the origin of this unidirectional 2PB:
with enhanced driving power $P_\mathrm{in}=0.3~\mathrm{pW}$, by
choosing $\Delta_L=-3U/2$ (i.e., $k=2.5$), 2PB emerges by driving
the device from the left ($\Delta_{F}>0$), while three-photon
resonance-induced PIT emerges by driving from the right side
($\Delta_{F}<0$). The other parameters are the same as those in
Fig.~\ref{Fig2}. } \label{Fig3}
\end{figure}

\emph{Nonreciprocal 2PB.}---The absorption of 2 photons can also
suppress the absorption of additional photons~\cite{adam2013two}.
This 2PB effect, featuring three-photon antibunching, but with
two-photon bunching, satisfies~\cite{hamsen2017two,SM}:
\begin{align}
g^{(3)}(0)<f\equiv&e^{-\langle\hat{n}\rangle},\nonumber\\
g^{(2)}(0)\geq
f^{(2)}\equiv&e^{-\langle\hat{n}\rangle}+\langle\hat{n}\rangle\cdot
g^{(3)}(0). \label{2PBC}
\end{align}
The third-order correlation function can be obtained analytically
as~\cite{SM}:
\begin{equation}\label{g3ana}
g^{(3)}(0)=\frac{6P_{3}}{(P_{1}+2P_{2}+3P_{3})^{3}}\simeq\frac{(\Delta^{2}+{\gamma^{2}}/{4})g^{(2)}(0)}{(\Delta+2U)^{2}+{\gamma^{2}}/{4}},
\end{equation}
with $\Delta=\Delta_{L}+\Delta_F$, also agreeing well with the
numerical results. Figures~\ref{Fig3}(a) and \ref{Fig3}(b) show
that 2PB emerges around $k=2.5$ by driving from the left side,
while we have PIT by driving from the right side, i.e.,
$g^{(2)}_\circlearrowleft(0)\sim36$,
$g^{(3)}_\circlearrowleft(0)\sim1003$. {By tuning the driving
frequency to the three-photon resonance [see Fig.~\ref{Fig3}(d)],
it is indeed possible to observe that $g^{(3)}(0)/g^{(2)}(0)\sim
100$, as shown in Fig.~\ref{Fig3}(a) for $\max\langle
n\rangle=0.0185$. This means that the probability of
simultaneously measuring three photons can be much larger than
that of two photons in this situation. Similar values of
$g^{(3)}(0)\sim 10^3$, $g^{(2)}(0)\sim 10$ were also predicted in
the PIT analysis in Ref.~\cite{Rundquist14}.}

Our results can be further confirmed by comparing the
photon-number distribution $P(n)$ with the Poisson distribution
$\mathcal{P}(n)$. Figure~\ref{Fig3}(c) shows that $P(2)$ is
enhanced while $P(n>2)$ are suppressed by driving from the left
side, which is in \emph{sharp} contrast to the case when driving
from the right side. This unidirectional 2PB effect can be
intuitively understood by considering the energy-level structure
of the system, as shown in Fig.~\ref{Fig3}(d). By choosing
$\Delta_L=-3U/2$ or $k=2.5$, the transition
$|0\rangle\to|2\rangle$ is resonantly driven by the left input
laser, but the transition $|2\rangle\to|3\rangle$ is detuned by
$4\hbar U$, which features the 2PB effect; in contrast, by driving
from the right side, three-photon resonance happens for the
transition $|0\rangle\to|3\rangle$, leading to PIT. Hence with
such a device, {sub-Poissonian} light can be achieved by driving
it from the left side, while {super-Poissonian} light is observed
by driving it from the right side.

\begin{figure}[b]
\centering
\includegraphics[width=0.48 \textwidth]{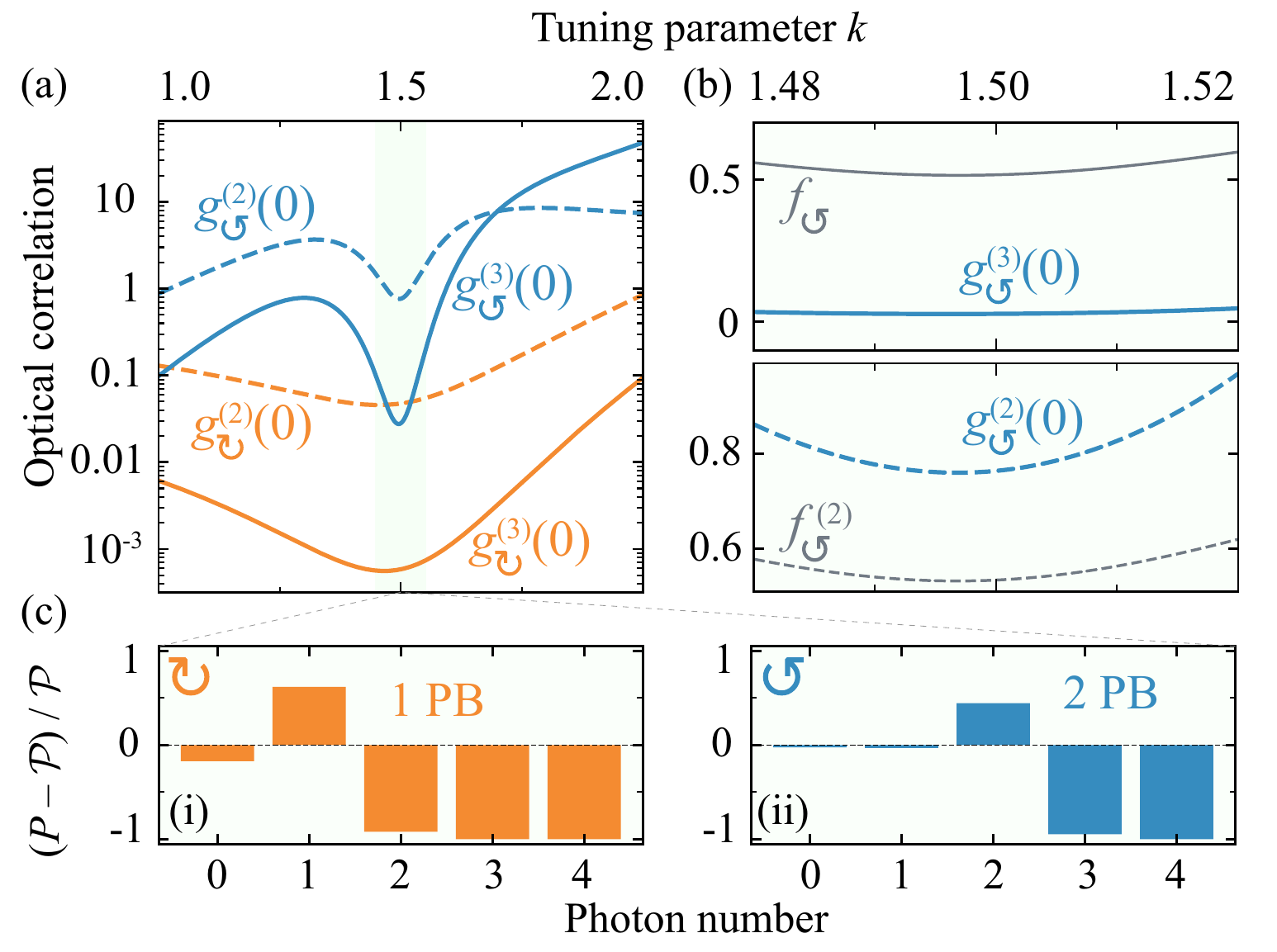}
\caption{(a) The correlation functions $g^{(3)}(0)$ (solid curves)
and $g^{(2)}(0)$ (dashed curves) versus the tuning parameter $k$
for different driving directions. 1PB can emerge around $k=1.5$ by
driving from the left side (orange), while 2PB occurs by driving
from the right side (blue). In (b), 2PB is confirmed by the
criteria given in Eq.~(\ref{2PBC}) for the CCW mode. (c) This
1PB-2PB nonreciprocity can also be recognized from the relative
photon population numbers in the resonator. For all plots, the
parameters are the same as those in Fig.~\ref{Fig3}.} \label{Fig4}
\end{figure}

\emph{Nonreciprocity of 1PB and 2PB.}---Figure~\ref{Fig4} shows
that at $k=1.5$, 1PB emerges by driving from the left side, due to
$g^{(2)}_\circlearrowright(0)\sim0.045$, while 2PB occurs by
driving from the right side since the criteria given in
Eq.~(\ref{2PBC}) are fulfilled for $\Delta_F<0$. This indicates a
purely quantum device with direction-dependent counting
statistics, a new nonreciprocal feature, which has not been
revealed previously. This 1PB-2PB nonreciprocity, as also clearly
seen in Fig.~\ref{Fig4}(c) for the populations of different Fock
states, provides a route for creating or processing different
quantum states in a single node of quantum
networks~\cite{bennett2000quantum,buluta2011natural}.
{Figures~\ref{Fig3}-\ref{Fig4} present our solutions of the
standard master equation, given in Eq.~(\ref{QME}), which
describes both a slow continuous nonunitary evolution and quantum
jumps occurring with a small probability~\cite{HarocheBook}. By
contrast, our approximate analytical solutions, based on the
complex Hamiltonian $H_\text{s}$ and the Schr\"odinger equation,
were obtained by ignoring these quantum jumps following the
standard approach of Ref.~\cite{Carmichael1991}.}

\emph{Conclusions.}---We have studied nonreciprocal PB effects in
a spinning Kerr resonator. By fixing the CCW rotation of the
resonator, we find the following: (i) for
$P_\mathrm{in}=2~\mathrm{fW}$, $\Delta_\mathrm{sag}=\pm U/2$ and
$k=1.5$, we have 1PB and PIT for the CW and CCW modes,
respectively. (ii) For $P_\mathrm{in}=0.3~\mathrm{pW}$,
$\Delta_\mathrm{sag}=\pm U/2$ and $k=2.5$, we have 2PB and PIT for
the CW and CCW modes, respectively. More interestingly, (iii) for
$P_\mathrm{in}=0.3~\mathrm{pW}$, $\Delta_\mathrm{sag}=\pm U/2$ and
$k=1.5$, we have 1 and 2PB for the CW and CCW modes, respectively
{(for more examples, see the Supplemental Material~\cite{SM}).}
These results can be useful in achieving, e.g., nonreciprocal
few-photon sources and quantum one-way devices.

The basic mechanism of this work can be generalized to a wide
range of systems, such as acoustic and electronic
devices~\cite{liang2010acoustic,popa2014non,kim2017dynamically,fleury2014sound,
Shabir2017mechanical,torrent2018nonreciprocal}, to achieve, e.g.,
nonreciprocal phonon blockade~\cite{Liu10,Didier11,adam16} as a
test of the quantumness of mechanical devices~\cite{adam10}. Our
work can also be extended to study, e.g., nonreciprocal photon
turnstiles~\cite{dayan2008a}, nonreciprocal photon
routers~\cite{aoki2009efficient,shomroni2014all,Liu14pb}, and
nonreciprocal extraction of a single photon from a laser
pulse~\cite{rosenblum2016extraction}, by considering a hybrid
device with atoms~\cite{aoki2006observation,junge2013strong},
quantum dots~\cite{michler2000a}, or nitrogen-vacancy
centers~\cite{faraon2011resonant}.

\begin{acknowledgements}
R.H. and H.J. are supported by the National Natural Science
Foundation of China (NSFC, 11474087 and 11774086). F.N. is
supported by the MURI Center for Dynamic Magneto-Optics via the
Air Force Office of Scientific Research (AFOSR)
(FA9550-14-1-0040), Army Research Office (ARO) (Grant
No.\,73315PH), Asian Office of Aerospace Research and Development
(AOARD) (Grant No.\,FA2386-18-1-4045), Japan Science and
Technology Agency (JST) (the ImPACT program and CREST Grant
No.\,JPMJCR1676), Japan Society for the Promotion of Science
(JSPS) (JSPS-RFBR Grant No.\,17-52-50023, and JSPS-FWO Grant No.
VS.059.18N), and the RIKEN-AIST Challenge Research Fund. A.M. and
F.N. are also supported by a grant from the John Templeton
Foundation. J.Q.L. is supported by the NSFC (11822501 and
11774087).
\end{acknowledgements}


%

\newpage

\onecolumngrid

\setcounter{equation}{0} \setcounter{figure}{0}
\setcounter{table}{0}
\setcounter{page}{1}\setcounter{secnumdepth}{3} \makeatletter
\renewcommand{\theequation}{S\arabic{equation}}
\renewcommand{\thefigure}{S\arabic{figure}}
\renewcommand{\bibnumfmt}[1]{[S#1]}
\renewcommand{\citenumfont}[1]{S#1}
\renewcommand\thesection{S\arabic{section}}

\begin{center}
{\large \bf Supplementary Material for ``Nonreciprocal Photon
Blockade''}
\end{center}

\begin{center}
Ran Huang$^{1}$, Adam Miranowicz$^{2,3}$, Jie-Qiao Liao$^1$,
Franco Nori$^{2,4}$, and Hui Jing$^{1,^*}$
\end{center}

\begin{minipage}[]{18cm}
\small{\it
\centering $^{1}$Key Laboratory of Low-Dimensional Quantum Structures and Quantum Control of Ministry of Education,  \\
\centering Department of Physics and Synergetic Innovation Center for Quantum  Effects and Applications, \\
\centering Hunan Normal University, Changsha 410081, China \\
\centering $^{2}$Theoretical Quantum Physics Laboratory, RIKEN Cluster for Pioneering Research, Wako-shi, Saitama 351-0198, Japan \\
\centering $^{3}$Faculty of Physics, Adam Mickiewicz University, 61-614 Pozna\' n, Poland \\
\centering $^{4}$Physics Department, The University of Michigan,
Ann Arbor, Michigan 48109-1040, USA \\}

\end{minipage}

\vspace{8mm}

Here, we present technical details on nonreciprocal photon
blockade (PB) in a driven Kerr-type model with a {Fizeau drag}.
Our discussion includes: (1) single- (1PB) and two-photon blockade
(2PB) effects; (2) our analytical solutions for the steady-state
optical-intensity correlation functions; and (3) rotation-induced
quantum nonreciprocity.

\section{Kerr-type interaction with the Fizeau drag}\label{KIWFSS}

To realize nonreciprocal photon blockade, we consider a rotating
optical resonator with a nonlinear Kerr medium which can be
described by a Kerr-type interaction with a {Fizeau drag} term,
\begin{equation}
\hat{H}_{\text{R}}=\hbar(\omega_{0}+\Delta_{{F}})\hat{a}^{\dagger}\hat{a}+\hbar
U\hat{a}^{\dagger}\hat{a}^{\dagger}\hat{a}\hat{a}.\label{HKerr}
\end{equation}

\begin{figure}[tbph]
\centering
\includegraphics[width=0.98 \textwidth]{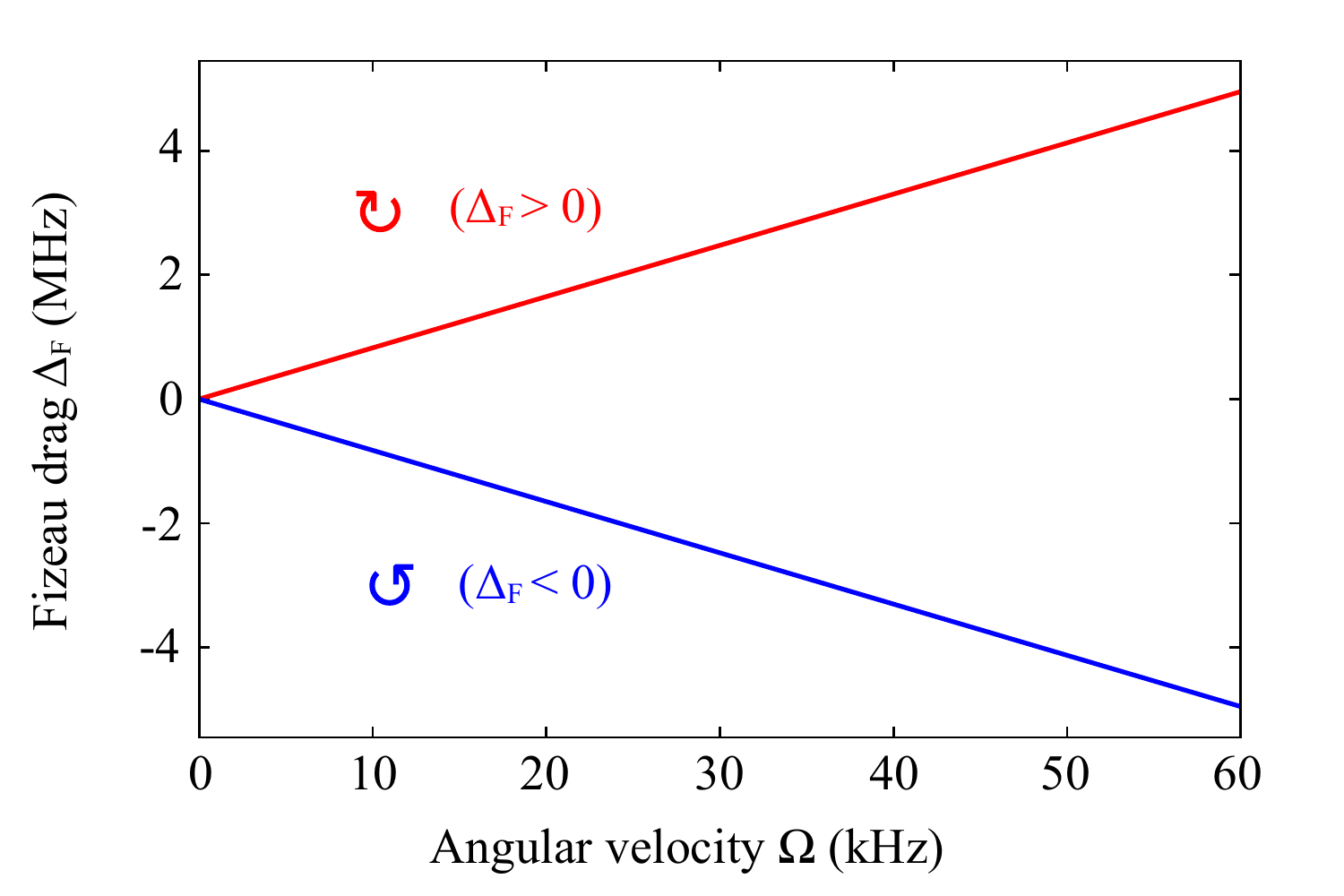}
\caption{{Fizeau drag $\Delta_{F}$ versus angular velocity of the
resonator for $\Delta_{{F}}>0$ (red line) and $\Delta_{{F}}<0$
(blue line) cases. The optical wavelength is
$\lambda=1550~\mathrm{nm}$, the radius of the resonator is
$R=30~\mu\mathrm{m}$, and the linear refractive index of the
resonator is $n=1.4$.}} \label{reply1}
\end{figure}
Here, $U\hat{a}^{\dagger}\hat{a}^{\dagger}\hat{a}\hat{a}$ is the
standard Kerr interaction term
~\cite{Smilburn1986quantum,Sleonski1994possibility,Simamoglu1997strongly,Sadam2013two},
$\hat{a}$ ($\hat{a}^\dag$) is the annihilation (creation) operator
for the cavity field, while
$U={\hbar\omega_0^2cn_2}/(n_0^2V_{\text{eff}})$ is the strength of
the nonlinear interaction with the nonlinear (linear) refraction
index $n_2$ ($n_0$), an effective cavity-mode volume
$V_{\text{eff}}$, and the speed of light in vacuum $c$ . Moreover,
$\omega_{0}$ is the resonance frequency of the nonspinning
resonator, and the rotation leads to a {Fizeau
shift}~\cite{Smalykin2000sagnac}:
\begin{equation}
\omega_{0}\to\omega_{\pm}=\omega_{0}+\Delta_{{F}},\label{Sagnaceffect}
\end{equation}
with
\begin{equation}
\Delta_{{F}}=\pm\frac{nr\Omega\omega_{0}}{c}\left(1-\frac{1}{n^{2}}-\frac{\lambda}{n}\frac{dn}{d\lambda}\right)=\pm\eta\Omega,
\label{Sagnacshift}
\end{equation}
where $\Delta_{{F}}>0$ ($\Delta_{{F}}<0$) denotes the light
propagating against (along) the direction of the spinning
resonator, $\lambda$ is the optical wavelength, $n$ is the
refractive index of the resonator, and $r$ is the radius of the
cavity. The dispersion term $dn/d\lambda$, characterizing the
relativistic origin of the Sagnac effect, is relatively small
($\sim1\%$)~\cite{Smalykin2000sagnac,Smaayani2018flying}.

{When the resonator is not spinning, the Fizeau drag is equal to
zero, owing to the same resonance frequency of light coming from
the left or right side. As implied by Eq.~(\ref{Sagnacshift}),
increasing the rotation frequency $\Omega$ results in an opposing
frequency linear shift of $\eta\Omega$ (see Fig.~\ref{reply1}) for
light coming from opposite directions~\cite{Smaayani2018flying}. }

\section{photon blockade effects}\label{PBE}

\subsection{Origin of photon blockade}\label{OPB}

In order to study conventional photon blockade (PB), we consider
the Hamiltonian~(\ref{HKerr}) including the driving term
\begin{equation}
\hat{H}=\hat{H}_{\text{R}}+\hbar\xi(\hat{a}^{\dagger}e^{-i\omega_{L}t}+\hat{a}e^{i\omega_{L}t}),
\end{equation}
where $\xi=\sqrt{\gamma P_\text{in}/(\hbar\omega_L)}$ is the
driving amplitude with the cavity loss rate $\gamma$, the driving
power $P_\text{in}$, and the driving frequency $\omega_{L}$
~\cite{Smajumdar2012probing}. In a frame rotating with the driving
frequency $\omega_{L}$, the Hamiltonian is transformed to
\[
\hat{H}_{\text{eff}}=i\hbar\frac{d\hat{D}^{\dagger}}{dt}\hat{D}+\hat{D}^{\dagger}\hat{H}\hat{D},
\]
with
$\hat{D}=\exp\left(-i\omega_{L}\hat{a}^{\dagger}\hat{a}t\right)$,
which leads to
\begin{align*}
\hat{H}_{\text{eff}} & =-\hbar\omega_{L}\hat{a}^{\dagger}\hat{a}+\hbar\omega_{\pm}\hat{a}^{\dagger}\hat{a}+\hbar U\hat{a}^{\dagger}\hat{a}^{\dagger}\hat{a}\hat{a}+\hbar\xi(\hat{a}^{\dagger}+\hat{a})\\
 & =\hbar(\omega_{0}+\Delta_{{F}}-\omega_{L})\hat{a}^{\dagger}\hat{a}+\hbar U\hat{a}^{\dagger}\hat{a}^{\dagger}\hat{a}\hat{a}+\hbar\xi(\hat{a}^{\dagger}+\hat{a}).
\end{align*}

\begin{figure}[tbp]
\centering
\includegraphics[width=0.98 \textwidth]{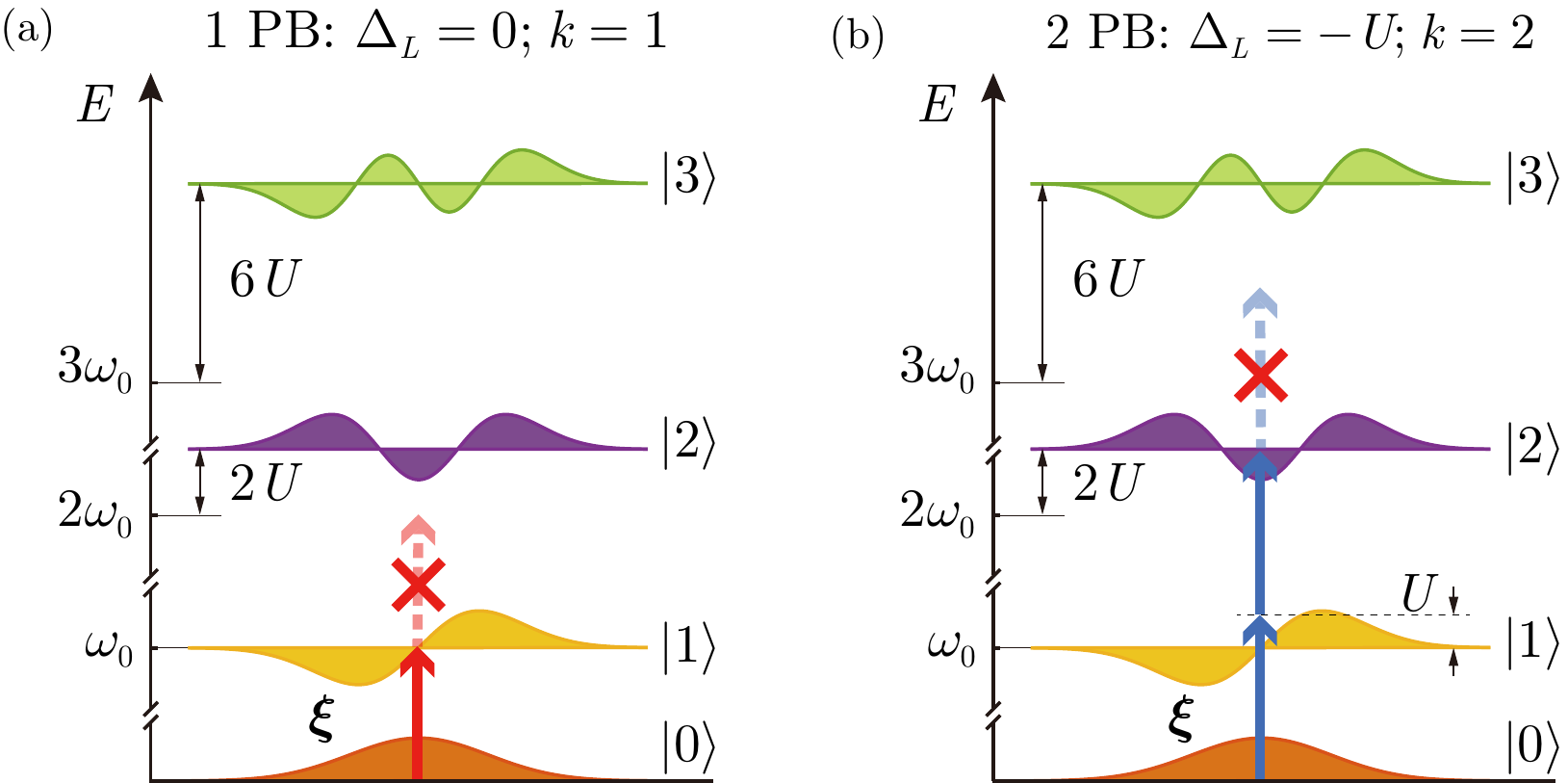}
\caption{Schematic energy-level diagram of the nonspinning
resonator. This explains the occurrence of $k$-photon blockade for
$\Delta_{F}=0$ in terms of $k$-photon transitions induced by the
driving field satisfying the resonance condition $\Delta_k=0$,
which corresponds to the driving-field frequency
$\omega_L=\omega_0+U(k-1)$. Here $\hbar=1$.} \label{PBmechanism}
\end{figure}

Thus, the effective Hamiltonian of this system becomes
\begin{equation}
\hat{H}_{\text{eff}}=\hbar(\Delta_{L}+\Delta_{{F}})\hat{a}^{\dagger}\hat{a}+\hbar
U\hat{a}^{\dagger}\hat{a}^{\dagger}\hat{a}\hat{a}+\hbar\xi(\hat{a}^{\dagger}+\hat{a}),\label{Heff}
\end{equation}
where $\Delta_{L}=\omega_{0}-\omega_{L}$ is the detuning between
the driving field and the cavity field for the nonspinning
resonator. The Hamiltonian of the isolated spinning system, i.e.,
\[
H_0=\hbar(\Delta_{L}+\Delta_{{F}})\hat{a}^{\dagger}\hat{a}+\hbar
U\hat{a}^{\dagger}\hat{a}^{\dagger}\hat{a}\hat{a},
\]
can be expressed as
\begin{align*}
\hat{H}_{0}|n\rangle & =[\hbar\Delta_{L}\hat{a}^{\dagger}\hat{a}+\hbar\Delta_{{F}}\hat{a}^{\dagger}\hat{a}+\hbar Ua^{\dagger}(\hat{a}\hat{a}^{\dagger}-1)\hat{a}]|n\rangle\\
 & =[\hbar\Delta_{L}\hat{a}^{\dagger}\hat{a}+\hbar\Delta_{{F}}\hat{a}^{\dagger}\hat{a}+\hbar U\hat{a}^{\dagger}\hat{a}\hat{a}^{\dagger}\hat{a}-\hbar U\hat{a}^{\dagger}\hat{a}]|n\rangle\\
 & =[\hbar\Delta_{L}\hat{a}^{\dagger}\hat{a}+\hbar(\Delta_{{F}}-U)\hat{a}^{\dagger}\hat{a}+\hbar U(\hat{a}^{\dagger}\hat{a})^{2}]|n\rangle\\
 & =[n\hbar\Delta_{L}+n\hbar(\Delta_{{F}}-U)+n^{2}\hbar U]|n\rangle\\
 & =E_{n}|n\rangle.
\end{align*}
Thus, we obtain the eigensystem for the weak-driving case,
\begin{align}
\hat{H}_{0}|n\rangle & =E_{n}|n\rangle,
\end{align}
with eigenvalues
\begin{equation}
E_{n}=n\hbar\Delta_{L}+n\hbar\Delta_{{F}}+(n^{2}-n)\hbar
U=n\hbar\Delta_{L}+(n^{2}-n)\hbar U\pm n\hbar|\Delta_{{F}}|,
\end{equation}
where $+n\hbar|\Delta_{{F}}|$ and $-n\hbar|\Delta_{{F}}|$ denote
the light propagating against ($\Delta_{F}>0$) and along
($\Delta_{F}<0$) the direction of the spinning resonator,
respectively.

\begin{figure}[tbp]
\centering
\includegraphics[width=0.90 \textwidth]{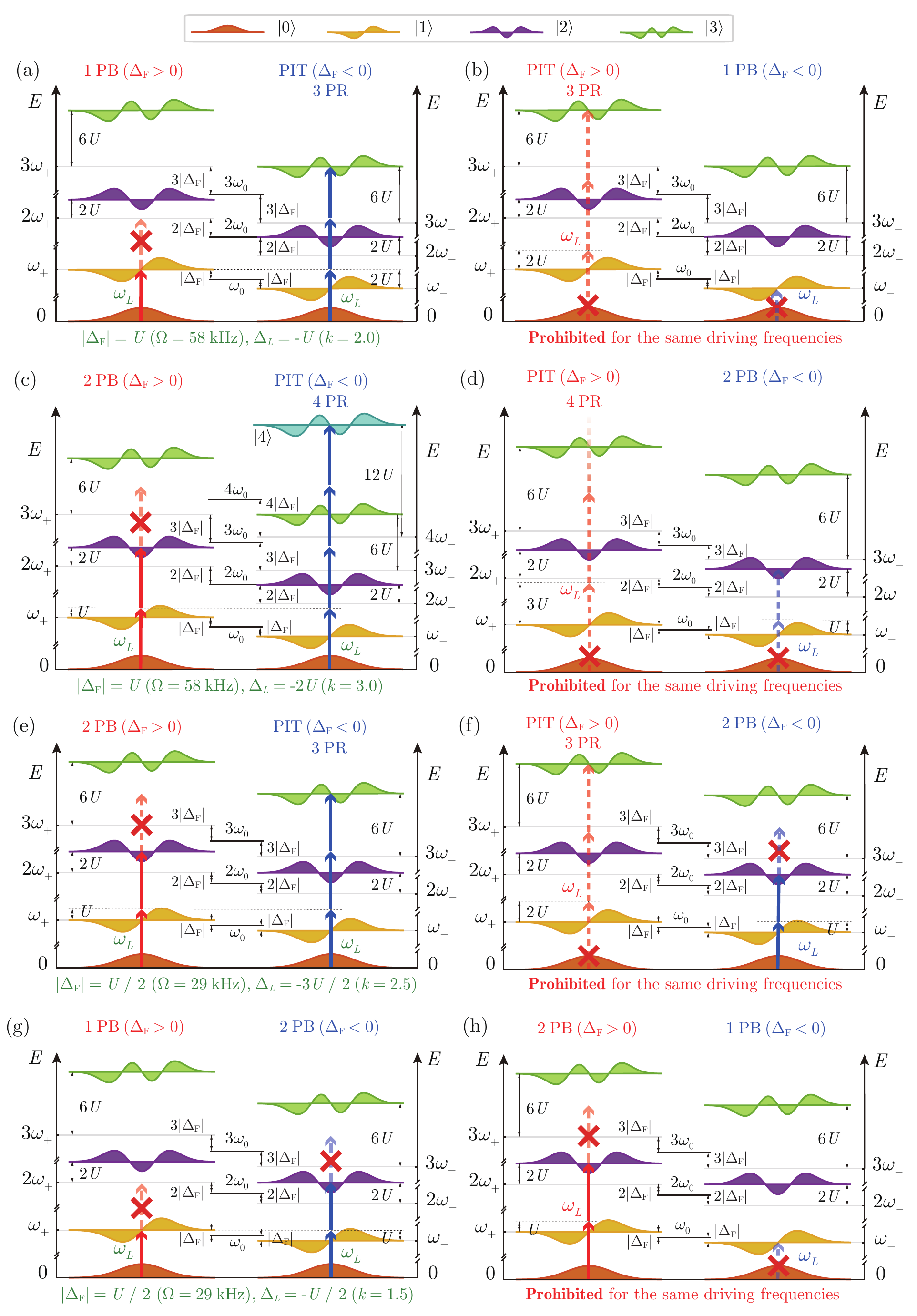}
\caption{Energy-level diagrams of the spinning resonator for
different cases of nonreciprocal PB effects. Here, photon-induced
tunneling (PIT) corresponds to an $n$-photon resonance ($n$ PR),
and $\hbar=1$. All of these diagrams correspond to the cases given
in Table~\ref{NRPBcases}.} \label{NRPBmechanism}
\end{figure}

The origin of conventional $n$-photon blockade can be understood
from the fact that due to the anharmonicity of the energy
structure, i.e., the energy difference between consecutive
manifolds is not constant, the Hilbert space of the system is
restricted to the states containing at most $n$ quanta. For
example, when the optical resonator is nonspinning
($|\Delta_{{F}}|$=0), single-photon blockade (1PB) is illustrated
in Fig.~\ref{PBmechanism}(a). If a coherent probe beam, tuned to
$\omega_0$ ($\Delta_L=0$), is coupled to the system, the probe is
on resonance with the $|0\rangle\to|1\rangle$ transition, but the
$|1\rangle\to|2\rangle$ transition is detuned by $2\hbar U$ and is
suppressed for $U>\gamma$ (where $\gamma$ denotes the optical loss
of the resonator). Consequently, once a photon is coupled to the
system, it suppresses the probability of coupling a second photon
with the same frequency. Similarly, two-photon blockade (2PB)
corresponds to a two-photon resonance (2PR) for a nonspinning
case, as shown in Fig.~\ref{PBmechanism}(b). Moreover, multi-PB
corresponds to a multi-photon resonance~\cite{Sshamailov2010multi,
Sadam2013two,Sadam14,Scarmichael2015breakdown,SZhu2017,Shamsen2017two}.
In addition to multi-PB, the energy-level diagrams of multi-photon
resonances in a Kerr-type system~\cite{Sadam2013two} also
correspond to photon-induced tunneling
(PIT)~\cite{SFaraon2008,Smajumdar2012probing,SMajumdar2012loss,SXu2013photon,SRundquist14}.
This indicates that the absorption of the first photon enhances
the absorption of subsequent photons~\cite{SFaraon2008}. The
distinction of 1PB, multi-PB, and PIT can be found by analysing
higher-order correlation functions $g^{(\mu)}(0)$ with $\mu\ge2$,
as discussed below.

Due to the rotation of the resonator, different cases of
nonreciprocal PB effects can be achieved. For example,
Table~\ref{NRPBcases} and Fig.~\ref{NRPBmechanism} summarize the
main results for $P_\mathrm{in}=0.3~\mathrm{pW}$, and these are
elaborated in detail later on in this Supplementary Material.

We observe that the Hamiltonian, given in Eq.~(\ref{Heff}), can be
rewritten as follows
\begin{equation}\label{Hk}
\hat{H}_k=\hbar(\Delta_{k}+\Delta_{{F}})\hat{a}^{\dag}\hat{a}+\hbar
U\hat{a}^{\dag}\hat{a}(\hat{a}^{\dag}\hat{a}-k)+\hbar\xi(\hat{a}^{\dag}+\hat{a}),
\end{equation}
where $\Delta_k=\Delta_L+U(k-1)$ is the frequency mismatch for the
nonspinning resonator. For convenience, we refer to $k$ as a
tuning parameter, as in Ref.~\cite{Sadam2013two}. Hereafter, we
analyze the resonant case of $\Delta_k=0$, which is related to the
resonant $k$-photon transitions in the nonspinning resonator, as
shown in Fig.~\ref{PBmechanism}. This condition implies that the
tuning parameter $k$ is related to the Kerr nonlinearity and the
driving-field and cavity frequencies as follows
\begin{equation}
k=-\Delta_L/U+1.
\end{equation}

\subsection{Criteria of photon blockade}\label{COPB}
We have studied the origin of conventional PB via the anharmonic
energy-level structure. In order to describe this picture
quantitatively, we apply two approaches. One is based on studying
the photon-number distribution of the system
~\cite{Sadam2013two,Shamsen2017two}, and the other is based on
investigating the optical intensity correlations
~\cite{SBirnbaum2005,SFaraon2008,Shamsen2017two}. Both can be
experimentally measured
~\cite{SBirnbaum2005,SFaraon2008,Shamsen2017two}.

Concerning the first method, in the case of an ideal $n$-photon
blockade, the cavity field shows the following photon-number
distribution~\cite{Sadam2013two}:
\begin{subequations}
\begin{align}
(i)\quad &P(m)=0\qquad\text{for}\ m>n, \label{PCa}\\
(ii)\quad &P(n)\neq0. \label{PCb}
\end{align}
\end{subequations}
with normalization $\sum_{m=0}^\infty P(m)=1$. While the first $n$
photons are resonantly absorbed in the system, the generation of
more photons is blockaded in the cavity. However, these
photon-number distribution conditions are hard to achieve in an
experiment, where $P(m)\neq0$ even for $m>n$. Thus, a comparison
with the Poissonian distribution was proposed by Hamsen \textit{et
al.}~\cite{Shamsen2017two}:
\begin{subequations}
\begin{align}
(i)\quad &P(m)<\mathcal{P}(m)\qquad\text{for}\ m>n, \label{PosCa} \\
(ii)\quad &P(n)\geq\mathcal{P}(n). \label{PosCb}
\end{align}
\end{subequations}
where $\mathcal{P}(m)$ is the Poissonian distribution
\begin{equation}\label{Poisson}
\mathcal{P}(m)=\frac{\langle\hat{m}\rangle^m}{m!}\exp\left({-\langle\hat{m}\rangle}\right),
\end{equation}
with the same average photon number $\langle\hat{m}\rangle$ as the
cavity field. The condition, given in Eq.~(\ref{PosCa}), indicates
that the first $n$ photons are effectively impenetrable to the
following photons; while the condition, given in
Eq.~(\ref{PosCb}), indicates that the coupling of an initial
photon to the system favors the coupling of the subsequent photons
within the first $n$ photons. This leads to the sub-Poissonian
photon-number statistics for ($n+1$) photons with the simultaneous
super-Poissonian statistics of the first $n$ photons. To show a
relative deviation of a given photon-number distribution from the
corresponding Poissonian distribution, we use the
formula~\cite{Shamsen2017two}:
\begin{equation}\label{deviation}
[P(n)-\mathcal{P}(n)]/\mathcal{P}(n).
\end{equation}

For the second approach, {correlation function
$G^{(n)}(t_1,t_2,\cdots,t_n)$ is the quantity measured at moments
$t_1,t_2,\cdots,t_n$ in extended Hanbury Brown-Twiss experiments
with $n$ detectors. Note that $g^{(n)}$ is $G^{(n)}$ normalized by
the $n$th power of the mean photon number. Thus,
$g^{(n)}(0)\equiv\lim_{t\to\infty}(t,t,\cdots,t)$ is related to
the probability of simultaneously measuring $n$ photons in their
steady state assuming photon detections at the same time
$t=t_1=t_2=\cdots=t_n$. The larger value of $g^{(n)}(0)>1$, the
higher probability of $n$-photon bunching (photon coalescence).
And the smaller value of $g^{(n)}(0)<1$, the lower probability of
$n$-photon bunching, which corresponds to the higher probability
of $n$-photon antibunching (photon anticoalescence). The case of
$g^{(n)}(0)=1$ is called photon unbunching, which is a typical
feature of coherent light for any $n$. These correlation functions
$G^{(n)}$ and $g^{(n)}$ are basic elements of the quantum
coherence theory of Glauber~\cite{SGlauberBook}.}

The normalized equal-time $\mu$th-order photon correlation is
given by
\begin{equation}
g^{(\mu)}(0)=\sum_{m=\mu}^\infty\frac{m!}{(m-\mu)!}\frac{P(m)}{\langle\hat{m}\rangle^\mu}={\langle\hat{m}\rangle^{-\mu}}\sum_{m=\mu}^\infty{m}(m-1)\cdots(m-\mu+1){P(m)}=\frac{\langle\hat{a}^{\dag\mu}\hat{a}^\mu\rangle}{\langle\hat{a}^{\dag}\hat{a}\rangle^\mu}.\label{gmu}
\end{equation}
In particular, the second-order photon correlation function is
\begin{equation}
g^{(2)}(0)=\sum_{m=2}^\infty{m(m-1)}\frac{P(m)}{\langle\hat{m}\rangle^2}=\frac{\langle\hat{m}(\hat{m}-1)\rangle}{\langle\hat{m}\rangle^2}=\frac{\langle\hat{a}^{\dag2}\hat{a}^2\rangle}{\langle\hat{a}^{\dag}\hat{a}\rangle^2},\label{g2}
\end{equation}
and the third-order photon correlation function is
\begin{equation}
g^{(3)}(0)=\sum_{m=3}^\infty{m(m-1)(m-2)}\frac{P(m)}{\langle\hat{m}\rangle^3}=\frac{\langle\hat{m}(\hat{m}-1)(\hat{m}-2)\rangle}{\langle\hat{m}\rangle^3}=\frac{\langle\hat{a}^{\dag3}\hat{a}^3\rangle}{\langle\hat{a}^{\dag}\hat{a}\rangle^3}.\label{g3}
\end{equation}

The photon-number distribution conditions for $n$-photon blockade,
given in Eqs.~(\ref{PCa}) and (\ref{PCb}), can be translated into
the following conditions:
\begin{subequations}
\begin{align}
(i)\quad &g^{(n+1)}(0)=0, \label{gnCa} \\
(ii)\quad &g^{(n)}(0)\neq0. \label{gnCb}
\end{align}
\end{subequations}
As aforementioned, these strict conditions can only be fulfilled
for an ideal case. The experimentally-realizable conditions can be
obtained based on Eqs.~(\ref{PosCa}) and (\ref{PosCb}). Since in
the weak-driving regime, the photon-number distribution fulfills
the condition ${P}(m) \gg P(m+1)$, it is sufficient to satisfy
$P(n+1)<\mathcal{P}(n+1)$ according to the condition in
Eq.~(\ref{PosCa}). Meanwhile, we can approximately express
$P(n+1)$ with $g^{(n+1)}(0)$ as follows:
\begin{align}
g^{(n+1)}(0)&=\sum_{m=n+1}^\infty\frac{m!}{(m-n-1)!}\frac{P(m)}{\langle\hat{m}\rangle^{n+1}}\approx\frac{(n+1)!}{\langle\hat{m}\rangle^{n+1}}P(n+1), \nonumber\\
P(n+1)&\approx\frac{\langle\hat{m}\rangle^{n+1}}{(n+1)!}\cdot
g^{(n+1)},
\end{align}
as the $P(m)$ have been neglected for all $m>(n+1)$. Thus, the
condition, given in Eq.~(\ref{PosCa}),
reads~\cite{Shamsen2017two}:
\begin{align}
{P}(n+1)&<\mathcal{P}(n+1), \nonumber \\
\frac{\langle\hat{m}\rangle^{n+1}}{(n+1)!}\cdot g^{(n+1)}(0)&<\frac{\langle\hat{m}\rangle^{n+1}}{(n+1)!}\exp\left({-\langle\hat{m}\rangle}\right), \nonumber \\
g^{(n+1)}(0)&<\exp\left({-\langle\hat{m}\rangle}\right).
\label{gnp1C}
\end{align}
We can also obtain an approximate $P(n)$ using a similar method as
follows:
\begin{align}
g^{(n)}(0)&=\sum_{m=n}^\infty\frac{m!}{(m-n)!}\frac{P(m)}{\langle\hat{m}\rangle^{n}}\approx\frac{n!}{\langle\hat{m}\rangle^{n}}P(n)+\frac{(n+1)!}{\langle\hat{m}\rangle^{n}}P(n+1), \nonumber \\
P(n)&\approx\frac{\langle\hat{m}\rangle^{n}}{n!}\cdot g^{(n)}{(0)}-{(n+1)}P(n+1), \nonumber \\
P(n)&\approx\frac{\langle\hat{m}\rangle^{n}}{n!}\cdot
g^{(n)}{(0)}-\frac{\langle\hat{m}\rangle^{n+1}}{n!}\cdot
g^{(n+1)}{(0)}.
\end{align}
Moreover, the condition, given in Eq.~(\ref{PosCb}), then reads:
\begin{align}
{P}(n)&\geq\mathcal{P}(n), \nonumber \\
\frac{\langle\hat{m}\rangle^{n}}{n!}\cdot g^{(n)}(0)-\frac{\langle\hat{m}\rangle^{n+1}}{n!}\cdot g^{(n+1)}(0)&\geq\frac{\langle\hat{m}\rangle^{n}}{n!}\exp\left({-\langle\hat{m}\rangle}\right), \nonumber \\
g^{(n)}(0)-{\langle\hat{m}\rangle}\cdot g^{(n+1)}(0)&\geq \exp\left({-\langle\hat{m}\rangle}\right), \nonumber \\
g^{(n)}(0)&\geq
\exp\left({-\langle\hat{m}\rangle}\right)+{\langle\hat{m}\rangle}\cdot
g^{(n+1)}(0), \label{gnC}
\end{align}
i.e., the experimentally-realizable conditions, given in
Eqs.~(\ref{PosCa}) and (\ref{PosCb}), can be translated into the
following conditions~\cite{Shamsen2017two}:
\begin{subequations}
\begin{align}
(i)\quad &g^{(n+1)}(0)<\exp\left({-\langle\hat{m}\rangle}\right),  \label{gCa} \\
(ii)\quad &g^{(n)}(0)\geq
\exp\left({-\langle\hat{m}\rangle}\right)+\langle\hat{m}\rangle\cdot
g^{(n+1)}(0),  \label{gCb}
\end{align}
\end{subequations}
{indicating a higher-order sub-Poissonian photon-number
statistics.}

Moreover, PIT can be quantified by photon-number correlation
functions. Table~\ref{PITcriterion} shows that more refined
criteria for PIT are sometimes applied based on higher-order
correlation functions $g^{(\mu)}(0)$ with
$\mu>2$~\cite{SRundquist14,SMeiWang18}. Here, we refer to PIT if
the following conditions are satisfied for $\mu\ge2$:
\begin{equation}
g^{(\mu)}(0)>\exp\left({-\langle\hat{m}\rangle}\right).
\end{equation}
For simplicity, in this work, we consider these conditions only
for $2\le\mu\le4$. This indicates {light with higher-order
super-Poissonian photon-number statistics}, i.e., once, a photon
is coupled in a resonator, it enhances the probabilities of more
photons entering the resonator. In the few-photon regime
($\langle\hat{m}\rangle\ll1$), these criteria become
\begin{equation}\label{PITc}
g^{(\mu)}(0)>1~~~\mathrm{for}~~~\mu=2,3,4.
\end{equation}
We provide a more basic criteria to identify multi-PB and PIT by
using $\mu$th-order correlation functions $g^{(\mu)}(0)$. These
criteria lead to the same conclusions as those based on
Eq.~(\ref{deviation}).

\begin{table}[htbp]
\renewcommand\arraystretch{1.5}
 \caption{\label{PITcriterion}Criteria of photon-induced tunneling (PIT) used in literature.}
 \begin{tabular}{p{18cm}}
 \rowcolor{darkblue}
   \ \ Reference{\qquad}{\qquad}{\qquad}{\qquad}{\qquad}{\qquad}{\quad}Criteria of PIT\\
\rowcolor{lightblue}
  \ \ Faraon \emph{et al}. (2008)~\cite{SFaraon2008}{\qquad}{\qquad}{\qquad}$g^{(2)}(0)$ is a local maximum \\
\rowcolor{lightblue}
  \ \ Majumdar \emph{et al}. (2012)~\cite{Smajumdar2012probing,SMajumdar2012loss}{\qquad}\ \,\,$g^{(2)}(0)>1$ \\
  \rowcolor{lightblue}
  \ \ Xu \emph{et al}. (2013)~\cite{SXu2013photon}{\qquad}{\qquad}{\qquad}{\quad}$\ \,\,g^{(2)}(0)>1$ (two-photon tunneling); $g^{(3)}(0)>g^{(2)}(0)>1$ (three-photon tunneling)\\
 \rowcolor{lightblue}
  \ \ Rundquist \emph{et al}. (2014)~\cite{SRundquist14}{\qquad}\quad\,\,\,\,\,\,\,\,$g^{(3)}(0)>g^{(2)}(0)$  \\
 \rowcolor{lightblue}
  \ \ Wang \emph{et al}. (2018)~\cite{SMeiWang18}{\qquad}{\qquad}\quad\ \ \,\,\,\,$g^{(4)}(0)>g^{(3)}(0)>g^{(2)}(0)>1$ (phonon-induced tunneling, an analogue of PIT)\\
  \end{tabular}
\end{table}

\subsection{Single- and Multi-photon blockade}\label{NPB}
\begin{figure}[tbp]
\centering
\includegraphics[width=0.98 \textwidth]{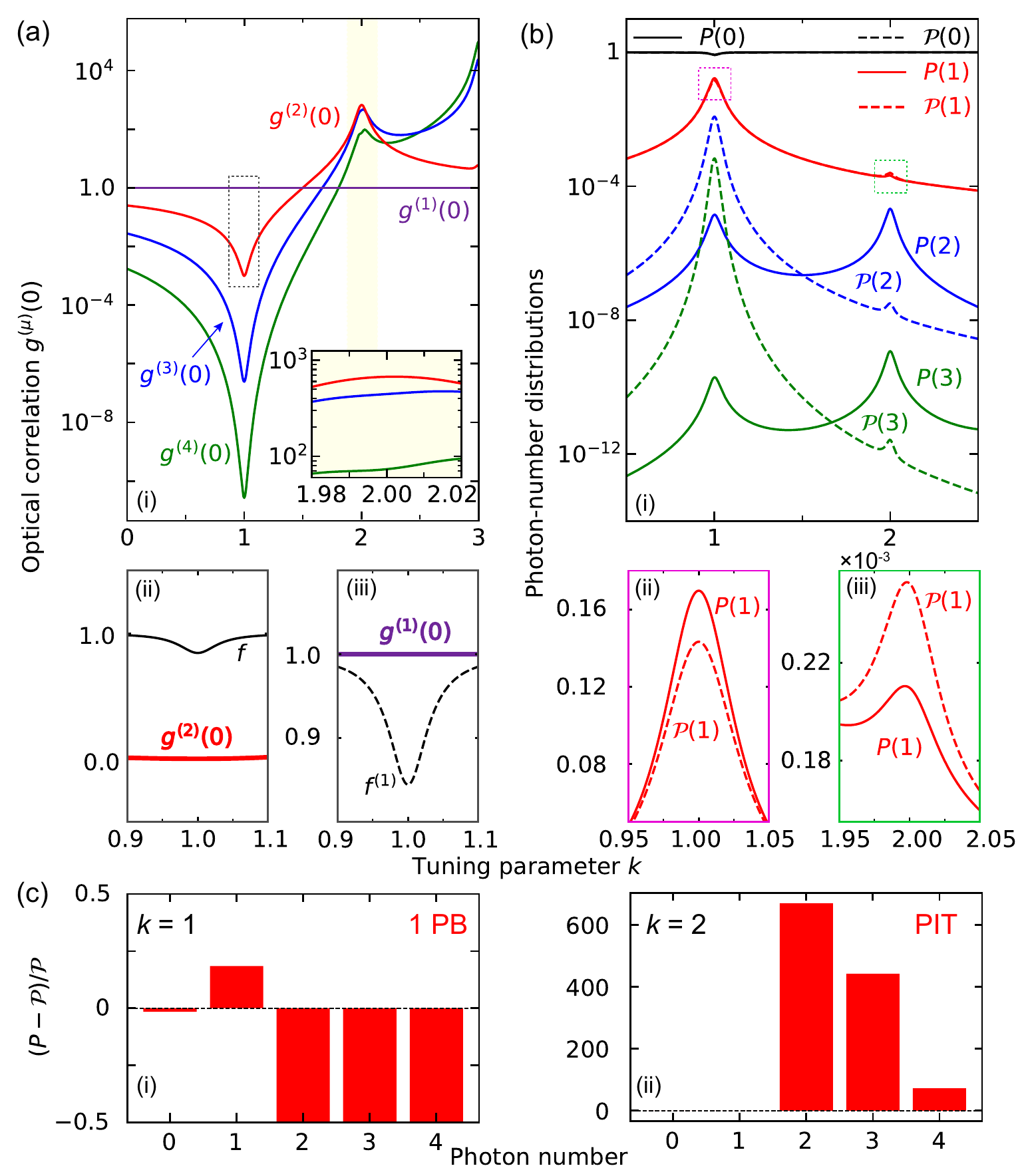}
\caption{(a) Correlation functions $g^{(\mu)}(0)$ versus the
tuning parameter $k$ for the nonspinning resonator
($\Delta_{F}=0$). Note that 1PB emerges at $k=1$, since (a-ii)
$g^{(2)}(0)<f$ and (a-iii) $g^{(1)}(0)>f^{(1)}$ fulfill the
criteria given in Eqs.~(\ref{1PBCa}) and (\ref{1PBCb}),
respectively. PIT occurs at $k=2$, since
$g^{(2)}(0)>g^{(3)}(0)>g^{(4)}(0)>1$ [see the inset in
panel~(a-i)] fulfills the condition given in Eq.~(\ref{PITc}).
These 1PB and PIT can also be recognized from (b) the
photon-number distributions and (c) the deviations given in
Eq.~(\ref{deviation}). At $k=1$, (b-ii) single-photon probability
is enhanced as $P(1)>\mathcal{P}(1)$, while $m$-photon ($m>1$)
probabilities are suppressed as $P(m)<\mathcal{P}(m)$ [see
panels~(b-i) and (c-i)]. These photon-number distributions fulfill
the conditions given in Eqs.~(\ref{PosCa}) and (\ref{PosCb}) for
$n=1$, i.e., resulting in 1PB. At $k=2$, (b-iii) single-photon
probability is suppressed as $P(1)<\mathcal{P}(1)$, while
$m$-photon ($m>1$) probabilities are enhanced as
$P(m)>\mathcal{P}(m)$ [see panels~(b-i) and (c-ii)], i.e.,
resulting in PIT. The parameters used here are: $\Omega=0$,
$n_2=3\times10^{-14}\,\mathrm{m}^2/\mathrm{W}$, $n_0=1.4$,
$V_{\text{eff}}=150\,\mu\mathrm{m}^3$, $Q=5\times10^9$,
$\lambda=1550\,\mathrm{nm}$, $P_{\text{in}}=2\,\mathrm{fW}$, and
$r=30\,\mu\mathrm{m}$.} \label{1PB}
\end{figure}

In this section, we only consider the nonspinning case
($\Delta_{F}$=0), while the spinning case is discussed in
Sec.~\ref{RINPB}. According to criteria, given in Eqs.~(\ref{gCa})
and (\ref{gCb}), 1PB has to fulfill the following conditions for
$n=1$:
\begin{subequations}
\begin{align}
(i)\quad &g^{(2)}(0)<\exp\left({-\langle\hat{m}\rangle}\right)\equiv f, \label{1PBCa} \\
(ii)\quad &g^{(1)}(0)\geq
\exp\left({-\langle\hat{m}\rangle}\right)+\langle\hat{m}\rangle\cdot
g^{(2)}(0)\equiv f^{(1)}. \label{1PBCb}
\end{align}
\end{subequations}
As expected from the intuitive picture discussed in
Sec.~\ref{OPB}, the strongest 1PB occurs at $\Delta_L=0$ ($k=1$),
since the correlation functions fulfill the criteria of 1PB given
in Eqs.~(\ref{1PBCa}) and (\ref{1PBCb}) [see Fig.~\ref{1PB}(a)].
In the weak-driving regime, $\langle\hat{m}\rangle\ll1$ implies
that $f\to1$ and $f^{(1)}\to1$. Then we obtain $g^{(2)}(0)<1$,
which corresponds to the usual criterion of 1PB, as known in the
published literature.

As aforementioned in 1PB, the first photon blocks the entrance of
a second photon, which indicates the enhancement of the
single-photon probability, and also the suppression of the two- or
more-photon probabilities. We can clearly see that
$P(1)>\mathcal{P}(1)$, while $P(2)<\mathcal{P}(2)$ and
$P(3)<\mathcal{P}(3)$ at $k=1$ in Fig.~\ref{1PB}(b). Moreover, 1PB
can be recognized from the deviations of the photon distribution
from the standard Poissonian distribution with the same mean
photon number [i.e., Eq.~(\ref{deviation})], as shown in
Fig.~\ref{1PB}(c-i).

At $k=2$, we find the correlation functions fulfill
$g^{(2)}(0)>g^{(3)}(0)>g^{(4)}(0)>1$, as shown in the inset in
Fig.~\ref{1PB}(a). {This shows that PIT corresponding to
super-Poissonian photon-number behavior of light, which occurs at
$k=2$,} since the correlation functions satisfy the conditions
given in Eq.~(\ref{PITc}). PIT can also be recognized from the
photon-number distributions and the deviations given in
Eq.~(\ref{deviation}). As shown in Figs.~\ref{1PB}(b) and
\ref{1PB}(c-ii), we find that $P(1)<\mathcal{P}(1)$,
$P(2)>\mathcal{P}(2)$, $P(3)>\mathcal{P}(3)$, and
$P(4)>\mathcal{P}(4)$ at $k=2$. This is a clear signature of PIT.
Since the case for $k=2$ corresponds to a two-photon resonance, we
refer to this PIT as two-photon resonance-induced PIT.

\begin{figure}[tbp]
\centering
\includegraphics[width=0.98 \textwidth]{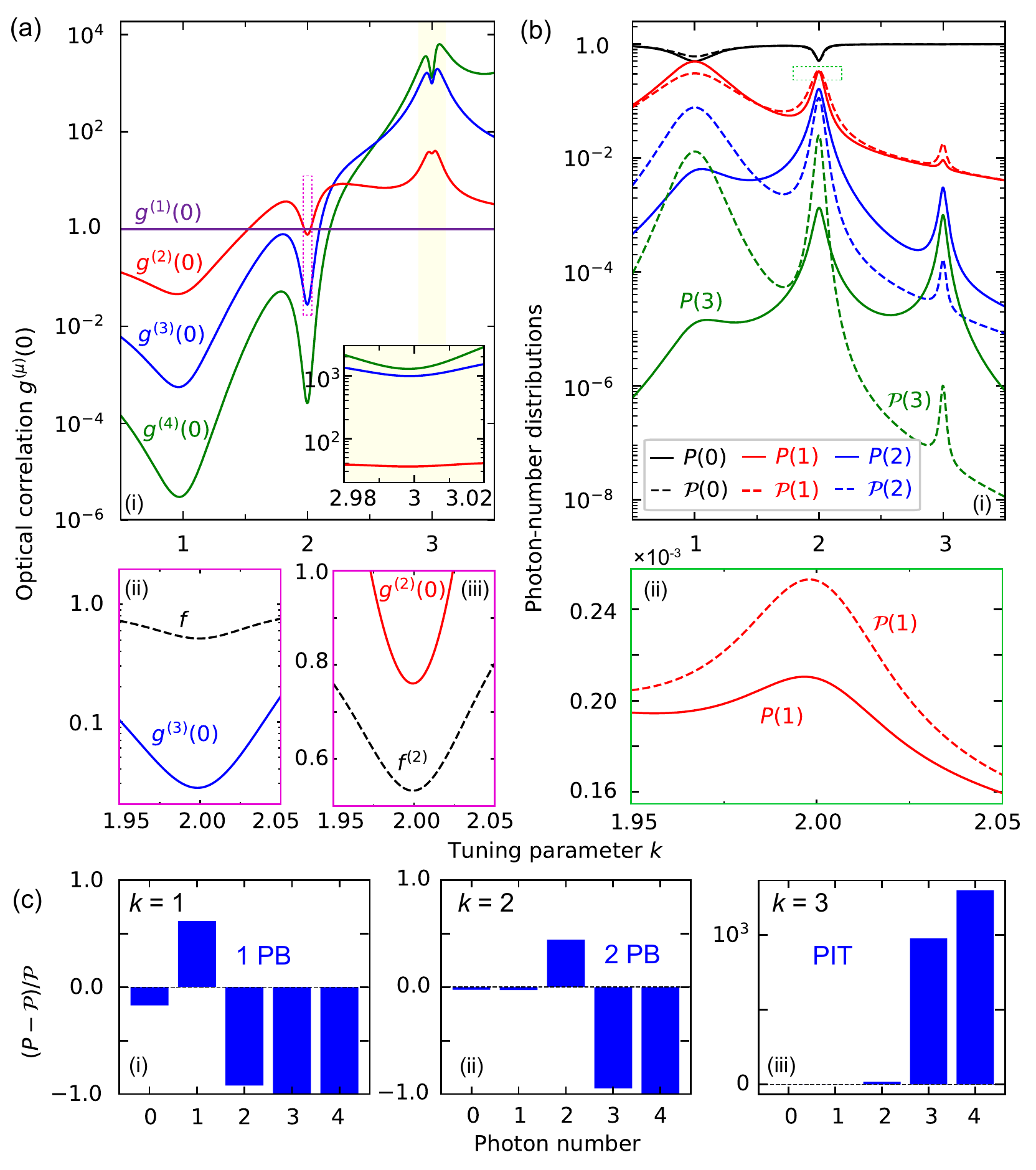}
\caption{(a) Correlation functions $g^{(\mu)}(0)$ versus the
tuning parameter $k$ for the nonspinning resonator
($\Delta_{F}=0$). Note that 2PB occurs at $k=2$, since (a-ii)
$g^{(3)}(0)<f$ and (a-iii) $g^{(2)}(0)>f^{(2)}$ fulfill the
criteria given in Eqs.~(\ref{2PBCa}) and (\ref{2PBCb}),
respectively. Also, 1PB emerges at $k=1$, since $g^{(2)}(0)<1$.
PIT occurs at $k=3$, since $g^{(4)}(0)>g^{(3)}(0)>g^{(2)}(0)>1$
fulfills the conditions given in Eq.~(\ref{PITc}) [see the inset
in panel~(a-i)]. These 1PB, 2PB, and PIT can also be recognized
from (b) the photon-number distributions and (c) the deviations
given in Eq.~(\ref{deviation}). At $k=1$, single-photon
probability is enhanced as $P(1)>\mathcal{P}(1)$, while $m$-photon
($m>1$) probabilities are suppressed as $P(m)<\mathcal{P}(m)$ [see
panels~(b-i) and (c-i)]. These photon-number distributions fulfill
the conditions given in Eqs.~(\ref{PosCa}) and (\ref{PosCb}) for
$n=1$, i.e., resulting in 1PB. At $k=2$, only two-photon
probability $P(2)$ is enhanced [see panels~(b-i), (b-ii) and
(c-ii)]. These photon-number distributions fulfill the conditions
given in Eqs.~(\ref{PosCa}) and (\ref{PosCb}) for $n=2$, i.e.,
resulting in 2PB. At $k=3$, single-photon probability is
suppressed as $P(1)<\mathcal{P}(1)$, while $m$-photon ($m>1$)
probabilities are enhanced as $P(m)>\mathcal{P}(m)$ [see
panels~(b-i) and (c-ii)], i.e., resulting in PIT. Here,
$P_{\text{in}}=0.3\,\mathrm{pW}$, and the other parameters are the
same as those in Fig.~\ref{1PB}.} \label{2PB}
\end{figure}

Similarly, the 2PB has to fulfill the criteria in Eqs.~(\ref{gCa})
and (\ref{gCb}) for $n=2$:
\begin{subequations}
\begin{align}
(i)\quad &g^{(3)}(0)<\exp\left({-\langle\hat{m}\rangle}\right)\equiv f, \label{2PBCa} \\
(ii)\quad &g^{(2)}(0)\geq
\exp\left({-\langle\hat{m}\rangle}\right)+\langle\hat{m}\rangle\cdot
g^{(3)}(0)\equiv f^{(2)}. \label{2PBCb}
\end{align}
\end{subequations}
As expected from the intuitive picture discussed in
Sec.~\ref{OPB}, 2PB occurs at $\Delta_L=-U$ ($k=2$), since the
correlation functions fulfill the conditions of 2PB given in
Eqs.~(\ref{2PBCa}) and (\ref{2PBCb}) [see Fig.~\ref{2PB}(a)]. We
find that, at $k=2$, $g^{(3)}(0)$ is smaller than $f$ defined in
the criterion given in Eq.~(\ref{2PBCa}), while $g^{(2)}(0)$ is
greater than $f^{(2)}$ defined in the criterion given in
Eq.~(\ref{2PBCb}). Here, 2PB indicates that the two-photon
probability is enhanced as $P(2)>\mathcal{P}(2)$, while the other
photon-number probabilities are suppressed, as shown in
Figs.~\ref{2PB}(b) and \ref{2PB}(c-ii). In Fig.~\ref{1PB}, there
is PIT at $k=2$. However, in Fig.~\ref{2PB}, there is 2PB at $k=2$
with an enhanced input power. We note that it is necessary to
properly increase the driving power to obtain a good-quality 2PB,
since we need a larger average photon number. Thus, we enhance the
input power from $P_{\text{in}}=2\,\mathrm{fW}$ (Fig.~\ref{1PB})
to $P_{\text{in}}=0.3\,\mathrm{pW}$ (Fig.~\ref{2PB}). Also, the
1PB still emerges at $k=1$, since the second-order correlation
function fulfills $g^{(2)}(0)<1$ [see Fig.~\ref{2PB}(a)], or only
the single-photon probability is enhanced at $k=1$ [see
Figs.~\ref{2PB}(b) and \ref{2PB}(c-i)].

At $k=3$, we find the correlation functions fulfill
$g^{(4)}(0)>g^{(3)}(0)>g^{(2)}(0)>1$, as shown in the inset in
Fig.~\ref{1PB}(a). It shows PIT occurs at $k=3$, since the
correlation functions satisfy the conditions given in
Eq.~(\ref{PITc}). PIT can also be recognized from the
photon-number distributions and the deviations given in
Eq.~(\ref{deviation}). As shown in Figs.~\ref{2PB}(b) and
\ref{2PB}(c-iii), we find that $P(1)<\mathcal{P}(1)$,
$P(2)>\mathcal{P}(2)$, $P(3)>\mathcal{P}(3)$, and
$P(4)>\mathcal{P}(4)$ at $k=3$. This is a clear signature of PIT.
Since the case for $k=3$ corresponds to a three-photon resonance,
we refer to this PIT as three-photon resonance-induced PIT.

{ In a sense, light with $g^{(3)}(0)\sim1000$ has three-photon
correlations 1000 stronger than those for coherent light. We note
that the ratio of $g^{(3)}(0)/g^{(2)}(0)$ can be quite large. For
example, $g^{(3)}(0)/g^{(2)}(0)\sim 100$ can be seen in
Fig.~\ref{2PB}(a). A similar prediction $g^{(3)}(0)/g^{(2)}(0)\sim
100$ has been reported in~\cite{SRundquist14}. This is possible
since the mean photon number is $\langle \hat n\rangle\ll 1$. For
example, if additionally $\langle\hat a^{\dag 3}\hat
a^3\rangle\approx \langle \hat a^{\dag 2}\hat a^2\rangle$, then
$g^{(3)}(0)/g^{(2)}(0)\approx 1/\langle \hat n\rangle\gg 1$. }

\section{analytic solution of the optical intensity correlation functions}\label{ASOICF}
\subsection{Second-order correlation function}\label{SCF}
According to the quantum trajectory method
~\cite{Splenio1998quantum}, we introduce an anti-Hermitian term to
the Hamiltonian in Eq.~(\ref{Heff}) to describe the dissipation of
the cavity photons. The effective non-Hermitian Hamiltonian is,
thus, given by
\begin{align}
\hat{H}_{\text{t}}=\hbar(\Delta_{L}+\Delta_{{F}})\hat{a}^{\dagger}\hat{a}+\hbar
U\hat{a}^{\dagger}\hat{a}^{\dagger}\hat{a}\hat{a}+\hbar\xi(\hat{a}^{\dagger}+\hat{a})-i\hbar\frac{\gamma}{2}\hat{a}^{\dagger}\hat{a},\label{Htotal}
\end{align}
where $\gamma$ is the rate of the cavity dissipation. Then the
Hamiltonian~(\ref{Htotal}) can be expressed in a \emph{spectral
representation} as
\begin{align*}
\hat{H}_{\text{t}} & =\sum_{n=0}^{\infty}\left(E_{n}-i\hbar\frac{\gamma}{2}n\right)|n\rangle\left\langle n\right|+\hbar\xi\sum_{n=0}^{\infty}|n\rangle\left\langle n\right|(\hat{a}^{\dagger}+\hat{a})\sum_{n'=0}^{\infty}|n'\rangle\left\langle n'\right|\\
& =\sum_{n=0}^{\infty}\left(E_{n}-i\hbar\frac{\gamma}{2}n\right)|n\rangle\left\langle n\right|+\hbar\xi\sum_{n=0}^{\infty}\sum_{n'=0}^{\infty}|n\rangle(\left\langle n\right|\hat{a}^{\dagger}|n'\rangle+\left\langle n\right|\hat{a}|n'\rangle)\left\langle n'\right|\\
& =\sum_{n=0}^{\infty}\left(E_{n}-i\hbar\frac{\gamma}{2}n\right)|n\rangle\left\langle n\right|+\hbar\xi\sum_{n=0}^{\infty}\sum_{n'=0}^{\infty}|n\rangle(\sqrt{n'+1}\langle n|n'+1\rangle+\sqrt{n'}\langle n|n'-1\rangle)\left\langle n'\right|\\
& =\sum_{n=0}^{\infty}\left(E_{n}-i\hbar\frac{\gamma}{2}n\right)|n\rangle\left\langle n\right|+\hbar\xi\sum_{n=0}^{\infty}\sum_{n'=0}^{\infty}|n\rangle(\sqrt{n'+1}\delta_{n,n'+1}+\sqrt{n'}\delta_{n,n'-1})\left\langle n'\right| \\
&=\sum_{n=0}^{\infty}\left(E_{n}-i\hbar\frac{\gamma}{2}n\right)|n\rangle\left\langle n\right|+\hbar\xi\sum_{n=0}^{\infty}\sum_{n'=0}^{\infty}|n\rangle(\sqrt{n'+1}\delta_{n,n'+1}+\sqrt{n'}\delta_{n',n+1})\left\langle n'\right| \tag{i} \\
& =\sum_{n=0}^{\infty}\left(E_{n}-i\hbar\frac{\gamma}{2}n\right)|n\rangle\left\langle n\right|+\hbar\xi\sum_{n'=0}^{\infty}\sqrt{n'+1}|n'+1\rangle\left\langle n'\right|+\hbar\xi\sum_{n=0}^{\infty}\sqrt{n+1}|n\rangle\left\langle n+1\right| \\
&
=\sum_{n=0}^{\infty}\left(E_{n}-i\hbar\frac{\gamma}{2}n\right)|n\rangle\left\langle
n\right|+\hbar\xi\sum_{n=0}^{\infty}\sqrt{n+1}|n+1\rangle\left\langle
n\right|+\hbar\xi\sum_{n=0}^{\infty}\sqrt{n+1}|n\rangle\left\langle
n+1\right|,\tag{ii}
\end{align*}
(i) To avoid negative $n$, we changed the subscript of the second
$\delta$; Also, (ii) we substituted $n$ for $n'$, for convenience.
Therefore, we obtain the Hamiltonian of the whole system as
\begin{equation}\label{Ht}
\hat{H}_{\text{t}}=\sum_{n=0}^{\infty}\left(E_{n}-i\hbar\frac{\gamma}{2}n\right)|n\rangle\left\langle
n\right|+\hbar\xi\sum_{n=0}^{\infty}\sqrt{n+1}|n+1\rangle\left\langle
n\right|+\hbar\xi\sum_{n=0}^{\infty}\sqrt{n+1}|n\rangle\left\langle
n+1\right|,
\end{equation}
with eigenenergies
\begin{equation}
E_{n}=n\hbar\Delta_{L}+ n\hbar\Delta_{{F}}+(n^{2}-n)\hbar
U,\label{energy}
\end{equation}
where $\Delta_{{F}}>0$ ($\Delta_{{F}}<0$) denotes the light
propagating against (along) the direction of the spinning
resonator.

For the weak-driving case, we restrict to a subspace spanned by
the basis states $\{|0\rangle,|1\rangle,|2\rangle\}$. Then, the
Hamiltonian in Eq.~(\ref{Ht}) becomes
\begin{align*}
\hat{H}_{\text{t}} & =E_{0}|0\rangle\left\langle 0\right|+\left(E_{1}-i\hbar\frac{\gamma}{2}\right)|1\rangle\left\langle 1\right|+(E_{2}-i\hbar\gamma)|2\rangle\left\langle 2\right|\\
 & \quad+\hbar\xi\sqrt{1}|1\rangle\left\langle 0\right|+\hbar\xi\sqrt{2}|2\rangle\left\langle 1\right|+\hbar\xi\sqrt{3}|3\rangle\left\langle 2\right|\\
 & \quad+\hbar\xi\sqrt{1}|0\rangle\left\langle 1\right|+\hbar\xi\sqrt{2}|1\rangle\left\langle 2\right|+\hbar\xi\sqrt{3}|2\rangle\left\langle 3\right|.
\end{align*}
Due to the limits of the basis states, the terms including
$|3\rangle$ can be neglected. Then we have
\begin{equation}
\hat{H}_{\text{t}}=E_{0}|0\rangle\left\langle
0\right|+\left(E_{1}-i\hbar\frac{\gamma}{2}\right)|1\rangle\left\langle
1\right|+(E_{2}-i\hbar\gamma)|2\rangle\left\langle
2\right|+\hbar\xi|1\rangle\left\langle
0\right|+\hbar\xi\sqrt{2}|2\rangle\left\langle
1\right|+\hbar\xi|0\rangle\left\langle
1\right|+\hbar\xi\sqrt{2}|1\rangle\left\langle
2\right|,\label{H2P}
\end{equation}
where:
\begin{align}
E_{0} & =0,\nonumber \\
E_{1} & =\hbar\Delta_{L}+\hbar\Delta_{{F}},\nonumber \\
E_{2} & =2\hbar\Delta_{L}+2\hbar\Delta_{{F}}+2\hbar U.\label{E2P}
\end{align}
In this subspace, a general state can be written as
\begin{equation}
|\varphi(t)\rangle=\sum_{n=0}^{2}C_{n}(t)|n\rangle=C_{0}(t)|0\rangle+C_{1}(t)|1\rangle+C_{2}(t)|2\rangle.\label{S2P}
\end{equation}
where $C_{n}$ are probability amplitudes. We substitute the
Hamiltonian~(\ref{H2P}) and the general state~(\ref{S2P}) into the
Schr\"{o}dinger equation
\begin{equation}
i\hbar|\dot{\varphi}(t)\rangle=\hat{H}_{\text{t}}|\varphi(t)\rangle.\label{SchroE}
\end{equation}
Then we have
\begin{equation}
i\hbar|\dot{\varphi}(t)\rangle
=i\hbar\dot{C}_{0}(t)|0\rangle+i\hbar\dot{C}_{1}(t)|1\rangle+i\hbar\dot{C}_{2}(t)|2\rangle,\label{psit2}
\end{equation}
and
\begin{equation}
\hat{H}_{\text{t}}|\varphi(t)\rangle
=\hat{H}_{\text{t}}C_{0}(t)|0\rangle+\hat{H}_{\text{t}}C_{1}(t)|1\rangle+\hat{H}_{\text{t}}C_{2}(t)|2\rangle,
\end{equation}
where:
\begin{equation*}
\hat{H}_{\text{t}}C_{0}(t)|0\rangle=(E_{0}|0\rangle\left\langle
0\right|+\hbar\xi|1\rangle\left\langle
0\right|)C_{0}(t)|0\rangle=E_{0}C_{0}(t)|0\rangle+\hbar\xi
C_{0}(t)|1\rangle,
\end{equation*}
\begin{align*}
\hat{H}_{\text{t}}C_{1}(t)|1\rangle & =\left[\left(E_{1}-i\hbar\frac{\gamma}{2}\right)|1\rangle\left\langle 1\right|+\hbar\xi\sqrt{2}|2\rangle\left\langle 1\right|+\hbar\xi|0\rangle\left\langle 1\right|\right]C_{1}(t)|1\rangle \\
 & =\hbar\xi C_{1}(t)|0\rangle+\left(E_{1}-i\hbar\frac{\gamma}{2}\right)C_{1}(t)|1\rangle+\hbar\xi\sqrt{2}C_{1}(t)|2\rangle,
\end{align*}
\begin{equation*}
\hat{H}_{\text{t}}C_{2}(t)|2\rangle
=[(E_{2}-i\hbar\gamma)|2\rangle\left\langle
2\right|+\hbar\xi\sqrt{2}|1\rangle\left\langle
2\right|]C_{2}(t)|2\rangle=\hbar\xi\sqrt{2}C_{2}(t)|1\rangle+(E_{2}-i\hbar\gamma)C_{2}(t)|2\rangle,
\end{equation*}
i.e.,
\begin{align}
\hat{H}_{\text{t}}|\varphi(t)\rangle & =[E_{0}C_{0}(t)+\hbar\xi C_{1}(t)]|0\rangle+\left[\left(E_{1}-i\hbar\frac{\gamma}{2}\right)C_{1}(t)+\hbar\xi C_{0}(t)+\hbar\xi\sqrt{2}C_{2}(t)\right]|1\rangle\nonumber \\
 & \quad+[(E_{2}-i\hbar\gamma)C_{2}(t)+\hbar\xi\sqrt{2}C_{1}(t)]|2\rangle.\label{Hpsi2}
\end{align}
By comparing the coefficients of the same basis states in
Eqs.~(\ref{psit2}) and (\ref{Hpsi2}), we have:
\begin{align*}
i\hbar\dot{C}_{0}(t)|0\rangle & =[E_{0}C_{0}(t)+\hbar\xi C_{1}(t)]|0\rangle,\\
i\hbar\dot{C}_{1}(t)|1\rangle & =\left[\left(E_{1}-i\hbar\frac{\gamma}{2}\right)C_{1}(t)+\hbar\xi C_{0}(t)+\hbar\xi\sqrt{2}C_{2}(t)\right]|1\rangle,\\
i\hbar\dot{C}_{2}(t)|2\rangle & =[(E_{2}-i\hbar\gamma)C_{2}(t)+\hbar\xi\sqrt{2}C_{1}(t)]|2\rangle,\\
\end{align*}
with $\nu_{n}=E_{n}/\hbar$. Then we obtain the following equations
of motion for the probability amplitudes $C_n(t)$:
\begin{align}
\dot{C}_{0}(t) & =-i\nu_{0}C_{0}(t)-i\xi C_{1}(t),\nonumber \\
\dot{C}_{1}(t) & =-i\left(\nu_{1}-i\frac{\gamma}{2}\right)C_{1}(t)-i\xi C_{0}(t)-i\xi\sqrt{2}C_{2}(t),\label{MEqs2} \\
\dot{C}_{2}(t) &
=-i(\nu_{2}-i\gamma)C_{2}(t)-i\xi\sqrt{2}C_{1}(t),\nonumber
\end{align}
where $\nu_{n}=E_{n}/\hbar$.

Weak driving means the driving strength is smaller than the cavity
damping rate $\xi<\gamma$. If there is no driving field, the
cavity field remains in the vacuum. When a weak-driving field is
applied to the cavity, it may excite a single photon or two
photons in the cavity. Thus, we have the following approximate
expressions: $C_{0}\sim1$, $C_{1}\sim{\xi}/{\gamma}$, and
$C_{2}\sim{\xi^{2}}/{\gamma^{2}}$. Then we can approximately solve
the equations in Eq.~(\ref{MEqs2}) using a perturbation method by
discarding higher-order terms in each equation for lower-order
variables. Thus, the Eq.~(\ref{MEqs2}) becomes:
\begin{align}
\dot{C}_{0}(t) & =-i\nu_{0}C_{0}(t),\nonumber \\
\dot{C}_{1}(t) & =-i\left(\nu_{1}-i\frac{\gamma}{2}\right)C_{1}(t)-i\xi C_{0}(t),\label{Eqs2}\\
\dot{C}_{2}(t) &
=-i(\nu_{2}-i\gamma)C_{2}(t)-i\xi\sqrt{2}C_{1}(t),\nonumber
\end{align}
where $\nu_{n}=E_{n}/\hbar$.

For the initially empty cavity, the initial conditions read as:
$C_{0}(0)=C_{0}(0)$, and $C_{1}(0)=C_{2}(0)=0$. Accordingly, the
solution of the zero-photon amplitude can be obtained as
\begin{equation}
C_{0}(t)=C_{0}(0)\exp\left({-i\nu_{0}t}\right).\label{C0_2}
\end{equation}
Hence, the equation for the single-photon amplitude in
Eq.~(\ref{Eqs2}) becomes
\begin{equation}
\dot{C}_{1}(t)=-i\left(\nu_{1}-i\frac{\gamma}{2}\right)C_{1}(t)-i\xi
C_{0}(t)\exp\left({-i\nu_{0}t}\right).\label{C1E_2}
\end{equation}
To solve this equation, we introduce a slowly-varying amplitude:
\begin{align}
C_{1}(t) & =c_{1}(t)\exp\left[{-i\left(\nu_{1}-i\frac{\gamma}{2}\right)t}\right],\nonumber \\
C_{1}(0) & =c_{1}(0).
\end{align}
Then we obtain
\begin{equation}
\dot{C}_{1}(t)=\dot{c}_{1}(t)\exp\left[{-i\left(\nu_{1}-i\frac{\gamma}{2}\right)t}\right]-i\left(\nu_{1}-i\frac{\gamma}{2}\right)c_{1}(t)\exp\left[{-i\left(\nu_{1}-i\frac{\gamma}{2}\right)t}\right],
\end{equation}
and Eq.~(\ref{C1E_2}) becomes:
\begin{align}
\dot{c}_{1}(t)e^{-i\left(\nu_{1}-i\frac{\gamma}{2}\right)t}-i\left(\nu_{1}-i\frac{\gamma}{2}\right)c_{1}(t)e^{-i\left(\nu_{1}-i\frac{\gamma}{2}\right)t} =& -i\left(\nu_{1}-i\frac{\gamma}{2}\right)c_{1}(t)e^{-i\left(\nu_{1}-i\frac{\gamma}{2}\right)t}-i\xi C_{0}(t)e^{-i\nu_{0}t},\nonumber \\
\dot{c}_{1}(t) =& -i\xi
C_{0}(t)\exp\left[{i\left(\nu_{1}-\nu_{0}-i\frac{\gamma}{2}\right)t}\right].\label{C1E_2v2}
\end{align}
The solution can be obtained by integrating both sides of
Eq.~(\ref{C1E_2v2}), as follows:
\begin{align*}
c_{1}(t)-c_{1}(0) & =-i\xi C_{0}(t)\int_{0}^{t}\exp\left[{i\left(\nu_{1}-\nu_{0}-i\frac{\gamma}{2}\right)t'}\right]\ dt',\\
c_{1}(t)-c_{1}(0) & =-i\xi\frac{C_{0}(t)}{i\left(\nu_{1}-\nu_{0}-i\frac{\gamma}{2}\right)}\left\{\exp\left[{i\left(\nu_{1}-\nu_{0}-i\frac{\gamma}{2}\right)t}\right]-1\right\},\\
c_{1}(t)\exp\left[{-i\left(\nu_{1}-i\frac{\gamma}{2}\right)t}\right] & =c_{1}(0)\exp\left[{-i\left(\nu_{1}-i\frac{\gamma}{2}\right)t}\right]-i\xi\frac{C_{0}(t)}{i\left(\nu_{1}-\nu_{0}-i\frac{\gamma}{2}\right)}\left\{\exp\left({-i\nu_{0}t}\right)-\exp\left[{-i\left(\nu_{1}-i\frac{\gamma}{2}\right)t}\right]\right\},\\
C_{1}(t) &
=C_{1}(0)\exp\left[{-i\left(\nu_{1}-i\frac{\gamma}{2}\right)t}\right]-i\xi\frac{C_{0}(t)}{i\left(\nu_{1}-\nu_{0}-i\frac{\gamma}{2}\right)}\left\{\exp\left({-i\nu_{0}t}\right)-\exp\left[{-i\left(\nu_{1}-i\frac{\gamma}{2}\right)t}\right]\right\}.
\end{align*}
With the initial condition $C_{1}(0)=0$, we have the solution for
the single-photon amplitude given by
\begin{equation}
C_{1}(t)=-i\xi\frac{C_{0}(t)}{i\left(\nu_{1}-\nu_{0}-i\frac{\gamma}{2}\right)}\left\{\exp\left({-i\nu_{0}t}\right)-\exp\left[{-i\left(\nu_{1}-i\frac{\gamma}{2}\right)t}\right]\right\}.\label{C1_2}
\end{equation}

Consider the solution of the single-photon amplitude in
Eq.~(\ref{C1_2}), the equation for the two-photon amplitude in
Eq.~(\ref{Eqs2}) becomes
\begin{equation}
\dot{C}_{2}(t)=-i(\nu_{2}-i\gamma)C_{2}(t)-\sqrt{2}\xi^{2}\frac{C_{0}(t)}{i\left(\nu_{1}-\nu_{0}-i\frac{\gamma}{2}\right)}\left\{\exp\left({-i\nu_{0}t}\right)-\exp\left[{-i\left(\nu_{1}-i\frac{\gamma}{2}\right)t}\right]\right\}.\label{C2E_2}
\end{equation}
To solve this equation, we introduce another slowly-varying
amplitude:
\begin{align}
C_{2}(t) & =c_{2}(t)\exp\left[{-i(\nu_{2}-i\gamma)t}\right],\nonumber \\
C_{2}(0) & =c_{2}(0),
\end{align}
and obtain
\begin{equation}
\dot{C}_{2}(t)=\dot{c}_{2}(t)\exp\left[{-i(\nu_{2}-i\gamma)t}\right]-i(\nu_{2}-i\gamma)c_{2}(t)\exp\left[{-i(\nu_{2}-i\gamma)t}\right],
\end{equation}
then Eq.~(\ref{C2E_2}) becomes:
\begin{align}
&\dot{c}_{2}(t)e^{-i(\nu_{2}-i\gamma)t}-i(\nu_{2}-i\gamma)c_{2}(t)e^{-i(\nu_{2}-i\gamma)t}=-i(\nu_{2}-i\gamma)c_{2}(t)e^{-i(\nu_{2}-i\gamma)t}-\sqrt{2}\xi^{2}\frac{C_{0}(t)}{i\left(\nu_{1}-\nu_{0}-i\frac{\gamma}{2}\right)}\left[e^{-i\nu_{0}t}-e^{-i\left(\nu_{1}-i\frac{\gamma}{2}\right)t}\right],\nonumber \\
&\dot{c}_{2}(t)=-\sqrt{2}\xi^{2}\frac{C_{0}(t)}{i\left(\nu_{1}-\nu_{0}-i\frac{\gamma}{2}\right)}\left\{\exp\left[{i(\nu_{2}-\nu_{0}-i\gamma)t}\right]-\exp\left[{i\left(\nu_{2}-\nu_{1}-i\frac{\gamma}{2}\right)t}\right]\right\}.\label{C2E_2v2}
\end{align}
The solution can also be obtained by integrating both sides of
Eq.~(\ref{C2E_2v2}), as follows:
\begin{align*}
c_{2}(t)-c_{2}(0) =& -\sqrt{2}\xi^{2}\frac{C_{0}(t)}{i\left(\nu_{1}-\nu_{0}-i\frac{\gamma}{2}\right)}\int_{0}^{t}\left\{\exp[{i(\nu_{2}-\nu_{0}-i\gamma)t'}]-\exp\left[{i\left(\nu_{2}-\nu_{1}-i\frac{\gamma}{2}\right)t'}\right]\right\}\ dt',\\
c_{2}(t)-c_{2}(0) =& -\sqrt{2}\xi^{2}\frac{C_{0}(t)}{i\left(\nu_{1}-\nu_{0}-i\frac{\gamma}{2}\right)}\left\{\frac{\exp\left[{i(\nu_{2}-\nu_{0}-i\gamma)t}\right]-1}{i(\nu_{2}-\nu_{0}-i\gamma)}-\frac{\exp\left[{i\left(\nu_{2}-\nu_{1}-i\frac{\gamma}{2}\right)t}\right]-1}{i\left(\nu_{2}-\nu_{1}-i\frac{\gamma}{2}\right)}\right\},\\
c_{2}(t)\exp\left[{-i(\nu_{2}-i\gamma)t}\right] =& c_{2}(0)\exp\left[{-i(\nu_{2}-i\gamma)t}\right]-\sqrt{2}\xi^{2}\frac{C_{0}(t)}{i\left(\nu_{1}-\nu_{0}-i\frac{\gamma}{2}\right)}\cdot\frac{\exp\left({-i\nu_{0}t}\right)-\exp\left[{-i(\nu_{2}-i\gamma)t}\right]}{i(\nu_{2}-\nu_{0}-i\gamma)}\\
& +\sqrt{2}\xi^{2}\frac{C_{0}(t)}{i\left(\nu_{1}-\nu_{0}-i\frac{\gamma}{2}\right)}\cdot\frac{\exp\left[{-i\left(\nu_{1}-i\frac{\gamma}{2}\right)t}\right]-\exp\left[{-i(\nu_{2}-i\gamma)t}\right]}{i\left(\nu_{2}-\nu_{1}-i\frac{\gamma}{2}\right)},\\
C_{2}(t) = & C_{2}(0)\exp\left[{-i(\nu_{2}-i\gamma)t}\right]-\sqrt{2}\xi^{2}\frac{C_{0}(t)}{i\left(\nu_{1}-\nu_{0}-i\frac{\gamma}{2}\right)}\cdot\frac{\exp\left({-i\nu_{0}t}\right)-\exp\left[{-i(\nu_{2}-i\gamma)t}\right]}{i(\nu_{2}-\nu_{0}-i\gamma)}\\
&+\sqrt{2}\xi^{2}\frac{C_{0}(t)}{i\left(\nu_{1}-\nu_{0}-i\frac{\gamma}{2}\right)}\cdot\frac{\exp\left[{-i\left(\nu_{1}-i\frac{\gamma}{2}\right)t}\right]-\exp\left[{-i(\nu_{2}-i\gamma)t}\right]}{i\left(\nu_{2}-\nu_{1}-i\frac{\gamma}{2}\right)}.
\end{align*}
With the initial condition $C_{2}(0)=0$, we have the following
solution of the two-photon amplitude
\begin{equation}
C_{2}(t)=\sqrt{2}\xi^{2}\frac{C_{0}(t)}{\left(\nu_{1}-\nu_{0}-i\frac{\gamma}{2}\right)}\left\{\frac{\exp\left({-i\nu_{0}t}\right)-\exp\left[{-i(\nu_{2}-i\gamma)t}\right]}{(\nu_{2}-\nu_{0}-i\gamma)}-\frac{\exp\left[{-i\left(\nu_{1}-i\frac{\gamma}{2}\right)t}\right]-\exp\left[{-i(\nu_{2}-i\gamma)t}\right]}{\left(\nu_{2}-\nu_{1}-i\frac{\gamma}{2}\right)}\right\}.\label{C2_2}
\end{equation}

Thus, for the initially empty resonator, the solutions of the
equations of motion for the probability amplitudes in the
equations in Eq.~(\ref{Eqs2}) can be obtained as:
\begin{align}
C_{0}(t) & =C_{0}(0)\exp\left({-i\nu_{0}t}\right),\nonumber \\
C_{1}(t) & =-\xi\frac{C_{0}(t)}{\left(\nu_{1}-\nu_{0}-i\frac{\gamma}{2}\right)}\left\{\exp\left({-i\nu_{0}t}\right)-\exp\left[{-i\left(\nu_{1}-i\frac{\gamma}{2}\right)t}\right]\right\},\nonumber \\
C_{2}(t) &
=\sqrt{2}\xi^{2}\frac{C_{0}(t)}{\left(\nu_{1}-\nu_{0}-i\frac{\gamma}{2}\right)}\left\{\frac{\exp\left({-i\nu_{0}t}\right)-\exp\left[{-i(\nu_{2}-i\gamma)t}\right]}{(\nu_{2}-\nu_{0}-i\gamma)}-\frac{\exp\left[{-i\left(\nu_{1}-i\frac{\gamma}{2}\right)t}\right]-\exp\left[{-i(\nu_{2}-i\gamma)t}\right]}{\left(\nu_{2}-\nu_{1}-i\frac{\gamma}{2}\right)}\right\},\label{Solution2}
\end{align}
where
\begin{align*}
\nu_{0} =0,~~~\nu_{1} =\Delta_{L}+\Delta_{{F}},~~~\nu_{2}
=2\Delta_{L}+2\Delta_{{F}}+2U.
\end{align*}
When the initial state of the system is the vacuum state
$\left|0\right\rangle $, i.e., the initial condition $C_{0}(0)=1$,
then the solutions in Eq.~(\ref{Solution2}) are reduced to:
\begin{align}
C_{0}(t)= & 1,\nonumber \\
C_{1}(t)= & -i\xi\frac{1}{i\left(\Delta_{L}+\Delta_{{F}}-i\frac{\gamma}{2}\right)}\left\{1-\exp\left[{-i\left(\Delta_{L}+\Delta_{{F}}-i\frac{\gamma}{2}\right)t}\right]\right\},\nonumber \\
C_{2}(t)= & \frac{\sqrt{2}\xi^{2}}{\left(\Delta_{L}+\Delta_{{F}}-i\frac{\gamma}{2}\right)}\left\{\frac{1-\exp[{-i(2\Delta_{L}+2\Delta_{{F}}+2U-i\gamma)t}]}{(2\Delta_{L}+2\Delta_{{F}}+2U-i\gamma)}-\frac{\exp\left[{-i\left(\Delta_{L}+\Delta_{{F}}-i\frac{\gamma}{2}\right)t}\right]}{(\Delta_{L}+\Delta_{{F}}+2U-i\frac{\gamma}{2})}\right\}\nonumber\\
&+\sqrt{2}\xi^{2}\cdot\frac{-\exp[{i(2\Delta_{L}+2\Delta_{{F}}+2U-i\gamma)t}]}{\left(\Delta_{L}+\Delta_{{F}}-i\frac{\gamma}{2}\right)(\Delta_{L}+\Delta_{{F}}+2U-i\frac{\gamma}{2})},
\end{align}
and for the infinite-time limit $\exp({-At})\to0\ (t\to\infty)$,
we have:
\begin{align}
C_{0}(\infty) & \equiv C_{0}=1,\nonumber \\
C_{1}(\infty) & \equiv C_{1}=\frac{-\xi}{\left(\Delta_{L}+\Delta_{{F}}-i\frac{\gamma}{2}\right)},\nonumber \\
C_{2}(\infty) & \equiv
C_{2}=\frac{-\sqrt{2}{\xi}C_{1}}{(2\Delta_{L}+2\Delta_{{F}}+2U-i\gamma)}.\label{GSolution2}
\end{align}

For the state given in Eq.~(\ref{S2P}), the infinite-time state
(steady state) of the system reads as
\begin{equation}
|\varphi(t\to\infty)\rangle=|0\rangle+\frac{-\xi}{\left(\Delta_{L}+\Delta_{{F}}-i\frac{\gamma}{2}\right)}|1\rangle+\frac{\sqrt{2}\xi^{2}}{\left(\Delta_{L}+\Delta_{{F}}-i\frac{\gamma}{2}\right)(2\Delta_{L}+2\Delta_{{F}}+2U-i\gamma)}|2\rangle,\label{steadystates2}
\end{equation}
and the normalization coefficient of the state is given by
\begin{equation}
N=1+\left|C_{1}\right|^{2}+\left|C_{2}\right|^{2},\label{Normalized2}
\end{equation}
where:
\begin{align}
\left|C_{1}\right|^{2} &
=\left|\frac{\xi}{\left(\Delta_{L}+\Delta_{{F}}-i\frac{\gamma}{2}\right)}\right|^{2}=\frac{\xi^{2}}{\left(\Delta_{L}+\Delta_{{F}}-i\frac{\gamma}{2}\right)\left(\Delta_{L}+\Delta_{{F}}+i\frac{\gamma}{2}\right)}=\frac{\xi^{2}}{\left[(\Delta_{L}+\Delta_{{F}})^{2}+\frac{\gamma^{2}}{4}\right]},
\end{align}
\begin{align}
\left|C_{2}\right|^{2} & =\left|\frac{\sqrt{2}\xi^{2}}{\left(\Delta_{L}+\Delta_{{F}}-i\frac{\gamma}{2}\right)(2\Delta_{L}+2\Delta_{{F}}+2U-i\gamma)}\right|^{2}\nonumber\\
&=\frac{2\xi^{4}}{\left(\Delta_{L}+\Delta_{{F}}-i\frac{\gamma}{2}\right)\left(\Delta_{L}+\Delta_{{F}}+i\frac{\gamma}{2}\right)(2\Delta_{L}+2\Delta_{{F}}+2U-i\gamma)(2\Delta_{L}+2\Delta_{{F}}+2U+i\gamma)}\nonumber\\
&
=\frac{2\xi^{4}}{\left[(\Delta_{L}+\Delta_{{F}})^{2}+\frac{\gamma^{2}}{4}\right][4(\Delta_{L}+\Delta_{{F}}+U)^{2}+\gamma^{2}]}.
\end{align}
The probabilities of finding single and two photons in the cavity
are, respectively, given by:
\begin{align}
P_{1}&=\frac{\left|C_{1}\right|^{2}}{N},\label{P1_2}\\
P_{2}&=\frac{\left|C_{2}\right|^{2}}{N}.\label{P2_2}
\end{align}

As mentioned in Sec.~\ref{COPB}, the equal-time (namely
zero-time-delay) second-order correlation function can be written
as
\[
g^{(2)}(0)\equiv\frac{\left\langle
\hat{a}^{\dagger2}\hat{a}^2\right\rangle }{\left\langle
\hat{a}^{\dagger}\hat{a}\right\rangle ^{2}}=\frac{\left\langle
\hat{a}^{\dagger}\hat{a}\hat{a}^{\dagger}\hat{a}\right\rangle
-\left\langle \hat{a}^{\dagger}\hat{a}\right\rangle}{\left\langle
\hat{a}^{\dagger}\hat{a}\right\rangle ^{2}}.
\]
When the cavity field is in the state given in~(\ref{S2P}), we
have
\begin{align*}
g^{(2)}(0) & =\frac{\sum_{n,n'=0}^{2}C_{n}^{*}C_{n}\left\langle n\right|\hat{a}^{\dagger}\hat{a}\hat{a}^{\dagger}\hat{a}|n'\rangle-\sum_{n,n'=0}^{2}C_{n}^{*}C_{n}\left\langle n\right|\hat{a}^{\dagger}\hat{a}|n'\rangle}{(\sum_{n,n'=0}^{2}C_{n}^{*}C_{n}\left\langle n\right|\hat{a}^{\dagger}\hat{a}|n'\rangle)^{2}}\\
 & =\frac{0+\left|C_{1}\right|^{2}+4\left|C_{2}\right|^{2}-(0+\left|C_{1}\right|^{2}+2\left|C_{2}\right|^{2})}{(0+\left|C_{1}\right|^{2}+2\left|C_{2}\right|^{2})^{2}}\\
 & =\frac{N(P_{1}+4P_{2}-P_{1}-2P_{2})}{N^{2}(P_{1}+2P_{2})^{2}}\\
 & =\frac{2P_{2}}{N(P_{1}+2P_{2})^{2}}.
\end{align*}
In the weak-driving regime, we have the following approximate
formulas: $C_{0}\sim1$, $C_{1}\sim{\xi}/{\gamma}$, and
$C_{2}\sim{\xi^{2}}/{\gamma^{2}}$, i.e., $N\sim1$ with
$\left|C_{2}\right|^{2}\ll\left|C_{1}\right|^{2}\ll1$. Hence, the
second-order correlation function can be written as
\begin{align}
g^{(2)}(0)\approx\frac{2P_{2}}{(P_{1}+2P_{2})^{2}}.\label{g2_2}
\end{align}
Because $P_{1}\gg P_{2}$, we have
\begin{equation}
g^{(2)}(0)\approx\frac{2P_{2}}{P_{1}^{2}}.\label{g2approx_2}
\end{equation}
Substituting Eqs.~(\ref{P1_2}) and (\ref{P2_2}) into
Eq.~(\ref{g2approx_2}), we can easily obtain
\begin{align}
g^{(2)}(0) & \approx\frac{4\xi^{4}}{\left[(\Delta_{L}+\Delta_{{F}})^{2}+\frac{\gamma^{2}}{4}\right][4(\Delta_{L}+\Delta_{{F}}+U)^{2}+\gamma^{2}]}\cdot\frac{\left[(\Delta_{L}+\Delta_{{F}})^{2}+\frac{\gamma^{2}}{4}\right]^{2}}{\xi^{4}}\nonumber \\
 & =\frac{(\Delta_{L}+\Delta_{{F}})^{2}+\gamma^{2}/4}{(\Delta_{L}+\Delta_{{F}}+U)^{2}+\gamma^{2}/4},\label{g2analytical_2}
\end{align}
where $\Delta_{{F}}>0$ ($\Delta_{{F}}<0$) denotes the light
propagating against (along) the direction of the spinning
resonator.

Here, we focus on the nonspinning case ($\Delta_{F}=0$), the
rotating case is discussed in Sec.~\ref{RINPB}. Then, the
second-order correlation function becomes
\begin{equation}
g^{(2)}_0(0) =
\frac{\Delta_{L}^{2}+\gamma^{2}/4}{(\Delta_{L}+U)^{2}+\gamma^{2}/4}.
\label{g2ana_0}
\end{equation}
When the driving laser tuned to a single-photon resonance,
$\Delta_L=0$ ($k=1$), the minimum of $g^{(2)}_0(0)$ is
$g_{0\mathrm{min}}^{(2)}=({\gamma^{2}/4})/({U^{2}+\gamma^{2}/4})=[{4(U/\gamma)^2+1}]^{-1}$.
We have $g_{0\mathrm{min}}^{(2)}<1$, when $U\not=0$. The larger
$U/\gamma$, the smaller is the correlation function
$g_{0\mathrm{min}}^{(2)}$. This indicates that 1PB can be
achieved. On the other hand, for the driving laser tuning to the
two-photon resonance, $\Delta_L=-U$ ($k=2$), there is
$g_{0\mathrm{max}}^{(2)}=({U^{2}+\gamma^{2}/4})/({\gamma^{2}/4})=4(U/\gamma)^2+1$.
We have $g_{0\mathrm{max}}^{(2)}>1$ when $U\not=0$. The larger
$U/\gamma$, the larger is the correlation function
$g_{0\mathrm{max}}^{(2)}$, which indicates a strong photon-induced
tunneling caused by two-photon resonance. In Sec.~\ref{TOCF}, we
find that this conclusion is completely confirmed by our numerical
results.

\subsection{Third-order correlation function}\label{TOCF}
Using a method similar to that in Sec.~\ref{SCF}, we calculate the
third-order photon-number correlation function. For the
weak-driving case, we restrict to a subspace spanned by the basis
states $\{|0\rangle,|1\rangle,|2\rangle,|3\rangle\}$. Then, the
Hamiltonian in Eq.~(\ref{Ht}) becomes
\begin{align*}
\hat{H}_{\text{t}} & =E_{0}|0\rangle\left\langle 0\right|+\left(E_{1}-i\hbar\frac{\gamma}{2}\right)|1\rangle\left\langle 1\right|+(E_{2}-i\hbar\gamma)|2\rangle\left\langle 2\right|+\left(E_{3}-i\hbar\frac{3\gamma}{2}\right)|3\rangle\left\langle 3\right|\\
 & \quad+\hbar\xi\sqrt{1}|1\rangle\left\langle 0\right|+\hbar\xi\sqrt{2}|2\rangle\left\langle 1\right|+\hbar\xi\sqrt{3}|3\rangle\left\langle 2\right|+\hbar\xi\sqrt{4}|4\rangle\left\langle 3\right|\\
 & \quad+\hbar\xi\sqrt{1}|0\rangle\left\langle 1\right|+\hbar\xi\sqrt{2}|1\rangle\left\langle 2\right|+\hbar\xi\sqrt{3}|2\rangle\left\langle 3\right|+\hbar\xi\sqrt{4}|3\rangle\left\langle 4\right|.\\
\end{align*}
Due to the limits of the basis states, the terms including
$|4\rangle$ can be neglected. Then we have
\begin{align}
\hat{H}_{\text{t}} & =E_{0}|0\rangle\left\langle 0\right|+\left(E_{1}-i\hbar\frac{\gamma}{2}\right)|1\rangle\left\langle 1\right|+(E_{2}-i\hbar\gamma)|2\rangle\left\langle 2\right|+\left(E_{3}-i\hbar\frac{3\gamma}{2}\right)|3\rangle\left\langle 3\right|\nonumber \\
 & \quad++\hbar\xi|1\rangle\left\langle 0\right|+\hbar\xi\sqrt{2}|2\rangle\left\langle 1\right|+\hbar\xi\sqrt{3}|3\rangle\left\langle 2\right|+\hbar\xi|0\rangle\left\langle 1\right|+\hbar\xi\sqrt{2}|1\rangle\left\langle 2\right|+\hbar\xi\sqrt{3}|2\rangle\left\langle 3\right|,\label{H3P}
\end{align}
where:
\begin{align}
E_{0} & =0,\nonumber \\
E_{1} & =\hbar\Delta_{L}+\hbar\Delta_{{F}},\nonumber \\
E_{2} & =2\hbar\Delta_{L}+2\hbar\Delta_{{F}}+2\hbar U,\nonumber \\
E_{3} & =3\hbar\Delta_{L}+3\hbar\Delta_{{F}}+6\hbar U.\label{E3P}
\end{align}

In this subspace, a general state can be written as
\begin{equation}
|\varphi(t)\rangle=\sum_{n=0}^{3}C_{n}(t)|n\rangle=C_{0}(t)|0\rangle+C_{1}(t)|1\rangle+C_{2}(t)|2\rangle+C_{3}(t)|3\rangle.\label{S3P}
\end{equation}
where $C_{n}$ are probability amplitudes. We substitute
Hamiltonian~(\ref{H3P}) and the general state~(\ref{S3P}) into the
Schr\"{o}dinger equation~(\ref{SchroE}) to obtain
\begin{equation}
i\hbar|\dot{\varphi}(t)\rangle=i\hbar\dot{C}_{0}(t)|0\rangle+i\hbar\dot{C}_{1}(t)|1\rangle+i\hbar\dot{C}_{2}(t)|2\rangle+i\hbar\dot{C}_{3}(t)|3\rangle;\label{psit3}
\end{equation}
and
\begin{equation}
\hat{H}_{\text{t}}|\varphi(t)\rangle=\hat{H}_{\text{t}}C_{0}(t)|0\rangle+\hat{H}_{\text{t}}C_{1}(t)|1\rangle+\hat{H}_{\text{t}}C_{2}(t)|2\rangle+\hat{H}_{\text{t}}C_{3}(t)|3\rangle,\label{Hpsit3}
\end{equation}
where:
\begin{equation*}
\hat{H}_{\text{t}}C_{0}(t)|0\rangle=[E_{0}|0\rangle\left\langle
0\right|+\hbar\xi|1\rangle\left\langle
0\right|]C_{0}(t)|0\rangle=E_{0}C_{0}(t)|0\rangle+\hbar\xi
C_{0}(t)|1\rangle,
\end{equation*}
\begin{align*}
\hat{H}_{\text{t}}C_{1}(t)|1\rangle & =\left[\left(E_{1}-i\hbar\frac{\gamma}{2}\right)|1\rangle\left\langle 1\right|+\hbar\xi\sqrt{2}|2\rangle\left\langle 1\right|+\hbar\xi|0\rangle\left\langle 1\right|\right]C_{1}(t)|1\rangle\nonumber \\
 & =\hbar\xi C_{1}(t)|0\rangle+\left(E_{1}-i\hbar\frac{\gamma}{2}\right)C_{1}(t)|1\rangle+\hbar\xi\sqrt{2}C_{1}(t)|2\rangle,
\end{align*}
\begin{align*}
\hat{H}_{\text{t}}C_{2}(t)|2\rangle & =[(E_{2}-i\hbar\gamma)|2\rangle\left\langle 2\right|+\hbar\xi\sqrt{2}|1\rangle\left\langle 2\right|+\hbar\xi\sqrt{3}|3\rangle\left\langle 2\right|]C_{2}(t)|2\rangle\nonumber \\
 & =\hbar\xi\sqrt{2}C_{2}(t)|1\rangle+(E_{2}-i\hbar\gamma)C_{2}(t)|2\rangle+\hbar\xi\sqrt{3}C_{2}(t)|3\rangle,
\end{align*}
\begin{equation*}
\hat{H}_{\text{t}}C_{3}(t)|0\rangle=\left[\left(E_{3}-i\hbar\frac{3\gamma}{2}\right)|3\rangle\left\langle
3\right|+\hbar\xi\sqrt{3}|2\rangle\left\langle
3\right|\right]C_{3}(t)|3\rangle=\hbar\xi\sqrt{3}C_{3}(t)|2\rangle+\left(E_{3}-i\hbar\frac{3\gamma}{2}\right)C_{3}(t)|3\rangle,
\end{equation*}
i.e.,
\begin{align}
\hat{H}_{\text{t}}|\varphi(t)\rangle & =[E_{0}C_{0}(t)+\hbar\xi C_{1}(t)]|0\rangle+\left[\hbar\xi C_{0}(t)+\left(E_{1}-i\hbar\frac{\gamma}{2}\right)C_{1}(t)+\hbar\xi\sqrt{2}C_{2}(t)\right]|1\rangle\nonumber \\
 & \quad+[\hbar\xi\sqrt{2}C_{1}(t)+(E_{2}-i\hbar\gamma)C_{2}(t)+\hbar\xi\sqrt{3}C_{3}(t)]|2\rangle+\left[\hbar\xi\sqrt{3}C_{2}(t)+\left(E_{3}-i\hbar\frac{3\gamma}{2}\right)C_{3}(t)\right]|3\rangle.\label{Hpsi3}
\end{align}
By comparing the coefficients of the same basis states in
Eqs.~(\ref{psit3}) and (\ref{Hpsi3}), we have:
\begin{align*}
i\hbar\dot{C}_{0}(t)|0\rangle & =[E_{0}C_{0}(t)+\hbar\xi C_{1}(t)]|0\rangle,\\
i\hbar\dot{C}_{1}(t)|1\rangle & =\left[\hbar\xi C_{0}(t)+\left(E_{1}-i\hbar\frac{\gamma}{2}\right)C_{1}(t)+\hbar\xi\sqrt{2}C_{2}(t)\right]|1\rangle,\\
i\hbar\dot{C}_{2}(t)|2\rangle & =[\hbar\xi\sqrt{2}C_{1}(t)+(E_{2}-i\hbar\gamma)C_{2}(t)+\hbar\xi\sqrt{3}C_{3}(t)]|2\rangle,\\
i\hbar\dot{C}_{3}(t)|3\rangle & =\left[\hbar\xi\sqrt{3}C_{2}(t)+\left(E_{3}-i\hbar\frac{3\gamma}{2}\right)C_{3}(t)\right]|3\rangle,\\
\end{align*}
with $\nu_{n}=E_{n}/\hbar$. Then we obtain the following equations
of motion for the probability amplitudes $C_n(t)$:
\begin{align}
\dot{C}_{0}(t) & =-i\nu_{0}C_{0}(t)-i\xi C_{1}(t),\nonumber \\
\dot{C}_{1}(t) & =-i\xi C_{0}(t)-i\left(\nu_{1}-i\frac{\gamma}{2}\right)C_{1}(t)-i\xi\sqrt{2}C_{2}(t),\nonumber \\
\dot{C}_{2}(t) & =-i\xi\sqrt{2}C_{1}(t)-i(\nu_{2}-i\gamma)C_{2}(t)-i\xi\sqrt{3}C_{3}(t),\nonumber\\
\dot{C}_{3}(t) &
=-i\xi\sqrt{3}C_{2}(t)-i\left(\nu_{3}-i\frac{3\gamma}{2}\right)C_{3}(t),\label{MEqs3}
\end{align}
where $\nu_{n}=E_{n}/\hbar$.

Similarly, due to the weak-driving case, we have the following
approximate formulas: $C_{0}\sim1$, $C_{1}\sim{\xi}/{\gamma}$,
$C_{2}\sim{\xi^{2}}/{\gamma^{2}}$, and
$C_{3}\sim{\xi^{3}}/{\gamma^{3}}$. Then we can approximately solve
the equations in Eq.~(\ref{MEqs3}) using a perturbation method by
discarding higher-order terms in each equation for lower-order
variables. Thus, the Eq.~(\ref{MEqs3}) becomes:
\begin{align}
\dot{C}_{0}(t) & =-i\nu_{0}C_{0}(t),\nonumber \\
\dot{C}_{1}(t) & =-i\left(\nu_{1}-i\frac{\gamma}{2}\right)C_{1}(t)-i\xi C_{0}(t),\nonumber \\
\dot{C}_{2}(t) & =-i(\nu_{2}-i\gamma)C_{2}(t)-i\xi\sqrt{2}C_{1}(t),\nonumber\\
\dot{C}_{3}(t) &
=-i\left(\nu_{3}-i\frac{3\gamma}{2}\right)C_{3}(t)-i\xi\sqrt{3}C_{2}(t),
\label{Eqs3}
\end{align}
where $\nu_{n}=E_{n}/\hbar$.

For an initially empty cavity, the initial conditions read as:
$C_{0}(0)=C_{0}(0)$, and $C_{1}(0)=C_{2}(0)=C_{3}(0)=0$. Then, the
solution of the zero-photon amplitude can be obtained as
\begin{equation}
C_{0}(t)=C_{0}(0)\exp\left({-i\nu_{0}t}\right).\label{C0_3}
\end{equation}
Hence, the equation for the single-photon amplitude in
Eq.~(\ref{Eqs3}) becomes
\begin{equation}
\dot{C}_{1}(t)=-i\left(\nu_{1}-i\frac{\gamma}{2}\right)C_{1}(t)-i\xi
C_{0}(0)\exp\left({-i\nu_{0}t}\right).\label{C1E_3}
\end{equation}
To solve this equation, we introduce a slowly-varying amplitude:
\begin{align}
C_{1}(t) & =c_{1}(t)\exp\left[{-i\left(\nu_{1}-i\frac{\gamma}{2}\right)t}\right],\nonumber \\
C_{1}(0) & =c_{1}(0),
\end{align}
then we obtain
\begin{equation}
\dot{C}_{1}(t)=\dot{c}_{1}(t)\exp\left[{-i\left(\nu_{1}-i\frac{\gamma}{2}\right)t}\right]-i\left(\nu_{1}-i\frac{\gamma}{2}\right)c_{1}(t)\exp\left[{-i\left(\nu_{1}-i\frac{\gamma}{2}\right)t}\right],
\end{equation}
and Eq.~(\ref{C1E_3}) becomes:
\begin{align}
\dot{c}_{1}(t)e^{-i\left(\nu_{1}-i\frac{\gamma}{2}\right)t}-i\left(\nu_{1}-i\frac{\gamma}{2}\right)c_{1}(t)e^{-i\left(\nu_{1}-i\frac{\gamma}{2}\right)t} =& -i\left(\nu_{1}-i\frac{\gamma}{2}\right)c_{1}(t)e^{-i\left(\nu_{1}-i\frac{\gamma}{2}\right)t}-i\xi C_{0}(0)e^{-i\nu_{0}t},\nonumber \\
\dot{c}_{1}(t) =& -i\xi
C_{0}(0)\exp\left[{i\left(\nu_{1}-\nu_{0}-i\frac{\gamma}{2}\right)t}\right].\label{C1E_3v2}
\end{align}
The solution can be obtained by integrating both sides of
Eq.~(\ref{C1E_3v2}), as follows:
\begin{align*}
c_{1}(t)-c_{1}(0) & =-i\xi C_{0}(0)\int_{0}^{t}\exp\left[{i\left(\nu_{1}-\nu_{0}-i\frac{\gamma}{2}\right)t'}\right]\ dt',\\
c_{1}(t)-c_{1}(0) & =-i\xi\frac{C_{0}(0)}{i\left(\nu_{1}-\nu_{0}-i\frac{\gamma}{2}\right)}\left\{\exp\left[{i\left(\nu_{1}-\nu_{0}-i\frac{\gamma}{2}\right)t}\right]-1\right\},\\
c_{1}(t)\exp\left[{-i\left(\nu_{1}-i\frac{\gamma}{2}\right)t}\right] & =c_{1}(0)\exp\left[{-i\left(\nu_{1}-i\frac{\gamma}{2}\right)t}\right]-i\xi\frac{C_{0}(0)}{i\left(\nu_{1}-\nu_{0}-i\frac{\gamma}{2}\right)}\left\{\exp\left({-i\nu_{0}t}\right)-\exp\left[{-i\left(\nu_{1}-i\frac{\gamma}{2}\right)t}\right]\right\},\\
C_{1}(t) &
=C_{1}(0)\exp\left[{-i\left(\nu_{1}-i\frac{\gamma}{2}\right)t}\right]-i\xi\frac{C_{0}(0)}{i\left(\nu_{1}-\nu_{0}-i\frac{\gamma}{2}\right)}\left\{\exp\left({-i\nu_{0}t}\right)-\exp\left[{-i\left(\nu_{1}-i\frac{\gamma}{2}\right)t}\right]\right\}.
\end{align*}
With the initial condition $C_{1}(0)=0$, we have the solution for
the single-photon amplitude given by
\begin{equation}
C_{1}(t)=-i\xi\frac{C_{0}(0)}{i\left(\nu_{1}-\nu_{0}-i\frac{\gamma}{2}\right)}\left\{\exp\left({-i\nu_{0}t}\right)-\exp\left[{-i\left(\nu_{1}-i\frac{\gamma}{2}\right)t}\right]\right\}.\label{C1_3}
\end{equation}

Consider the solution of the single-photon amplitude in
Eq.~(\ref{C1_3}), the equation for the two-photon amplitude in
Eq.~(\ref{Eqs3}) becomes
\begin{equation}
\dot{C}_{2}(t)=-i(\nu_{2}-i\gamma)C_{2}(t)-\sqrt{2}\xi^{2}\frac{C_{0}(0)}{i\left(\nu_{1}-\nu_{0}-i\frac{\gamma}{2}\right)}\left\{\exp\left({-i\nu_{0}t}\right)-\exp\left[{-i\left(\nu_{1}-i\frac{\gamma}{2}\right)t}\right]\right\}.\label{C2E_3}
\end{equation}
To solve this equation, we introduce another slowly-varying
amplitude:
\begin{align}
C_{2}(t) & =c_{2}(t)\exp\left[{-i(\nu_{2}-i\gamma)t}\right],\nonumber \\
C_{2}(0) & =c_{2}(0),
\end{align}
and obtain
\begin{equation}
\dot{C}_{2}(t)=\dot{c}_{2}(t)\exp\left[{-i(\nu_{2}-i\gamma)t}\right]-i(\nu_{2}-i\gamma)c_{2}(t)\exp\left[{-i(\nu_{2}-i\gamma)t}\right],
\end{equation}
then Eq.~(\ref{C2E_3}) becomes:
\begin{align}
&\dot{c}_{2}(t)e^{-i(\nu_{2}-i\gamma)t}-i(\nu_{2}-i\gamma)c_{2}(t)e^{-i(\nu_{2}-i\gamma)t}=-i(\nu_{2}-i\gamma)c_{2}(t)e^{-i(\nu_{2}-i\gamma)t}-\sqrt{2}\xi^{2}\frac{C_{0}(0)}{i\left(\nu_{1}-\nu_{0}-i\frac{\gamma}{2}\right)}\left[e^{-i\nu_{0}t}-e^{-i\left(\nu_{1}-i\frac{\gamma}{2}\right)t}\right],\nonumber \\
& \dot{c}_{2}(t)
=-\sqrt{2}\xi^{2}\frac{C_{0}(0)}{i\left(\nu_{1}-\nu_{0}-i\frac{\gamma}{2}\right)}\left\{\exp\left[{i(\nu_{2}-\nu_{0}-i\gamma)t}\right]-\exp\left[{i\left(\nu_{2}-\nu_{1}-i\frac{\gamma}{2}\right)t}\right]\right\}.\label{C2E_3v2}
\end{align}
The solution can also be obtained by integrating both sides of
Eq.~(\ref{C2E_3v2}), as follows:
\begin{align*}
c_{2}(t)-c_{2}(0) =& -\sqrt{2}\xi^{2}\frac{C_{0}(0)}{i\left(\nu_{1}-\nu_{0}-i\frac{\gamma}{2}\right)}\int_{0}^{t}\left\{\exp[{i(\nu_{2}-\nu_{0}-i\gamma)t'}]-\exp\left[{i\left(\nu_{2}-\nu_{1}-i\frac{\gamma}{2}\right)t'}\right]\right\}\ dt',\\
c_{2}(t)-c_{2}(0) =& -\sqrt{2}\xi^{2}\frac{C_{0}(0)}{i\left(\nu_{1}-\nu_{0}-i\frac{\gamma}{2}\right)}\left\{\frac{\exp\left[{i(\nu_{2}-\nu_{0}-i\gamma)t}\right]-1}{i(\nu_{2}-\nu_{0}-i\gamma)}-\frac{\exp\left[{i\left(\nu_{2}-\nu_{1}-i\frac{\gamma}{2}\right)t}\right]-1}{i\left(\nu_{2}-\nu_{1}-i\frac{\gamma}{2}\right)}\right\},\\
c_{2}(t)\exp\left[{-i(\nu_{2}-i\gamma)t}\right] =& c_{2}(0)\exp\left[{-i(\nu_{2}-i\gamma)t}\right]-\sqrt{2}\xi^{2}\frac{C_{0}(0)}{i\left(\nu_{1}-\nu_{0}-i\frac{\gamma}{2}\right)}\cdot\frac{\exp\left({-i\nu_{0}t}\right)-\exp\left[{-i(\nu_{2}-i\gamma)t}\right]}{i(\nu_{2}-\nu_{0}-i\gamma)}\\
&+\sqrt{2}\xi^{2}\frac{C_{0}(0)}{i\left(\nu_{1}-\nu_{0}-i\frac{\gamma}{2}\right)}\cdot\frac{\exp\left[{-i\left(\nu_{1}-i\frac{\gamma}{2}\right)t}\right]-\exp\left[{-i(\nu_{2}-i\gamma)t}\right]}{i\left(\nu_{2}-\nu_{1}-i\frac{\gamma}{2}\right)},\\
C_{2}(t) =& C_{2}(0)\exp\left[{-i(\nu_{2}-i\gamma)t}\right]-\sqrt{2}\xi^{2}\frac{C_{0}(0)}{i\left(\nu_{1}-\nu_{0}-i\frac{\gamma}{2}\right)}\cdot\frac{\exp\left({-i\nu_{0}t}\right)-\exp\left[{-i(\nu_{2}-i\gamma)t}\right]}{i(\nu_{2}-\nu_{0}-i\gamma)}\\
&+\sqrt{2}\xi^{2}\frac{C_{0}(0)}{i\left(\nu_{1}-\nu_{0}-i\frac{\gamma}{2}\right)}\cdot\frac{\exp\left[{-i\left(\nu_{1}-i\frac{\gamma}{2}\right)t}\right]-\exp\left[{-i(\nu_{2}-i\gamma)t}\right]}{i\left(\nu_{2}-\nu_{1}-i\frac{\gamma}{2}\right)}.
\end{align*}
With the initial condition $C_{2}(0)=0$, we have the following
solution of the two-photon amplitude
\begin{equation}
C_{2}(t)=\sqrt{2}\xi^{2}\frac{C_{0}(0)}{\left(\nu_{1}-\nu_{0}-i\frac{\gamma}{2}\right)}\left\{\frac{\exp\left({-i\nu_{0}t}\right)-\exp\left[{-i(\nu_{2}-i\gamma)t}\right]}{(\nu_{2}-\nu_{0}-i\gamma)}-\frac{\exp\left[{-i\left(\nu_{1}-i\frac{\gamma}{2}\right)t}\right]-\exp\left[{-i(\nu_{2}-i\gamma)t}\right]}{\left(\nu_{2}-\nu_{1}-i\frac{\gamma}{2}\right)}\right\}.\label{C2_3}
\end{equation}

Consider the solution of the two-photon amplitude in
Eq.~(\ref{C2_3}), the equation for the three-photon amplitude in
Eq.~(\ref{Eqs3}) becomes
\begin{align}
\dot{C}_{3}(t)=&-i\left(\nu_{3}-i\frac{3\gamma}{2}\right)C_{3}(t)+\sqrt{6}\xi^{3}C_{0}(0)\frac{\exp\left({-i\nu_{0}t}\right)-\exp\left[{-i(\nu_{2}-i\gamma)t}\right]}{i\left(\nu_{1}-\nu_{0}-i\frac{\gamma}{2}\right)(\nu_{2}-\nu_{0}-i\gamma)}\nonumber\\
&-\sqrt{6}\xi^{3}C_{0}(0)\frac{\exp\left[{-i\left(\nu_{1}-i\frac{\gamma}{2}\right)t}\right]-\exp\left[{-i(\nu_{2}-i\gamma)t}\right]}{i\left(\nu_{1}-\nu_{0}-i\frac{\gamma}{2}\right)\left(\nu_{2}-\nu_{1}-i\frac{\gamma}{2}\right)}.\label{C3E_3}
\end{align}
To solve this equation, we introduce the slowly-varying amplitude:
\begin{align}
C_{3}(t) & =c_{3}(t)\exp\left[{-i\left(\nu_{3}-i\frac{3\gamma}{2}\right)t}\right],\nonumber \\
C_{3}(0) & =c_{3}(0),
\end{align}
and obtain
\begin{equation}
\dot{C}_{3}(t)=\dot{c}_{3}(t)\exp\left[{-i\left(\nu_{3}-i\frac{3\gamma}{2}\right)t}\right]-i\left(\nu_{3}-i\frac{3\gamma}{2}\right)c_{3}(t)\exp\left[{-i\left(\nu_{3}-i\frac{3\gamma}{2}\right)t}\right],
\end{equation}
then Eq.~(\ref{C3E_3}) becomes:
\begin{align*}
\dot{c}_{3}(t)e^{-i\left(\nu_{3}-i\frac{3\gamma}{2}\right)t}-i\left(\nu_{3}-i\frac{3\gamma}{2}\right)c_{3}(t)e^{-i\left(\nu_{3}-i\frac{3\gamma}{2}\right)t}=&-i\left(\nu_{3}-i\frac{3\gamma}{2}\right)C_{3}(t)+\sqrt{6}\xi^{3}C_{0}(0)\frac{e^{-i\nu_{0}t}-e^{-i(\nu_{2}-i\gamma)t}}{i\left(\nu_{1}-\nu_{0}-i\frac{\gamma}{2}\right)(\nu_{2}-\nu_{0}-i\gamma)}\nonumber\\
&-\sqrt{6}\xi^{3}C_{0}(0)\frac{e^{-i\left(\nu_{1}-i\frac{\gamma}{2}\right)t}-e^{-i(\nu_{2}-i\gamma)t}}{i\left(\nu_{1}-\nu_{0}-i\frac{\gamma}{2}\right)\left(\nu_{2}-\nu_{1}-i\frac{\gamma}{2}\right)},
\end{align*}
\begin{align}
\dot{c}_{3}(t)=&\sqrt{6}\xi^{3}C_{0}(0)\frac{\exp\left[{i(\nu_{3}-\nu_{0}-i\frac{3\gamma}{2})t}\right]-\exp\left[{i(\nu_{3}-\nu_{2}-i\frac{\gamma}{2})t}\right]}{i\left(\nu_{1}-\nu_{0}-i\frac{\gamma}{2}\right)(\nu_{2}-\nu_{0}-i\gamma)}\nonumber\\
&-\sqrt{6}\xi^{3}C_{0}(0)\frac{\exp[{i(\nu_{3}-\nu_{1}-i\gamma)t}]-\exp\left[{-i(\nu_{3}-\nu_{2}-i\frac{\gamma}{2})t}\right]}{i\left(\nu_{1}-\nu_{0}-i\frac{\gamma}{2}\right)\left(\nu_{2}-\nu_{1}-i\frac{\gamma}{2}\right)}.\label{C3E_3v2}
\end{align}
The solution can also be obtained by integrating both sides of
Eq.~(\ref{C3E_3v2}), as follows:
\begin{align*}
c_{3}(t)-c_{3}(0) & =\sqrt{6}\xi^{3}\frac{C_{0}(0)}{i\left(\nu_{1}-\nu_{0}-i\frac{\gamma}{2}\right)(\nu_{2}-\nu_{0}-i\gamma)}\int_{0}^{t}\left\{\exp\left[{i(\nu_{3}-\nu_{0}-i\frac{3\gamma}{2})t'}\right]-\exp\left[{i(\nu_{3}-\nu_{2}-i\frac{\gamma}{2})t'}\right]\right\}\ dt'\\
 & \quad-\sqrt{6}\xi^{3}\frac{C_{0}(0)}{i\left(\nu_{1}-\nu_{0}-i\frac{\gamma}{2}\right)\left(\nu_{2}-\nu_{1}-i\frac{\gamma}{2}\right)}\int_{0}^{t}\left\{\exp[{i(\nu_{3}-\nu_{1}-i\gamma)t'}]-\exp\left[{-i(\nu_{3}-\nu_{2}-i\frac{\gamma}{2})t'}\right]\right\}\ dt'\\
& =\sqrt{6}\xi^{3}\frac{C_{0}(0)}{i\left(\nu_{1}-\nu_{0}-i\frac{\gamma}{2}\right)(\nu_{2}-\nu_{0}-i\gamma)}\left\{\frac{\exp\left[{i(\nu_{3}-\nu_{0}-i\frac{3\gamma}{2})t}\right]-1}{i(\nu_{3}-\nu_{0}-i\frac{3\gamma}{2})}-\frac{\exp\left[{i(\nu_{3}-\nu_{2}-i\frac{\gamma}{2})t}\right]-1}{i(\nu_{3}-\nu_{2}-i\frac{\gamma}{2})}\right\}\\
 & \quad-\sqrt{6}\xi^{3}\frac{C_{0}(0)}{i\left(\nu_{1}-\nu_{0}-i\frac{\gamma}{2}\right)\left(\nu_{2}-\nu_{1}-i\frac{\gamma}{2}\right)}\left\{\frac{\exp[{i(\nu_{3}-\nu_{1}-i\gamma)t}]-1}{i(\nu_{3}-\nu_{1}-i\gamma)}-\frac{\exp\left[{i(\nu_{3}-\nu_{2}-i\frac{\gamma}{2})t}\right]-1}{i(\nu_{3}-\nu_{2}-i\frac{\gamma}{2})}\right\},
\end{align*}
\begin{align*}
c_{3}(t)\exp\left[{-i\left(\nu_{3}-i\frac{3\gamma}{2}\right)t}\right] & =c_{3}(0)\exp\left[{-i\left(\nu_{3}-i\frac{3\gamma}{2}\right)t}\right]-\sqrt{6}\xi^{3}\frac{C_{0}(0)\left\{\exp\left({-i\nu_{0}t}\right)-\exp\left[{-i\left(\nu_{3}-i\frac{3\gamma}{2}\right)t}\right]\right\}}{\left(\nu_{1}-\nu_{0}-i\frac{\gamma}{2}\right)(\nu_{2}-\nu_{0}-i\gamma)(\nu_{3}-\nu_{0}-i\frac{3\gamma}{2})}\\
 & \quad+\sqrt{6}\xi^{3}\frac{C_{0}(0)\left\{\exp\left[{-i(\nu_{2}-i\gamma)t}\right]-\exp\left[{-i\left(\nu_{3}-i\frac{3\gamma}{2}\right)t}\right]\right\}}{\left(\nu_{1}-\nu_{0}-i\frac{\gamma}{2}\right)(\nu_{2}-\nu_{0}-i\gamma)(\nu_{3}-\nu_{2}-i\frac{\gamma}{2})}\\
 & \quad+\sqrt{6}\xi^{3}\frac{C_{0}(0)\left\{\exp\left[{-i\left(\nu_{1}-i\frac{\gamma}{2}\right)t}\right]-\exp\left[{-i\left(\nu_{3}-i\frac{3\gamma}{2}\right)t}\right]\right\}}{\left(\nu_{1}-\nu_{0}-i\frac{\gamma}{2}\right)\left(\nu_{2}-\nu_{1}-i\frac{\gamma}{2}\right)(\nu_{3}-\nu_{1}-i\gamma)}\\
 & \quad-\sqrt{6}\xi^{3}\frac{C_{0}(0)\left\{\exp\left[{-i(\nu_{2}-i\gamma)t}\right]-\exp\left[{-i\left(\nu_{3}-i\frac{3\gamma}{2}\right)t}\right]\right\}}{\left(\nu_{1}-\nu_{0}-i\frac{\gamma}{2}\right)\left(\nu_{2}-\nu_{1}-i\frac{\gamma}{2}\right)(\nu_{3}-\nu_{2}-i\frac{\gamma}{2})},\\
C_{3}(t) & =C_{3}(0)\exp\left[{-i\left(\nu_{3}-i\frac{3\gamma}{2}\right)t}\right]-\sqrt{6}\xi^{3}\frac{C_{0}(0)\left\{\exp\left({-i\nu_{0}t}\right)-\exp\left[{-i\left(\nu_{3}-i\frac{3\gamma}{2}\right)t}\right]\right\}}{\left(\nu_{1}-\nu_{0}-i\frac{\gamma}{2}\right)(\nu_{2}-\nu_{0}-i\gamma)(\nu_{3}-\nu_{0}-i\frac{3\gamma}{2})}\\
 & \quad+\sqrt{6}\xi^{3}\frac{C_{0}(0)\left\{\exp\left[{-i(\nu_{2}-i\gamma)t}\right]-\exp\left[{-i\left(\nu_{3}-i\frac{3\gamma}{2}\right)t}\right]\right\}}{\left(\nu_{1}-\nu_{0}-i\frac{\gamma}{2}\right)(\nu_{2}-\nu_{0}-i\gamma)(\nu_{3}-\nu_{2}-i\frac{\gamma}{2})}\\
 & \quad+\sqrt{6}\xi^{3}\frac{C_{0}(0)\left\{\exp\left[{-i\left(\nu_{1}-i\frac{\gamma}{2}\right)t}\right]-\exp\left[{-i\left(\nu_{3}-i\frac{3\gamma}{2}\right)t}\right]\right\}}{\left(\nu_{1}-\nu_{0}-i\frac{\gamma}{2}\right)\left(\nu_{2}-\nu_{1}-i\frac{\gamma}{2}\right)(\nu_{3}-\nu_{1}-i\gamma)}\\
 & \quad-\sqrt{6}\xi^{3}\frac{C_{0}(0)\left\{\exp\left[{-i(\nu_{2}-i\gamma)t}\right]-\exp\left[{-i\left(\nu_{3}-i\frac{3\gamma}{2}\right)t}\right]\right\}}{\left(\nu_{1}-\nu_{0}-i\frac{\gamma}{2}\right)\left(\nu_{2}-\nu_{1}-i\frac{\gamma}{2}\right)(\nu_{3}-\nu_{2}-i\frac{\gamma}{2})}.
\end{align*}
With the initial condition $C_{3}(0)=0$, we have the following
solution of the three-photon amplitude
\begin{align}
C_{3}(t) & =-\sqrt{6}\xi^{3}\frac{C_{0}(0)\left\{\exp\left({-i\nu_{0}t}\right)-\exp\left[{-i\left(\nu_{3}-i\frac{3\gamma}{2}\right)t}\right]\right\}}{\left(\nu_{1}-\nu_{0}-i\frac{\gamma}{2}\right)(\nu_{2}-\nu_{0}-i\gamma)(\nu_{3}-\nu_{0}-i\frac{3\gamma}{2})}\nonumber \\
 & \quad+\sqrt{6}\xi^{3}\frac{C_{0}(0)\left\{\exp\left[{-i(\nu_{2}-i\gamma)t}\right]-\exp\left[{-i\left(\nu_{3}-i\frac{3\gamma}{2}\right)t}\right]\right\}}{\left(\nu_{1}-\nu_{0}-i\frac{\gamma}{2}\right)(\nu_{2}-\nu_{0}-i\gamma)(\nu_{3}-\nu_{2}-i\frac{\gamma}{2})}\nonumber \\
 & \quad+\sqrt{6}\xi^{3}\frac{C_{0}(0)\left\{\exp\left[{-i\left(\nu_{1}-i\frac{\gamma}{2}\right)t}\right]-\exp\left[{-i\left(\nu_{3}-i\frac{3\gamma}{2}\right)t}\right]\right\}}{\left(\nu_{1}-\nu_{0}-i\frac{\gamma}{2}\right)\left(\nu_{2}-\nu_{1}-i\frac{\gamma}{2}\right)(\nu_{3}-\nu_{1}-i\gamma)}\nonumber \\
 & \quad-\sqrt{6}\xi^{3}\frac{C_{0}(0)\left\{\exp\left[{-i(\nu_{2}-i\gamma)t}\right]-\exp\left[{-i\left(\nu_{3}-i\frac{3\gamma}{2}\right)t}\right]\right\}}{\left(\nu_{1}-\nu_{0}-i\frac{\gamma}{2}\right)\left(\nu_{2}-\nu_{1}-i\frac{\gamma}{2}\right)(\nu_{3}-\nu_{2}-i\frac{\gamma}{2})}.\label{C3_3}
\end{align}

Thus, for the initially empty resonator, the solutions of the
equations of motion for the probability amplitudes in the
equations in Eq.~(\ref{Eqs3}) can be obtained as:
\begin{align}
C_{0}(t) & =C_{0}(0)\exp\left({-i\nu_{0}t}\right),\nonumber \\
C_{1}(t) & =-\xi\frac{C_{0}(0)}{\left(\nu_{1}-\nu_{0}-i\frac{\gamma}{2}\right)}\left\{\exp\left({-i\nu_{0}t}\right)-\exp\left[{-i\left(\nu_{1}-i\frac{\gamma}{2}\right)t}\right]\right\},\nonumber \\
C_{2}(t) & =\sqrt{2}\xi^{2}\frac{C_{0}(0)}{\left(\nu_{1}-\nu_{0}-i\frac{\gamma}{2}\right)}\left\{\frac{\exp\left({-i\nu_{0}t}\right)-\exp\left[{-i(\nu_{2}-i\gamma)t}\right]}{(\nu_{2}-\nu_{0}-i\gamma)}-\frac{\exp\left[{-i\left(\nu_{1}-i\frac{\gamma}{2}\right)t}\right]-\exp\left[{-i(\nu_{2}-i\gamma)t}\right]}{\left(\nu_{2}-\nu_{1}-i\frac{\gamma}{2}\right)}\right\},\nonumber \\
C_{3}(t) & =-\sqrt{6}\xi^{3}\frac{C_{0}(0)\left\{\exp\left({-i\nu_{0}t}\right)-\exp\left[{-i\left(\nu_{3}-i\frac{3\gamma}{2}\right)t}\right]\right\}}{\left(\nu_{1}-\nu_{0}-i\frac{\gamma}{2}\right)(\nu_{2}-\nu_{0}-i\gamma)(\nu_{3}-\nu_{0}-i\frac{3\gamma}{2})}\nonumber \\
 & \quad+\sqrt{6}\xi^{3}\frac{C_{0}(0)\left\{\exp\left[{-i(\nu_{2}-i\gamma)t}\right]-\exp\left[{-i\left(\nu_{3}-i\frac{3\gamma}{2}\right)t}\right]\right\}}{\left(\nu_{1}-\nu_{0}-i\frac{\gamma}{2}\right)(\nu_{2}-\nu_{0}-i\gamma)(\nu_{3}-\nu_{2}-i\frac{\gamma}{2})}\nonumber \\
 & \quad+\sqrt{6}\xi^{3}\frac{C_{0}(0)\left\{\exp\left[{-i\left(\nu_{1}-i\frac{\gamma}{2}\right)t}\right]-\exp\left[{-i\left(\nu_{3}-i\frac{3\gamma}{2}\right)t}\right]\right\}}{\left(\nu_{1}-\nu_{0}-i\frac{\gamma}{2}\right)\left(\nu_{2}-\nu_{1}-i\frac{\gamma}{2}\right)(\nu_{3}-\nu_{1}-i\gamma)}\nonumber \\
 & \quad-\sqrt{6}\xi^{3}\frac{C_{0}(0)\left\{\exp\left[{-i(\nu_{2}-i\gamma)t}\right]-\exp\left[{-i\left(\nu_{3}-i\frac{3\gamma}{2}\right)t}\right]\right\}}{\left(\nu_{1}-\nu_{0}-i\frac{\gamma}{2}\right)\left(\nu_{2}-\nu_{1}-i\frac{\gamma}{2}\right)(\nu_{3}-\nu_{2}-i\frac{\gamma}{2})},\label{S3}
\end{align}
where
\begin{align*}
\nu_{0} =0,~~~\nu_{1} =\Delta_{L}+\Delta_{{F}},~~~\nu_{2}
=2\Delta_{L}+2\Delta_{{F}}+2U,~~~\nu_{3}
=3\Delta_{L}+3\Delta_{{F}}+6U.
\end{align*}
When the initial state of the system is the vacuum state
$\left|0\right\rangle $, i.e., the initial condition $C_{0}(0)=1$,
the solutions in Eq.~(\ref{S3}) are reduced to:
\begin{align*}
C_{0}(t) =&1,\\
C_{1}(t) =&-\xi\frac{1}{\left(\Delta_{L}+\Delta_{{F}}-i\frac{\gamma}{2}\right)}\left\{1-\exp\left[{-i\left(\Delta_{L}+\Delta_{{F}}-i\frac{\gamma}{2}\right)t}\right]\right\},\\
C_{2}(t) =&\frac{\sqrt{2}\xi^{2}}{\left(\Delta_{L}+\Delta_{{F}}-i\frac{\gamma}{2}\right)}\left\{\frac{1-\exp[{-i(2\Delta_{L}+2\Delta_{{F}}+2U-i\gamma)t}]}{(2\Delta_{L}+2\Delta_{{F}}+2U-i\gamma)}-\frac{\exp\left[{-i\left(\Delta_{L}+\Delta_{{F}}-i\frac{\gamma}{2}\right)t}\right]}{(\Delta_{L}+\Delta_{{F}}+2U-i\frac{\gamma}{2})}\right\}\\
&+\sqrt{2}\xi^{2}\frac{\exp[{-i(2\Delta_{L}+2\Delta_{{F}}+2U-i\gamma)t}]}{\left(\Delta_{L}+\Delta_{{F}}-i\frac{\gamma}{2}\right)(\Delta_{L}+\Delta_{{F}}+2U-i\frac{\gamma}{2})},\\
\end{align*}
\begin{align*}
C_{3}(t) & =-\sqrt{6}\xi^{3}\frac{\left\{1-\exp\left[{-i\left(3\Delta_{L}+3\Delta_{F}+6U-i\frac{3\gamma}{2}\right)t}\right]\right\}}{\left(\Delta_{L}+\Delta_{{F}}-i\frac{\gamma}{2}\right)(2\Delta_{L}+2\Delta_{{F}}+2U-i\gamma)\left(3\Delta_{L}+3\Delta_{F}+6U-i\frac{3\gamma}{2}\right)}\\
 & \quad+\sqrt{6}\xi^{3}\frac{\left\{\exp[{-i(2\Delta_{L}+2\Delta_{{F}}+2U-i\gamma)t}]-\exp\left[{-i\left(3\Delta_{L}+3\Delta_{F}+6U-i\frac{3\gamma}{2}\right)t}\right]\right\}}{\left(\Delta_{L}+\Delta_{{F}}-i\frac{\gamma}{2}\right)(2\Delta_{L}+2\Delta_{{F}}+2U-i\gamma)(\Delta_{L}+\Delta_{F}+4U-i\frac{\gamma}{2})}\\
 & \quad+\sqrt{6}\xi^{3}\frac{\left\{\exp\left[{-i\left(\Delta_{L}+\Delta_{{F}}-i\frac{\gamma}{2}\right)t}\right]-\exp\left[{-i\left(3\Delta_{L}+3\Delta_{F}+6U-i\frac{3\gamma}{2}\right)t}\right]\right\}}{\left(\Delta_{L}+\Delta_{{F}}-i\frac{\gamma}{2}\right)(\Delta_{L}+\Delta_{{F}}+2U-i\frac{\gamma}{2})(2\Delta_{L}+2\Delta_{F}+6U-i\gamma)}\\
 & \quad-\sqrt{6}\xi^{3}\frac{\left\{\exp[{-i(2\Delta_{L}+2\Delta_{{F}}+2U-i\gamma)t}]-\exp\left[{-i\left(3\Delta_{L}+3\Delta_{F}+6U-i\frac{3\gamma}{2}\right)t}\right]\right\}}{\left(\Delta_{L}+\Delta_{{F}}-i\frac{\gamma}{2}\right)(\Delta_{L}+\Delta_{{F}}+2U-i\frac{\gamma}{2})(\Delta_{L}+\Delta_{F}+4U-i\frac{\gamma}{2})},
\end{align*}
and for the infinite-time limit $\exp({-At})\to0\ (t\to\infty)$,
we have:
\begin{align}
C_{0}(\infty) & \equiv C_{0}=1,\nonumber \\
C_{1}(\infty) & \equiv C_{1}=\frac{-\xi}{\left(\Delta_{L}+\Delta_{{F}}-i\frac{\gamma}{2}\right)},\nonumber \\
C_{2}(\infty) & \equiv C_{2}=\frac{-\sqrt{2}{\xi}C_{1}}{(2\Delta_{L}+2\Delta_{{F}}+2U-i\gamma)},\nonumber \\
C_{3}(\infty) & \equiv
C_{3}=\frac{-\sqrt{3}{\xi}C_{2}}{\left(3\Delta_{L}+3\Delta_{F}+6U-i\frac{3\gamma}{2}\right)}.\label{Solution}
\end{align}

For the state given in Eq.~(\ref{S3P}), the infinite-time state
(steady state) of the system reads as
\begin{align}
\left|\varphi(t\to\infty)\right\rangle  & =|0\rangle+\frac{-\xi}{\left(\Delta_{L}+\Delta_{{F}}-i\frac{\gamma}{2}\right)}|1\rangle+\frac{\sqrt{2}\xi^{2}}{\left(\Delta_{L}+\Delta_{{F}}-i\frac{\gamma}{2}\right)(2\Delta_{L}+2\Delta_{{F}}+2U-i\gamma)}|2\rangle\nonumber \\
 & \quad+\frac{-\sqrt{6}\xi^{3}}{\left(\Delta_{L}+\Delta_{{F}}-i\frac{\gamma}{2}\right)(2\Delta_{L}+2\Delta_{{F}}+2U-i\gamma)\left(3\Delta_{L}+3\Delta_{F}+6U-i\frac{3\gamma}{2}\right)}|3\rangle,\label{steadystates3}
\end{align}
and the normalization constant of the state is given by
\begin{equation}
N=1+\left|C_{1}\right|^{2}+\left|C_{2}\right|^{2}+\left|C_{3}\right|^{2},\label{Normalized3}
\end{equation}
where:
\begin{align}
\left|C_{1}\right|^{2} &
=\left|\frac{\xi}{\left(\Delta_{L}+\Delta_{{F}}-i\frac{\gamma}{2}\right)}\right|^{2}=\frac{\xi^{2}}{\left(\Delta_{L}+\Delta_{{F}}-i\frac{\gamma}{2}\right)\left(\Delta_{L}+\Delta_{{F}}+i\frac{\gamma}{2}\right)}=\frac{\xi^{2}}{\left[(\Delta_{L}+\Delta_{{F}})^{2}+\frac{\gamma^{2}}{4}\right]},
\end{align}
\begin{align}
\left|C_{2}\right|^{2} & =\left|\frac{\sqrt{2}\xi^{2}}{\left(\Delta_{L}+\Delta_{{F}}-i\frac{\gamma}{2}\right)(2\Delta_{L}+2\Delta_{{F}}+2U-i\gamma)}\right|^{2} \nonumber\\
&=\frac{2\xi^{4}}{\left(\Delta_{L}+\Delta_{{F}}-i\frac{\gamma}{2}\right)\left(\Delta_{L}+\Delta_{{F}}+i\frac{\gamma}{2}\right)(2\Delta_{L}+2\Delta_{{F}}+2U-i\gamma)(2\Delta_{L}+2\Delta_{{F}}+2U+i\gamma)}\nonumber\\
&
=\frac{2\xi^{4}}{\left[(\Delta_{L}+\Delta_{{F}})^{2}+\frac{\gamma^{2}}{4}\right][4(\Delta_{L}+\Delta_{{F}}+U)^{2}+\gamma^{2}]},
\end{align}
\begin{align}
\left|C_{3}\right|^{2} & =\left|\frac{-\sqrt{6}\xi^{3}}{\left(\Delta_{L}+\Delta_{{F}}-i\frac{\gamma}{2}\right)(2\Delta_{L}+2\Delta_{{F}}+2U-i\gamma)\left(3\Delta_{L}+3\Delta_{F}+6U-i\frac{3\gamma}{2}\right)}\right|^{2}\nonumber\\
 & =\frac{6\xi^{6}}{\left|\Delta_{L}+\Delta_{{F}}-i\frac{\gamma}{2}\right|^2|(2\Delta_{L}+2\Delta_{{F}}+2U-i\gamma)|^2\left|3\Delta_{L}+3\Delta_{F}+6U-i\frac{3\gamma}{2}\right|^2}\nonumber\\
 & =\frac{6\xi^{6}}{\left[(\Delta_{L}+\Delta_{{F}})^{2}+\frac{\gamma^{2}}{4}\right][4(\Delta_{L}+\Delta_{{F}}+U)^{2}+\gamma^{2}]\left[9(\Delta_{L}+\Delta_{F}+2U)^{2}+\frac{9\gamma^{2}}{4}\right]}.
\end{align}
The probabilities of finding single, two and three photons in the
cavity are, respectively, given by:
\begin{align}
P_{1}&=\frac{\left|C_{1}\right|^{2}}{N},\label{P1_3}\\
P_{2}&=\frac{\left|C_{2}\right|^{2}}{N},\label{P2_3}\\
P_{3}&=\frac{\left|C_{3}\right|^{2}}{N}.\label{P3_3}
\end{align}

As mentioned in Sec.~\ref{COPB}, the equal-time third-order
correlation function can be written as
\begin{align*}
g^{(3)}(0,0) \equiv g^{(3)}(0)\equiv\frac{\left\langle
a^{\dagger3}a^3\right\rangle }{\left\langle
\hat{a}^{\dagger}\hat{a}\right\rangle ^{3}}=\frac{\left\langle
\hat{a}^{\dagger}\hat{a}(\hat{a}^{\dagger}\hat{a}-1)(\hat{a}^{\dagger}\hat{a}-2)\right\rangle}{\left\langle
\hat{a}^{\dagger}\hat{a}\right\rangle ^{3}}=\frac{\left\langle
(\hat{a}^{\dagger}\hat{a})^3-3(\hat{a}^{\dagger}\hat{a})^2+2\hat{a}^{\dagger}\hat{a}\right\rangle
}{\left\langle \hat{a}^{\dagger}\hat{a}\right\rangle ^{3}}
\end{align*}
When the cavity field is in the state~(\ref{S3P}), we have
\begin{align*}
g^{(3)}(0) & =\frac{\sum_{n,n'=0}^{3}C_{n'}^{*}C_{n}\left\langle n'\right|(\hat{a}^{\dagger}\hat{a})^3|n\rangle-3\sum_{n,n'=0}^{3}C_{n'}^{*}C_{n}\left\langle n'\right|(\hat{a}^{\dagger}\hat{a})^2|n\rangle+2\sum_{n,n'=0}^{3}C_{n'}^{*}C_{n}\left\langle n'\right|\hat{a}^{\dagger}\hat{a}|n\rangle}{(\sum_{n,n'=0}^{3}C_{n'}^{*}C_{n}\left\langle n'\right|\hat{a}^{\dagger}\hat{a}|n\rangle)^{3}}\\
 & =\frac{\left|C_{1}\right|^{2}+8\left|C_{2}\right|^{2}+27\left|C_{3}\right|^{2}-3(\left|C_{1}\right|^{2}+4\left|C_{2}\right|^{2}+9\left|C_{3}\right|^{2})+2(\left|C_{1}\right|^{2}+2\left|C_{2}\right|^{2}+3\left|C_{3}\right|^{2})}{(\left|C_{1}\right|^{2}+2\left|C_{2}\right|^{2}+3\left|C_{3}\right|^{2})^{3}}\\
 & =\frac{N(P_{1}+8P_{2}+27P_{3}-3P_{1}-12P_{2}-27P_{3}+2P_{1}+4P_{2}+6P_{3})}{N^{2}(P_{1}+2P_{2}+3P_{3})^{3}}\\
 & =\frac{6P_{3}}{N(P_{1}+2P_{2}+3P_{3})^{3}}\\
\end{align*}
In the weak-driving regime, we have the following approximate
amplitudes: $C_{0}\sim1$, $C_{1}\sim{\xi}/{\gamma}$,
$C_{2}\sim{\xi^{2}}/{\gamma^{2}}$, and
$C_{3}\sim{\xi^{3}}/{\gamma^{3}}$, i.e., $N\sim1$ with
$\left|C_{3}\right|^{2}\ll\left|C_{2}\right|^{2}\ll\left|C_{1}\right|^{2}\ll1$.
Hence, the third-order correlation function can be written as
\begin{align}
g^{(3)}(0)\approx\frac{6P_{3}}{P_{1}^{3}}.\label{g3approx}
\end{align}

Substituting Eqs.~(\ref{P1_3}) and (\ref{P3_3}) into
Eq.~(\ref{g3approx}), we can easily obtain
\begin{align}
g^{(3)}(0) & \approx\frac{36\xi^{6}}{\left[(\Delta_{L}+\Delta_{{F}})^{2}+\frac{\gamma^{2}}{4}\right][4(\Delta_{L}+\Delta_{{F}}+U)^{2}+\gamma^{2}]\left[9(\Delta_{L}+\Delta_{F}+2U)^{2}+\frac{9\gamma^{2}}{4}\right]}\cdot\frac{\left[(\Delta_{L}+\Delta_{{F}})^{2}+\frac{\gamma^{2}}{4}\right]^{3}}{\xi^{6}}\nonumber \\
 & =\frac{\left[(\Delta_{L}+\Delta_{{F}})^{2}+\frac{\gamma^{2}}{4}\right]^{2}}{[(\Delta_{L}+\Delta_{{F}}+U)^{2}+\frac{\gamma^{2}}{4}]\left[(\Delta_{L}+\Delta_{F}+2U)^{2}+\frac{\gamma^{2}}{4}\right]},\label{g3analytical}
\end{align}
where $\Delta_{{F}}>0$ ($\Delta_{{F}}<0$) denotes the light
propagating against (along) the direction of the spinning
resonator.

\begin{figure}[b]
\centering
\includegraphics[width=0.98 \textwidth]{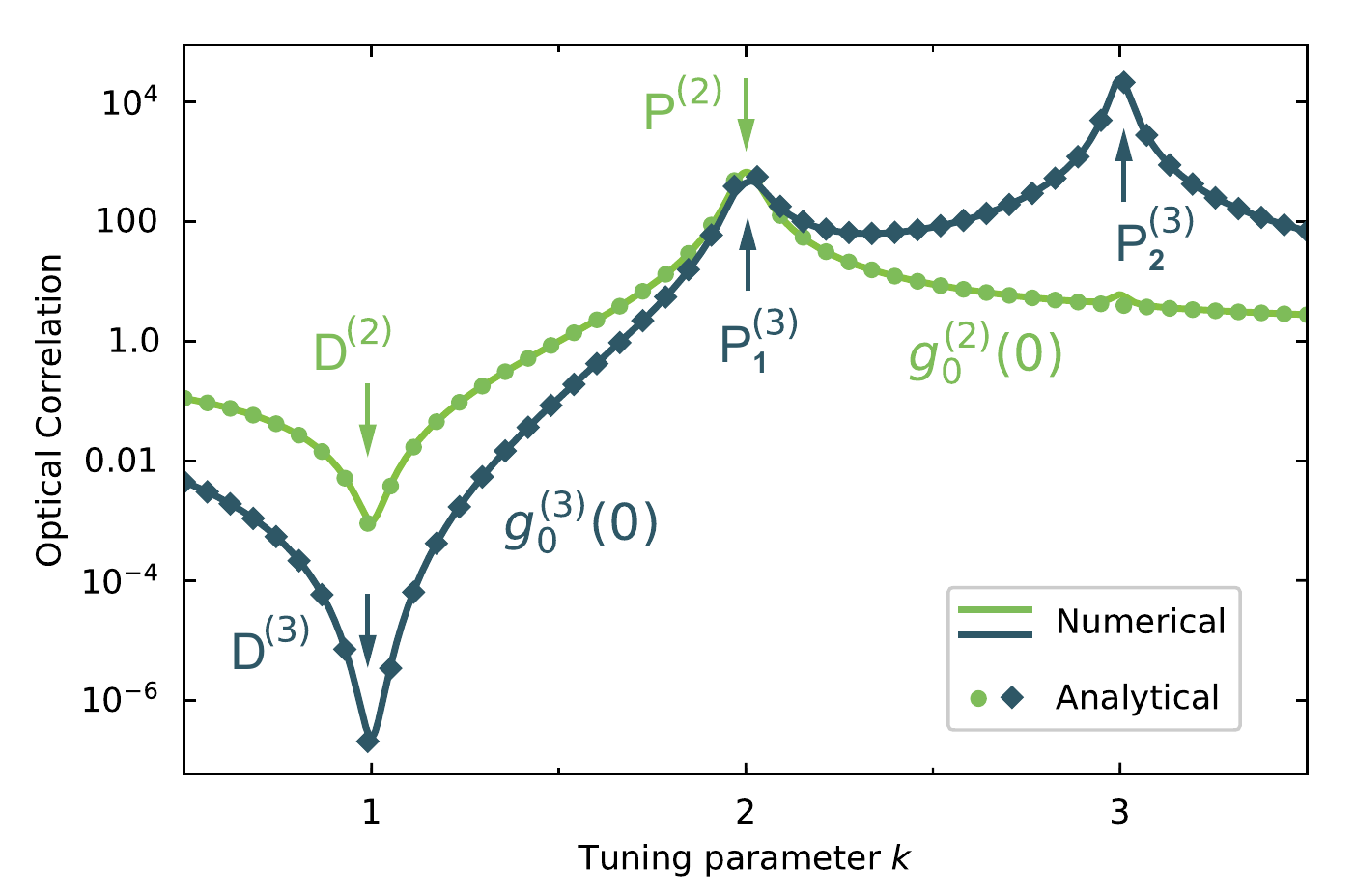}
\caption{The second- and third-order correlation functions versus
the tuning parameter $k$ for the nonspinning resonator case. The
symbols denote our approximate analytical results [$g^{(2)}_0(0)$
given in Eq.~(\ref{g2ana_0}), $g^{(3)}_0(0)$ given in
Eq.~(\ref{g3ana_0})], while the solid curves correspond to our
numerical results. Here, $D^{(2)}$ [$D^{(3)}$] is the dip in the
$g^{(2)}_0(0)$ [$g^{(3)}_0(0)$] curves; $P^{(2)}$ and $P^{(3)}$
are the peaks in the $g^{(2)}_0(0)$ and $g^{(3)}_0(0)$ curves,
respectively. The parameters used here are the same as those in
Fig.~\ref{1PB}.} \label{ana-num}
\end{figure}

Here, we focus on the nonspinning case ($\Delta_{F}=0$), the
rotating case is discussed in Sec.~\ref{RINPB}. For this case, the
third-order correlation function becomes
\begin{equation}
g_{\text{0}}^{(3)}(0)=\frac{(\Delta_{L}^{2}+{\gamma^{2}}/{4})^{2}}{[(\Delta_{L}+U)^{2}+\gamma^{2}/{4}][(\Delta_{L}+2U)^{2}+{\gamma^{2}}/{4}]}.\label{g3ana_0}
\end{equation}
Including the second-order correlation function, we can
quantitatively compare our analytical results with numerical
calculations~\cite{Sjohansson2012qutip,Sjohansson2013qutip2}. We
find an excellent agreement between the numerical calculations and
the approximate analytical solutions, as shown in
Fig.~\ref{ana-num}. Here, the solid curves are plotted using the
numerical solution, while the curves with symbols are based on the
analytical solution given in Eqs.~(\ref{g2ana_0}) and
(\ref{g3ana_0}). As for the $g_{\text{0}}^{(2)}(0)\approx
2P(2)/P(1)^2$, given in Eq.~(\ref{g2approx_2}), the dip $D^{(2)}$
and the peak $P^{(2)}$ in the light green curves correspond to the
single- and two-photon resonant driving cases, respectively. In
the single-photon resonant driving case ($k=1$), a single photon
can be resonantly injected into the cavity, while the probability
of finding two photons in the cavity is largely suppressed due to
the energy restriction; this represents 1PB. We find that the
analytical value of $g_{\text{0}}^{(2)}(0)\sim0.0008$ at this dip
$D^{(2)}$, which is well-matched with our numerical value
$g_{\text{0}}^{(2)}(0)\sim0.0009$. In the two-photon resonant
driving case ($k=2$), the probability for finding two photons
inside the cavity is resonantly enhanced, and this corresponds to
a peak in the curve of $g^{(2)}(0)$. We find that the analytical
value of $g^{(2)}(0)\sim974$ at this peak $P^{(2)}$ is above the
numerical solution $g^{(2)}(0)\sim673$, since we neglected the
two-photon probability in the denominator of the analytical
formula [this can be seen more clearly in Eqs.~(\ref{g2_2}) and
(\ref{g2ana_0})]. As for the $g_{\text{0}}^{(3)}(0)\approx
6P(3)/P(1)^3$, given in Eq.~(\ref{g3approx}), the dip $D^{(3)}$
and the peaks $P^{(3)}_{1}$ and $P^{(3)}_{2}$ in the dark green
curves correspond to the single-, two-, and three-photon
resonant-driving cases, respectively. In the single-photon
resonant-driving case ($k=1$), $P(1)\gg P(2)\gg P(3)$, thus, there
is a dip [i.e., $D^{(3)}$] in the $g_{\text{0}}^{(3)}(0)$ curve.
For the two-photon resonant-driving case ($k=2$), the
single-photon probability is suppressed, which causes the
occurrence of the peak $P^{(3)}_{1}$. However, the peak
$P^{(3)}_{1}$ is lower than the peak $P^{(3)}_{2}$ at $k=3$, since
the three-photon probability is enhanced at $k=3$ (i.e.,
three-photon resonant-driving case), but still suppressed at $k=2$
(i.e., two-photon resonant-driving case).

\section{rotation-induced quantum nonreciprocity}\label{RINPB}
\subsection{Rotation-induced shifts}\label{S_RIS}

For the optical microtoroid resonator, an input-laser light
applied from the left or right side of the cavity causes a
clockwise (CW) circulating mode or a counterclockwise (CCW)
circulating mode. When the microresonator is rotating,
$\Delta_{{F}}>0$ and $\Delta_{{F}}<0$ denote the cases with the
light propagating against and along the spinning direction of the
resonator, respectively, i.e., for the CCW spinning resonator,
$\Delta_{{F}}>0$ ($\Delta_{{F}}<0$) indicates an input-laser
applied from the left (right) side; for the CW spinning resonator,
$\Delta_{{F}}>0$ ($\Delta_{{F}}<0$) indicates an input-laser used
from the right (left) side.

When the resonator is rotating, the second-order correlation
function in Eq.~(\ref{g2analytical_2}) can be written as
\begin{equation}
g_{\pm}^{(2)}(0)
=\frac{(\Delta_{L}\pm\left|\Delta_{{F}}\right|)^{2}+\gamma^{2}/4}{(\Delta_{L}\pm\left|\Delta_{{F}}\right|+U)^{2}+\gamma^{2}/4},\label{g2pm}
\end{equation}
where $g_{-}^{(2)}(0)$ [$g_{+}^{(2)}(0)$] denotes the equal-time
second-order correlation function for $\Delta_{{F}}<0$
($\Delta_{{F}}>0$).

\begin{figure}[bp]
\centering
\includegraphics[width=0.98 \textwidth]{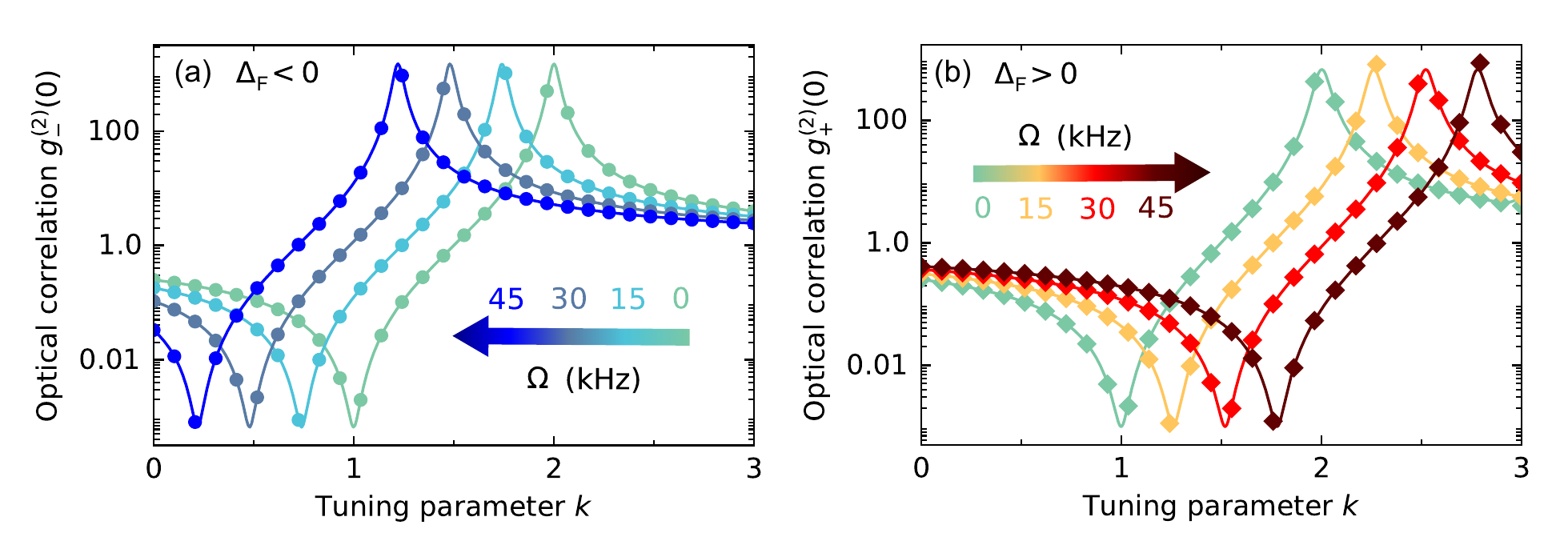}
\caption{Dependence of the equal-time second-order correlation
functions $g^{(2)}_\pm(0)$ on the tuning parameter $k$ for various
values of the angular speed $\Omega$. The symbols are our
approximate analytical results given in Eq.~(\ref{g2pm}), while
the solid curves are our numerical results. The other parameters
used here are the same as those in Fig.~\ref{1PB}.} \label{RIS}
\end{figure}

For the $\Delta_{{F}}<0$ case, 1PB emerges at
$\Delta_{L}=\left|\Delta_{{F}}\right|$ with
$g^{(2)}_-(0)=({\gamma^{2}/4})/({U^{2}+\gamma^{2}/4})=[{4(U/\gamma)^{2}+1}]^{-1}$.
This minimum value of $g_{-}^{(2)}(0)$ is independent of the
angular speed $\Omega$; thus, the minimum value of
$g_{-}^{(2)}(0)$ is a constant. Since $\left|\Delta_{{F}}\right|$
is an amount proportional to the angular speed $\Omega$, the dip
$D^{(2)}$ experiences linearly shifts with $\Omega$. Also,
$D^{(2)}$ experiences linearly shifts to the opposite direction
for the $\Delta_{{F}}<0$ case, since now 1PB emerges at
$\Delta_{L}=-\left|\Delta_{{F}}\right|$. The shifts of the curve
can also be understood from an energy-level structure, where the
rotation of the resonator causes upper or lower shifts of energy
levels, as shown in Fig.~\ref{NRPBmechanism}.

Here, we plot the correlation function $g^{(2)}(0)$ as a function
of $k$ when the angular speed $\Omega$ takes various values, as
shown in Fig.~\ref{RIS}. For the $\Delta_{{F}}<0$ case, a blue
shift of the $g^{(2)}(0)$ curve can be clearly seen in
Fig.~\ref{RIS}(a). For the $\Delta_{{F}}>0$ case, a red shift can
be seen in Fig.~\ref{RIS}(b). This indicates a highly-tunable
nonreciprocal PB device, i.e., {\emph{sub-Poissonian} light can be
achieved by driving from one side; \emph{super-Poissonian} light}
emerges by driving from the opposite side (see Fig.~2 in the main
article).

\begin{figure}[t]
\centering
\includegraphics[width=0.98 \textwidth]{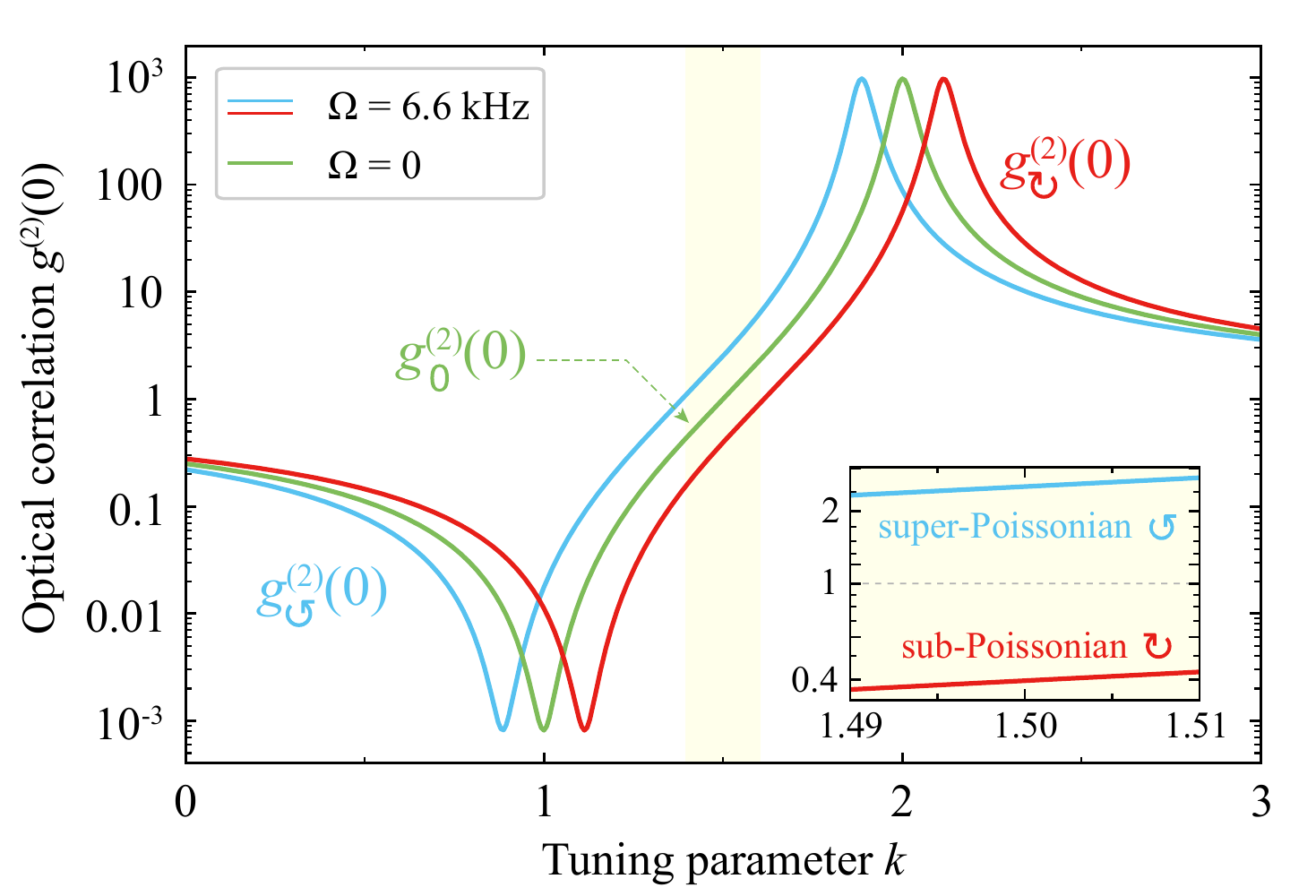}
\caption{{Second-order correlation function $g^{(2)}(0)$ versus
the tuning parameter $k$ for different input directions. At
$k=1.5$, sub- and super-Poissonian light can be achieved by
driving the device from its left (red curve) and right (blue
curve) sides, respectively. Here, we assume that the angular
velocity is $\Omega=6.6~\mathrm{kHz}$~\cite{Smaayani2018flying}
($\Omega=0$) for the spinning (nonspinning) resonator. The other
parameter values are the same as those in the main text.}}
\label{figR1}
\end{figure}

{For example, let us now fix the CCW rotation of the resonator;
hence $\Delta_{F}>0$ ($\Delta_{F}<0$) corresponds to the situation
of driving the resonator from its left (right) side, i.e., the CW
(CCW) mode frequency is
$\omega_\circlearrowright\equiv\omega_0+|\Delta_{F}|$
($\omega_\circlearrowleft\equiv\omega_0-|\Delta_{F}|$), as
aforementioned. When the optical resonator rotates with an angular
velocity $\Omega=6.6~\mathrm{kHz}$~\cite{Smaayani2018flying}, we
find $g^{(2)}_\circlearrowright(0)\sim0.39$ and
$g^{(2)}_\circlearrowleft(0)\sim2.53$, i.e., \emph{sub-Poissonian}
light can be achieved by driving the device from its left side,
while \emph{super-Poissonian} light emerges by driving from the
right side, as shown in Fig.~{\ref{figR1}}.}

The third-order correlation function Eq.~(\ref{g3analytical}) in
the rotating resonator becomes
\begin{align}
g_{\pm}^{(3)}(0) &
=\frac{[(\Delta_{L}\pm\left|\Delta_{{F}}\right|)^{2}+{\gamma^{2}}/{4}]^{2}}{[(\Delta_{L}\pm\left|\Delta_{{F}}\right|+U)^{2}+\gamma^{2}/{4}][(\Delta_{L}\pm\left|\Delta_{{F}}\right|+2U)^{2}+{\gamma^{2}}/{4}]},\label{g3pm}
\end{align}
where $g_{-}^{(3)}(0)$ ($g_{+}^{(3)}(0)$) denotes the third-order
optical intensity correlation for the $\Delta_{{F}}<0$
($\Delta_{{F}}>0$) case. Similarly, the curve of $g^{(3)}(0)$ also
experiences opposite shifts for different driving directions.

\subsection{Nonreciprocal photon blockade}

We have investigated PB effects (witnessing sub-Poissonian light)
and photon-induced tunneling (PIT, corresponding to
{super-Poissonian light)} for the nonspinning case in the former
Sections. Note that PB and PIT always emerge at fixed locations of
the tuning parameter $k$, no matter if the input-laser comes from
the left or right side (see Figs.~\ref{1PB} and~\ref{2PB}).
However, the rotation of the resonator can lead to upper or lower
shifts of energy levels for different driving directions, as
discussed in Sec.~\ref{S_RIS}. Therefore, using a spinning
nonlinear optical resonator, under the same driving frequencies,
PIT can emerge by driving from one side and 1PB/2PB can emerge by
driving from the other direction, i.e., \emph{unidirectional}
1PB/2PB. Furthermore, 1PB for driving from one side and 2PB for
driving from the opposite direction can also be realized with this
spinning device.

\begin{figure}[tbp]
\centering
\includegraphics[width=0.98 \textwidth]{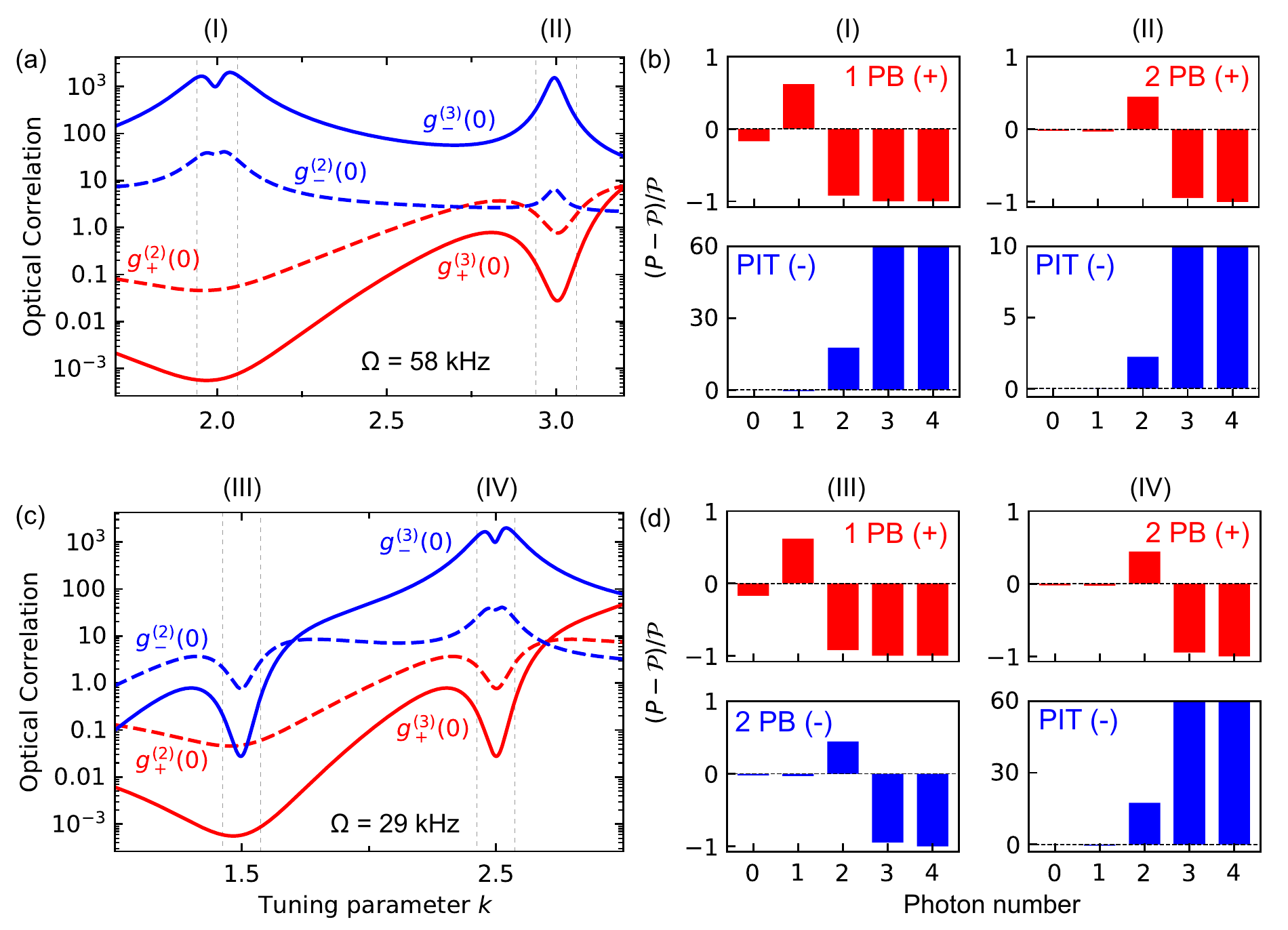}
\caption{Optical intensity correlation functions $g^{(2)}_\pm(0)$
(dashed curves) and $g^{(3)}_\pm(0)$ (solid curves) versus the
tuning parameter $k$ for different driving directions. Different
cases of nonreciprocal PB can be achieved for different angular
speeds (a,b) $\Omega=58\,\mathrm{kHz}$ and (c,d)
$\Omega=29\,\mathrm{kHz}$. These effects can also be recognized
from (b,d) the deviations of the photon distribution to the
standard Poissonian distribution with the same mean photon number
[i.e., Eq.~(\ref{deviation})]. The panels (b) and (d) correspond
to panels (a) and (b), respectively. Here, `PIT' is photon-induced
tunneling, and the other parameters used here are the same as
those in Fig.~\ref{2PB}.} \label{NRPB}
\end{figure}

As shown in Figs.~\ref{NRPB}(a) and~\ref{NRPB}(b), when the
angular speed of the resonator is $\Omega=58\,\mathrm{kHz}$, we
find (i) 1PB for $\Delta_{{F}}>0$ and PIT for $\Delta_{{F}}<0$, at
$k=2.0$; (ii) 2PB for $\Delta_{{F}}>0$ and PIT for
$\Delta_{{F}}<0$, at $k=3.0$. These nonreciprocal 1PB and 2PB can
also be confirmed by comparing the photon-number distribution
$P(n)$ with the Poissonian distribution $\mathcal{P}(n)$.
Figure~\ref{NRPB}(b) shows that: (i) single-photon probability
$P(1)$ is enhanced while two- and more-photon probabilities
$P(m>1)$ are suppressed for the $\Delta_{{F}}>0$ case, leading to
1PB; in contrast, $P(1)$ is suppressed while $P(m>1)$ are enhanced
for the $\Delta_{{F}}<0$ case, leading to PIT. (ii) only
two-photon probability $P(2)$ is enhanced for $\Delta_{{F}}>0$,
which corresponds to 2PB; in contrast, PIT emerges for
$\Delta_{{F}}<0$. The unidirectional 2PB can also be achieved at
$k=2.5$ when $\Omega=29\,\mathrm{kHz}$, as shown in
Figs.~\ref{NRPB}(c) and \ref{NRPB}(d). Such \emph{quantum}
nonreciprocities indicate one-way quantum devices at the
few-photon level, and open up exciting prospects for applications
in nonreciprocal quantum technologies, such as nonreciprocal
quantum information processing or few-photon topological
devices~\cite{Sbennett2000quantum,
Sbuluta2011natural,Slodahl2017chiral}.

More interestingly, when the angular speed of the nonlinear
optical resonator is $\Omega=29$~kHz, 2PB emerges at $k=1.5$ for
$\Delta_{{F}}<0$, while 1PB emerges with the same driving strength
for $\Delta_{{F}}>0$, as shown in Figs.~\ref{NRPB}(c)
and~\ref{NRPB}(d). In contrast to the nonreciprocities of the
former cases between the {sub- and super-Poissonian states of
light}, this is a new kind of nonreciprocal PB between two
{sub-Poissonian states of light}, indicating possible applications
for few-photon nonreciprocal devices with direction-dependent
counting-statics.

All of the cases of nonreciprocal PB can be intuitively understood
by considering the energy-level structure of the system. As shown
in Fig.~\ref{NRPBmechanism}(a), for the $\Delta_{{F}}>0$ case,
when angular speed fulfills $|\Delta_{{F}}|=U$ and the probe light
with frequency $\omega_0+|\Delta_{{F}}|$ ($k=2.0$), the light is
resonantly coupled to the transition $|0\rangle\to|1\rangle$. The
transition $|1\rangle\to|2\rangle$ is detuned by $2\hbar U$ and,
thus, suppressed for $U>\gamma$, i.e., once, a photon is coupled
into the resonator, it suppresses the probability of the second
photon with the same frequency going into the resonator. In
contrast, for the $\Delta_{{F}}<0$ case, there is a three-photon
resonance with the transition $|0\rangle\to|3\rangle$, hence the
absorption of the first photon favors also that of the second or
subsequent photons, i.e., resulting in PIT. This is a clear
signature of nonreciprocal 1PB, i.e., {\emph{sub-Poissonian} light
emerges for $\Delta_{{F}}>0$, while \emph{super-Poissonian} light}
can be observed for $\Delta_{{F}}<0$.

As shown in Figs.~\ref{NRPBmechanism}(c) [\ref{NRPBmechanism}(e)],
for the $\Delta_{{F}}>0$ case, by choosing $|\Delta_{{F}}|=U$
($|\Delta_{{F}}|=U/2$) and $\Delta_L=-2U$ ($\Delta_L=-3U/2$), the
transition $|0\rangle\to|2\rangle$ is resonantly driven by the
input laser, but the transition $|2\rangle\to|3\rangle$ is detuned
by $4\hbar U$, which features the 2PB effect; in contrast, for the
$\Delta_{{F}}<0$ case, four-photon resonance (three-photon
resonance) happens for the transition $|0\rangle\to|4\rangle$
($|0\rangle\to|3\rangle$), leading to PIT. This is also a
nonreciprocal PB.

As shown in Fig.~\ref{NRPBmechanism}(g), for the $\Delta_{{F}}>0$
case, when $|\Delta_{{F}}|=U/2$ and $\Delta_L=-U/2$ ($k=1.5$), the
input light is resonantly coupled to the transition
$|0\rangle\to|1\rangle$, and the transition
$|1\rangle\to|2\rangle$ is detuned by $2\hbar U$, leading to 1PB.
More interestingly, for the $\Delta_{{F}}<0$ case, the input light
is just resonantly coupled to the transition
$|0\rangle\to|2\rangle$, and the transition
$|2\rangle\to|3\rangle$ is detuned by $4\hbar U$, i.e., resulting
in 2PB. This 1PB-2PB nonreciprocity can suggest an application for
a purely quantum device with \emph{direction-dependent counting
statistics}. This new nonreciprocal feature, which (to our
knowledge) has not been revealed previously.

\begin{table}[tbp]
\renewcommand\arraystretch{1.5}
 \caption{\label{NRPBcases}Different cases of nonreciprocal PB effects in a spinning resonator for $P_{\text{in}}=0.3\,\mathrm{pw}$. Here, photon-induced tunneling (PIT) corresponds to an $n$-photon resonance ($n$ PR).}
 \begin{tabular}{p{17cm}}
 \rowcolor{darkblue}
   \ \ No.\quad\ \ $\Delta_{F}>0$\qquad\quad$\Delta_{F}<0${\qquad}{\qquad}{\qquad}Conditions{\qquad}{\qquad}{\qquad}{\qquad}{\qquad}Parameters\\
\rowcolor{lightblue}
  \ \ (1)\quad\ \ \ 1PB{\qquad}{\qquad}\ \ \ PIT (3PR)\qquad\qquad\ \ \ $\Delta_{{F}}=\pm U$, $\Delta_L=-U${\qquad}{\qquad}\ \ \ $\Omega=58\,\mathrm{kHz}$, $k=2.0$ \\
\rowcolor{lightblue}
  \ \ (2)\qquad PIT (3PR){\quad}\ \ \ 1PB{\qquad}{\qquad}{\qquad}{\quad}\ \ \ prohibited \\
 \rowcolor{lightblue}
  \ \ (3)\quad\ \ \ 2PB{\qquad}{\qquad}\ \ \ PIT (4PR)\qquad\qquad\ \ \ $\Delta_{{F}}=\pm U$, $\Delta_L=-2U${\qquad}{\qquad}\ \,$\Omega=58\,\mathrm{kHz}$, $k=3.0$ \\
  \rowcolor{lightblue}
  \ \ (4)\qquad PIT (4PR){\quad}\ \ \ 2PB{\qquad}{\qquad}{\qquad}{\quad}\ \ \ prohibited \\
 \rowcolor{lightblue}
  \ \ (5)\quad\ \ \ 2PB{\qquad}{\qquad}\ \ \ PIT (3PR)\qquad\qquad\ \ \ $\Delta_{{F}}=\pm U/2$, $\Delta_L=-3U/2${\qquad}\ \ $\Omega=29\,\mathrm{kHz}$, $k=2.5$ \\
   \rowcolor{lightblue}
  \ \ (6)\qquad PIT (3PR){\quad}\ \ \ 2PB{\qquad}{\qquad}{\qquad}{\quad}\ \ \ prohibited \\
 \rowcolor{lightblue}
  \ \ (7)\quad\ \ \ 1PB{\qquad}{\qquad}\ \ \,2PB \qquad\qquad\qquad\quad\ \ $\Delta_{{F}}=\pm U/2$, $\Delta_L=-U/2${\qquad}\ \ \ \ $\Omega=29\,\mathrm{kHz}$, $k=1.5$ \\
     \rowcolor{lightblue}
  \ \ (8)\quad\ \ \ 2PB{\qquad}{\qquad}\ \ \,1PB{\qquad}{\qquad}{\qquad}{\quad}\ \ \ \,prohibited \\
  \end{tabular}
\end{table}

Table~\ref{NRPBcases} shows different cases of nonreciprocal PB.
Interestingly, both PB-PIT and 1PB-2PB nonreciprocities can only
occur in an irreversible way. Unidirectional 1PB for
$\Delta_{{F}}>0$, i.e., 1PB emerges for $\Delta_{{F}}>0$ and PIT
emerges for $\Delta_{{F}}<0$, can occur with the same angular
speeds ($\Delta_{{F}}=\pm U$), and the same driving frequencies
($\Delta_L=-U$). However, the case of PIT for $\Delta_{{F}}>0$ and
1PB for $\Delta_{{F}}<0$ cannot be observed with the same angular
speeds and driving frequencies, i.e., one-way 1PB is an
irreversible quantum nonreciprocal effect. Also, 1PB-2PB
nonreciprocity can only happen in the case of 1PB for
$\Delta_{{F}}>0$ and 2PB for $\Delta_{{F}}<0$, but $not$ vice
versa.

Note that 1PB and 2PB correspond to the single- and two-photon
resonances, respectively. PIT is also caused by a multi-photon
resonance. The multi-photon resonance can be clearly seen in
energy-level diagrams, thus, the origin of this irreversible
feature can be understood from the energy-level diagrams for
$\Delta_{{F}}>0$ and $\Delta_{{F}}<0$. Without the rotation, the
energy-level diagrams for the $\Delta_{{F}}>0$ and
$\Delta_{{F}}<0$ cases are symmetric. Due to the rotation, energy
levels experience shifts to different directions for
$\Delta_{{F}}>0$ and $\Delta_{{F}}<0$, leading to asymmetries of
energy-level diagrams, as shown in Fig.~\ref{NRPBmechanism}. From
Sec.~\ref{OPB}, the energy levels of this spinning system are
$E_{n}=n\hbar\Delta_{L}+n\hbar\Delta_{F}+(n^{2}-n)\hbar U$. Thus,
we have
\begin{equation}
E_{n}/n=\hbar(\Delta_{L}+\Delta_{F})+(n-1)\hbar U.
\end{equation}
Then the driving frequency of an $n$-photon resonance for the
$\Delta_{{F}}>0$ case is
\begin{equation}
\omega_L=\omega_{0}+|\Delta_{F}|+nU-U,
\end{equation}
and the driving frequency of an $m$-photon resonance for the
$\Delta_{{F}}<0$ case is
\begin{equation}
\omega_L'=\omega_{0}-|\Delta_{F}|+mU-U.
\end{equation}
Under the same driving frequency, we have
\begin{align}
\omega_{0}+|\Delta_{F}|+nU-U&=\omega_{0}-|\Delta_{F}|+mU-U \nonumber \\
|\Delta_{F}|+nU&=-|\Delta_{F}|+mU \nonumber \\
2|\Delta_{F}|&=(m-n)U.
\end{align}
Because $|\Delta_{F}|>0$ (i.e., $\Omega\neq0$) and $U>0$, we have
the following condition for the allowed cases of nonreciprocal PB
\begin{equation}
n<m.
\end{equation}
When the driving frequencies for $\Delta_{{F}}>0$ and
$\Delta_{{F}}<0$ are the same, an $n$-photon resonance for
$\Delta_{{F}}>0$ and an $m$-photon resonance for $\Delta_{{F}}<0$
can only happen under the condition $n<m$. In contrast to this,
the cases of $n>m$ are prohibited, as shown in
Figs.~\ref{NRPBmechanism}(b),~\ref{NRPBmechanism}(d),
\ref{NRPBmechanism}(f), and~\ref{NRPBmechanism}(h).

\vspace{8mm} \small{$^*$Corresponding author.

$~$$\,$jinghui73@foxmail.com}


%

\end{document}